\documentclass{aa} 

\usepackage{natbib} 
\usepackage{graphicx}
\usepackage{txfonts}
\usepackage{hyperref}
\usepackage{siunitx}

\newcommand{\Teff}{\ensuremath{T_{\rm eff}}}                      
\newcommand{\logg}{\ensuremath{\log g}}                           

\newcommand{\cmmnt}[1]{}

\newcommand{\orcid}[1]{\protect\href{https://orcid.org/#1}{\protect\includegraphics[width=8pt]{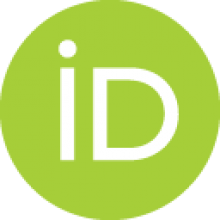}}}

\makeatletter
\renewcommand*\aa@pageof{, page \thepage{} of \pageref*{LastPage}}
\makeatother

\newcounter{lenote}

\makeatletter
\renewcommand*\maketitle{%
  \thispagestyle{firstpage}
\begingroup
    \if@wideboxfn
    \setlength\bibindent{1.4\parindent}
    \else
    \setlength\bibindent{\parindent}
    \fi
    \renewcommand*\thefootnote{\@fnsymbol\c@footnote}%
    \renewcommand\@makefntext[1]{%
    \ifaa@longfn\hsize\textwidth\fi
    \noindent
    \hb@xt@\bibindent{\hss\@makefnmark\enspace}##1}
  \ifaa@twocolumn
  \begin{aa@strip}
    \aa@maketitle
    \@thanks
  \end{aa@strip}
  \else
    \begingroup
      \let\thanks\footnote
      \aa@maketitle
    \endgroup
  \fi
\endgroup
  \setcounter{footnote}{0}%
}
\makeatother

\begin{document}

\title{Absolute dimensions of solar-type eclipsing binaries}

\subtitle{NY\,Hya: A test for magnetic stellar evolution models\thanks{Partially based on observations
carried out with the Str\"omgren Automatic Telescope (SAT), the FEROS spectrograph at the 1.52m and Danish 1.54m telescopes at ESO, La Silla, Chile.}}

\author{T. C. Hinse$^{\orcid{0000-0001-8870-3146}}$ \inst{1 \thanks{\email{tchinse@gmail.com}} }
       \and
       {\"O}. Ba{\c s}t{\"u}rk$^{\orcid{0000-0002-4746-0181}}$ \inst{2,3}
       \and
       J. Southworth$^{\orcid{0000-0002-3807-3198}}$ \inst{4}
       \and
       G. A. Feiden$^{\orcid{0000-0002-2012-7215}}$ \inst{5}
       \and
       J. Tregloan-Reed$^{\orcid{0000-0002-9024-4185}}$ \inst{6}
       \and
       V. B. Kostov$^{\orcid{0000-0001-9786-1031}}$ \inst{7,8}
       \and
       J. Livingston$^{\orcid{0000-0002-4881-3620}}$ \inst{9,10,11}
       \and
       E. M. Esmer$^{\orcid{0000-0002-6191-459X}}$ \inst{2,3}
       \and
       Mesut Y{\i}lmaz$^{\orcid{0000-0002-3276-0704}}$ \inst{2,3}
       \and
       Sel\c{c}uk Yal\c{c}{\i}nkaya$^{\orcid{0000-0002-5224-247X}}$ \inst{2,3,12}
       \and
       \c{S}eyma Torun \inst{12} 
       \and
       J. Vos$^{\orcid{0000-0001-6172-1272}}$ \inst{13}
       \and
       D. F. Evans$^{\orcid{0000-0002-5383-0919}}$ \inst{4}
       \and
       J. C. Morales$^{\orcid{0000-0003-0061-518X}}$ \inst{14,15}
       \and
       J. C. A. Wolf$^{\orcid{0000-0002-7755-6504}}$ \inst{16}
       \and
       E. H. Olsen \inst{16}
       \and
       J. V. Clausen \inst{16}
       \and
       B. E. Helt \inst{16}
       \and
       C. T. K. L{\'y}$^{\orcid{0000-0002-1887-6858}}$ \inst{17}
       \and
       O. Stahl$^{\orcid{0000-0001-5455-9967}}$ \inst{18}
       \and
       R. Wells$^{\orcid{0000-0002-7240-8473}}$ \inst{19}
       \and
       M. Herath$^{\orcid{0009-0004-3980-8143}}$ \inst{20,21}
       \and
       U. G. J{\o}rgensen$^{\orcid{0000-0001-7303-914X}}$ \inst{22}
       \and
       M. Dominik$^{\orcid{0000-0002-3202-0343}}$\inst{23}
       \and
       J. Skottfelt$^{\orcid{0000-0003-1310-8283}}$\inst{24}
       \and
       N. Peixinho$^{\orcid{0000-0002-6830-476X}}$ \inst{25}
       \and
       P. Longa-Pe{\~n}a$^{\orcid{0000-0001-9330-5003}}$ \inst{26}
       \and
       Y. Kim \inst{27}
       \and
       H.-E. Kim \inst{27}
       \and
       T. S. Yoon \inst{28}
       \and
       H. I. Alrebdi$^{\orcid{0000-0002-2271-1060}}$ \inst{29}
       \and
       E. E. Zotos$^{\orcid{0000-0002-1565-4467}}$ \inst{30}
       }

\institute{
          University of Southern Denmark, Department of Physics, Chemistry and Pharmacy, SDU-Galaxy, Campusvej 55, 5230 Odense M, Denmark
          \and
          Ankara University, Faculty of Science, Department of Astronomy and Space Sciences, 
          Tando{\c g}an, TR-06100, Ankara, T\"urkiye
          \and
          Ankara University, Astronomy and Space Sciences Research and Application Center (Kreiken Observatory),{\.I}ncek Blvd., TR-06837, Ahlatl{\i}bel, Ankara, T\"urkiye
          \and
          Astrophysics Group, Keele University, Staffordshire, ST5 5BG, UK
          \and
          Department of Physics \& Astronomy, University of North Georgia, Dahlonega, GA 30597, USA
          \and
          Instituto de Investigaci{\'o}n en Astronomia y Ciencias Planetarias, Universidad de 
          Atacama, Avenida Copayapu 485, Copiap{\'o}, Atacama, Chile
          \and
          NASA Goddard Space Flight Center, 8800 Greenbelt Road, Greenbelt, MD 20771, USA
          \and
          SETI Institute, 189 Bernardo Ave, Suite 200, Mountain View, CA 94043, USA
          \and
          Astrobiology Center, NINS, 2-21-1 Osawa, Mitaka, Tokyo 181-8588, Japan
          \and
          National Astronomical Observatory of Japan, NINS, 2-21-1 Osawa, Mitaka, Tokyo 181-8588, Japan
          \and
          Astronomical Science Program, Graduate University for Advanced Studies, SOKENDAI, 2-21-1, Osawa, Mitaka, Tokyo, 181-8588, Japan
          \and
          Ankara University, Graduate School of Natural and Applied Sciences,  Department of Astronomy and Space Sciences, Tando\u{g}an, TR-06100, Ankara, T\"urkiye
          \and
          Astronomical Institute of the Czech Academy of Sciences, CZ-251\,65, Ond\v{r}ejov, Czech Republic
          \and
          Institut de Ci{\`e}ncies de l'Espai (ICE, CSIC), Campus UAB, c/ Can Magrans s/n, E-08193 Bellaterra (Barcelona), Spain
          \and
          Institut d'Estudis Espacials de Catalunya (IEEC), Edifici RDIT, Campus UPC, E-08860 Castelldefels (Barcelona), Spain
          \and
          Niels Bohr Institute, University of Copenhagen, Jagtvej 155, DK-2200 Copenhagen, Denmark {\O}
          \and
          Chungnam National University, Department of Astronomy, Space Science and Geology, Daejeon, Republic of Korea.
          \and
          Landessternwarte, Zentrum f{\"u}r Astronomie der Universit{\"a}t Heidelberg, D-69117 Heidelberg, Germany
          \and
          Center for Space and Habitability, University of Bern, Gesellschaftsstrasse 6, 3012, Bern, Switzerland
          \and
          McGill Space Institute, McGill University, 3550 University Street, Montreal, QC H3A 2A7
          \and
          Department of Astronomy, Arthur C. Clarke Institute for Modern Technologies, 0272 Moratuwa, Sri Lanka
          \and
          Centre for ExoLife Sciences, Niels Bohr Institute, University of Copenhagen, {\O}ster Voldgade 5, DK-1350 Copenhagen, Denmark
          \and
          University of St Andrews, Centre for Exoplanet Science, SUPA School of Physics \& Astronomy, North Haugh, St Andrews, KY16 9SS, United Kingdom
          \and
          Centre for Electronic Imaging, School of Physical Sciences, The Open University, Milton Keynes MK7 6AA, UK
          \and
          Instituto de Astrof\'{\i}sica e Ci\^{e}ncias do Espa\c{c}o, Departamento de F\'{\i}sica, Universidade de Coimbra, PT3040-004 Coimbra, Portugal
          \and
          Centro de Astronom{\'i}a, Universidad de Antofagasta, Avenida Angamos 601, Antofagasta 1270300, Chile
          \and
          Chungbuk National University Observatory, Chungbuk National University, 28644 Cheongju, South Korea
          \and
          Department of Astronomy and Atmospheric Sciences, Kyungpook National University, 41566 Daegu, South Korea
          \and
          Department of Physics, College of Science, Princess Nourah bint Abdulrahman University, P.O. Box 84428, Riyadh 11671, Saudi Arabia
          \and
          Department of Physics, School of Science, Aristotle University of Thessaloniki, GR-541 24, Thessaloniki, Greece
          }

\date{Received March 1, 2021; accepted June 01, 2021}

\titlerunning{Absolute dimensions of the eclipsing binary NY\,Hya}

\abstract
   {The binary star NY\,Hya is a bright, detached, double-lined eclipsing system with an orbital period of just under five days with two components each nearly identical to the Sun and located in the solar neighbourhood.}
   {The objective of this study is to test and confront various stellar evolution models for solar-type stars based on accurate measurements of stellar mass and radius.}
   {We present new ground-based spectroscopic and photometric as well as high-precision space-based
   photometric and astrometric data from which we derive orbital as well as physical properties of the
   components via the method of least-squares minimisation based on a standard binary model valid for two
   detached components. Classic statistical techniques were invoked to test the significance of 
   model parameters. Additional empirical evidence was compiled from the public domain; the derived system
   properties were compared with archival broad-band photometry data enabling a measurement of the system's
   spectral energy distribution that  allowed an independent estimate of stellar properties. We also utilised semi-empirical calibration methods to derive atmospheric properties from Str\"omgren photometry and related colour indices.}
   {We measured (percentages are fractional uncertainties) masses, radii, and effective temperatures of the two stars in NY\,Hya and found them to be $M_A = 1.1605 \pm 0.0090\,M_{\sun}\,(0.78\%)$, $R_A =
   1.407 \pm 0.015\,R_{\sun}\,(1.1\%)$, $T_{\rm eff, A} = 5595 \pm 61$\,K\,(1.09\%), $M_B = 1.1678 \pm
   0.0096\,M_{\sun}\,(0.82\%)$, $R_B = 1.406 \pm 0.017\,R_{\sun}\,(1.2\%)$, and $T_{\rm eff, B} = 
   5607 \pm 61$\,K\,(1.09\%). The atmospheric properties from Str\"omgren photometry agree well with
   spectroscopic results. No evidence was found for nearby companions from high-resolution imaging. A
   detailed analysis of space-based data revealed a small but significant eccentricity $(e\cos\omega)$ of
   the orbit. The spectroscopic and  frequency analysis on photometric time series data reveal evidence of clear photospheric activity on both components likely in the form of star spots caused by magnetic activity.}
  {We confronted the observed physical properties with classic and magnetic stellar evolution models. Classic models yielded both young pre-main-sequence and old main-sequence turn-off solutions with the two components at super-solar metallicities, in disagreement with observations. Based on chromospheric activity and X-ray observations, we invoke magnetic models. While magnetic fields are likely to play an important role, we still encounter problems in explaining adequately the observed properties. To reconcile the observed tensions we also considered the effects of star spots known to mimic magnetic inhibition of convection. Encouraging results were obtained, although  unrealistically large spots were required  on each component. Overall we conclude that NY\,Hya proves to be complex in nature, and  requires additional follow-up work aiming at a more accurate determination of stellar effective temperature and metallicity.}

\keywords{stars: fundamental parameters -- binaries: eclipsing -- stars: individual: HD\,80747, 
HD\,82074, HD\,80446, HD\,80633 -- stars: solar-type -- stars: binaries: spectroscopic}
   
\maketitle
\section{Introduction}
\label{introduction}
Stellar components in a detached eclipsing binary (dEB) system provide a direct empirical measure of their
physical properties with a minimum number of assumptions. From photometric and double-lined spectroscopic data, the masses and radii can be determined providing an empirical test of stellar evolution models. The most strict model tests are for those systems where elemental abundances are determined spectroscopically, reducing the number of free parameters to the fractional helium content, age and model-internal parameters \citep{andersen1991}. Secure astrophysical information about system properties is further substantiated with recent space-based astrometric distance measurements \citep{StassunTorres:2021} allowing a reliable measurement of system properties without limiting assumptions; in pre-\textit{Gaia} times nearby eclipsing binaries were used as reliable distance estimators. Detached systems are therefore of fundamental astrophysical importance and constitute the primary source of knowledge of properties of individual stars to form the test bed when confronting stellar evolution theory. 
From high-quality photometry and spectroscopy, the stellar masses and radii can be determined to better than 1\% precision from in-depth analyses of eclipse light curves and double-lined radial velocity (RV) curves with good phase coverage \citep{torres1997,southworth2005b}.
The reliable measurement of accurate stellar properties is further increased for systems located in the solar neighbourhood where interstellar reddening effects are small.

\begin{table*}[hbt!]
\caption{Photometric data for NY\,Hya and the comparison stars HD\,82074, HD\,80446, and HD\,80633.}
\label{nyhya_photometric_data}
\centering          
\begin{tabular}{llllrrrrrrrrrrrr}    
\hline
  Object        & Phase   & Spec. Type  &    $V$ &   $\sigma$   &       $b-y$   &   $\sigma$      &       $m_1$   &   $\sigma$    &       $c_1$   &    $\sigma$   &   $N$($uvby$)   &       $\beta$ &   $\sigma$    &     $N$($\beta$)\\

\hline
\hline
NY\,Hya         & 0.000   &             &       8.874   &       16      &       0.444   &       4       &       0.246   &       5       &       0.353   &       7       &       69      &               &               &               \\
                & 0.250   & G5          &       8.584   &       16      &       0.444   &       4       &       0.245   &       7       &       0.350   &       7       &       42      &       2.598   &       9       &       54      \\
                & 0.500   &             &       8.874   &       13      &       0.445   &       4       &       0.244   &       7       &       0.352   &       8       &       38      &               &               &               \\
\hline
HD\,82074       &         & G5  &       6.253   &       6       &       0.504   &       3       &       0.287   &       6       &       0.306   &       6       &       592     &       2.557   &       6       &       28      \\
HD\,80446       &         & G5          &       9.144   &       7       &       0.381   &       5       &       0.199   &       8       &       0.314   &       9       &       499     &       2.596   &       6       &       24      \\
HD\,80633       &         & F8          &       7.388   &       6       &       0.307   &       4       &       0.162   &       7       &       0.424   &       7       &       472     &       2.643   &       9       &       29      \\
\hline                  
\end{tabular} 
\tablefoot{For reasons of completeness we reproduce the data from \cite{clausen2001}, but with an updated spectral type for HD\,82074 from the Henry Draper Catalog \citep{cannonpickering1918}. Johnson $V$ apparent magnitudes and Str{\"o}mgren $(b-y), m_1, c_1$, and $\beta$ colour indices are in the standard photometric systems (not corrected for interstellar extinction or reddening). The unit is magnitudes (mag). Uncertainties $(\sigma)$ are in units of milli-magnitudes ($m$mag). For details on the transformation we refer to \cite{olsen1994} and references therein. For NY\,Hya the $uvby$ measurements are taken at maximum light level outside eclipse (phase 0.25), at primary eclipse (phase 0.00), and at secondary eclipse (phase 0.50). The $\beta$ measurement is the mean value taken for measurements outside eclipses (cf. fig. 1 in \citealt{clausen2001}). $N$ is the total number of observations used to form the mean and $\sigma$ is the root mean square (per observation) in units of $m$mag. We note that fig. 1 in \citet{clausen2001} shows a single $\beta$ measurement during the primary eclipse and five during the secondary.}
\end{table*}

As pointed out by \cite{southworthclausen2007}, stellar evolution models are generally good at predicting the physical properties of stars (see \citealt{maxted2015}) for two reasons. First, the effects of improved input physics often result in only small adjustments in the evolutionary phases for the two components. Second, some model parameters that are not known independently (or are harder to infer empirically) can be freely adjusted in order to match the observed properties. The additional parameters are helium abundance and age, as well as input-physics parameters describing convective core overshooting, mixing length, mass loss, and various treatments of opacity \citep{cassisi2005}. While some parameters, such as stellar abundances (from a careful abundance analysis) and age (in the case of cluster membership or by other means such as chromospheric activity, gyrochronology, or depletion of certain age-dependent species) can be inferred empirically, thus providing additional constraints on the investigated model, this is not always the case. Therefore, the second aspect is worrying, since it is hard to identify and test physical mechanisms governing stellar evolution given the extra degrees of freedom in adjusting and tweaking one or more parameters.

For lower-mass detached eclipsing binary stars, discrepancies between the observed properties and model predictions have been noted in the literature for stars of spectral types G to M \citep{popper1997,torresribas2002,ribas2003,clausen2009,Me22obs6}. This implies that the principles of single-star stellar evolution do not fully apply to stars in a detached eclipsing system; often the predicted effective temperature (\Teff) are too high for the observed masses or, for a fixed age, the predicted radii are too small, whilst masses and radii are measured with fractional uncertainties to 1\% or better. Further, in the case when no independent age constraint is available, the models fail in placing the two components on the same isochrone in the stellar radius-mass ($R-M$) plane.

In an attempt to reconcile certain stellar evolution models with observations, the usual approach is the computation of dedicated model grids for which the mixing length parameter ($l/H_p$) is adjusted for each component and compared to observations. Often this results in a success in matching the observed properties \citep{vos2012}, but highlights the aforementioned worries: input physics parameters are adjusted until a good match was found, which concludes the analysis. However, while the observations, within observational errors, were adequately described, the explanation was only described at a phenomenological level without properly addressing an adequate physical mechanism.

The short-coming of solar-type evolution models to properly describe solar-type components is to be found in the fast rotation of short-period detached binaries. For short-period eclipsing binaries tidal interactions have resulted in an increased stellar rotation spin-up to a pseudo-synchronous state (1:1 spin-orbit resonance)  \citep{hut1981,li2018}. Rotation velocities ($v\sin i$) for single solar-type stars are of the order of 1 km/s or lower. The rotation velocities of similar stars in an eclipsing system are found to be 10 km/s or higher. The fast rotation is thought to have a significant effect on the convective energy transport \citep{chabrier2007}. Qualitatively, in mixing-length theory, convection is described by the mixing-length parameter and is capable of phenomenologically addressing the often neglected and complicated effects of magnetic fields; for fast rotators a strong magnetic field is generated, which  triggers the Lorentz force to act on the outer material region of the convection zone,   thereby decreasing its convective height. Thus, by adjusting $l/H_p$ one has the option to indirectly describe the effect of a magnetic field on the convective region. However, a more realistic treatment of this problem is to directly include the effect of a magnetic field in stellar evolution models. Magnetic fields are often ignored and, as shown recently, are capable of providing a viable physical mechanism to explain observed discrepancies \citep{feidenchaboyer2012}.

In this work we present and analyse for the first time high-quality light curves ({\it TESS}) and high-resolution spectroscopic (FEROS) observations of NY\,Hya. The bright ($V=8.58$ mag) stellar object HD\,80747 (NY\,Hya, BD-06 2891, HIP\,45887) was discovered by the \textit{Hipparcos} mission \citep{ESA1997} to be an eclipsing binary and subsequently named NY\,Hya by \cite{kazarovets1999}. The \textit{Hipparcos} parallax places NY\,Hya at a distance of $81.8 \pm 8.6$ pc. In the Henry Draper Catalogue, \citet{cannonpickering1918} classified NY\,Hya as a G5 star without any remarks to its binary nature. 

Dedicated photometric follow-up observations in the Str\"omgren system were reported in \cite{clausen2001}, establishing the first accurate ephemeris and announcing a significant update on the orbital period (4.77 days) compared to the period determined from the  \textit{Hipparcos} data. The ephemeris was further updated by \cite{kreiner2004} and \cite{pojmanski1997} with the latter providing instrumental $V$-band data as part of their All Sky Automated Survey (ASAS). The Str\"omgren-Crawford $uvby\beta$ photometry catalogue by \cite{paunzen2015} provides entries for the Str\"omgren indices of NY\,Hya. No $\beta$ measurement was given. The data are based on a single measurement, and are therefore questionable. No reliable uncertainties are provided. Therefore, we suggest that the  Str\"omgren data should be taken from \cite{clausen2001} (see Table \ref{nyhya_photometric_data}). For a $(b-y) = 0.444$ mag (without reddening correction) we determine the spectral type of NY\,Hya to be G6V, or G5V \citep[Table V]{Olsen1984} with a correction for extinction (see Sect.~\ref{interstellarextinctionandreddening}). The Michigan catalogue of HD stars \citep{houk1999} lists a spectral type of G5V.

The eclipsing system NY\,Hya belongs to the class of fast rotators and classic stellar models (as investigated here) are not able to describe the observed physical properties of this system. A closer look at the X-ray emission of NY\,Hya indicates a high chromospheric activity level indicating the presence of a magnetic field. This finding is further supported by a significant intrinsic photometric variability most readily seen during out-of-eclipse phases mostly due to the  changing appearance of star spots. However, significant progress has been made in recent years in the development of stellar evolution models considering the effect of a magnetic field. We therefore apply state-of-the-art evolution models \citep{feidenchaboyer2012} to the observed physical properties, providing a more rigorous benchmark to test formerly neglected input physics.

The paper is structured as follows. In Sect.~\ref{observations} we present new observations including {\it TESS} photometry and high-resolution FEROS spectra. In Sect.~\ref{spectralanalysis} we analyze the spectra to derive empirical spectroscopic light ratios and measure radial velocities. In Sect.~\ref{timesofminimumlightandnewephemeris} we present an updated and improved ephemeris thanks to the {\it TESS} data. A complete analysis of the light and radial velocity curves is presented in Sect.~\ref{Light-andRVcurveanalysis}. Then we determine the physical parameters of the system and its components through a rigorous spectroscopic analysis of the spectra we disentangled in Sect. \ref{physicalproperties}. In Sect.~\ref{spectralenergydistributionmodeling} we compile intermediate- and broad-band photometry from the literature (which we corrected for the interstellar extinction in Sect.~\ref{interstellarextinctionandreddening}) and present the results obtained from an analysis of the spectral energy distribution (SED) considering NY\,Hya as a single star. This provides a mean estimate for the \Teff\ providing an independent consistency check on the spectroscopy.  In Sect.~\ref{stellaractivity} we carry out a period analysis based on the out-of-eclipse variability. We were able to independently determine the orbital period from the presence of surface spots; this  allowed us to conclude synchronous rotation via tidal locking of the two components. In Sect.~\ref{Comparisonwithstellarevolutionmodels} we compare the observed properties of the two components with stellar evolution models. We invoke both classic and     non-standard magnetic models. We substantiate the importance of magnetic fields by providing a semi-empirical estimate of magnetic field strength from observations of the total X-ray luminosity. Finally, we present our conclusions  in Sect.~\ref{Summaryandconclusions}.

\section{Observations}
\label{observations}

We present Str\"omgren photometry from the literature, high-precision and high-cadence space-based observations from the {\it TESS} telescope, high-resolution lucky-imaging, and high-quality spectra from the FEROS spectrograph. Additional archival photometric data are presented in Sect.~\ref{spectralenergydistributionmodeling} as part of a SED analysis.

\subsection{Str{\"o}mgren photometry}
\label{stromgrenphotometry}

Photometric data were obtained by \cite{clausen2001} using the (now decommissioned) 0.5m Str{\"o}mgren Automatic Telescope (SAT) at the ESO/La Silla observatory \citep{fn1987} during two observing seasons: Season 1,  spanning from November 23, 1997 (JD 2,450,775.8), to April 28, 1998 (JD 2,450,931.6), and  Season 2, spanning from October 12, 1998 (JD 2,451,098.9), to March 16, 1999 (JD 2,451,253.7). All comparison stars, relevant photometric data for which are provided in Table \ref{nyhya_photometric_data}, were found to be constant with respect to the check star, HD\,80633. In particular, \cite{lockwood1997} found HD\,82074 to be stable in their long-term photometric variability program of solar-type stars.

The SAT data reduction follows an equivalent procedure as described in \citet{clausen2001,clausen2008}. Here we add a few more details. To form the final photometry for the comparison stars (C1, C2) a $2.5\sigma-$clipping was applied to remove 44 outliers. A total of 19 measurements were discarded by this criterion when forming the C1-C2 magnitude level. Linear extinction coefficients were determined from the comparison stars for each night and apparent magnitudes corrected accordingly. The maximum and minimum root-mean-square errors between the comparison stars was 7.6 $m$mag and 4.6 $m$mag, respectively depending on moon illumination, airmass and sky conditions. Photometric root-mean-square errors for NY\,Hya SAT observations were found to be 3.7 $m$mag ($u$ band) and 2.2 $m$mag ($vby$ bands).

The instrumental $uvby\beta$ system at the SAT was found to be long-term stable \citep{clausen2001} from observing $uvby\beta$ standard stars. This implies that no transformation to the standard $uvby\beta$ system was needed to form differential ($uvby$) light curves in the instrument system having the additional advantage of avoiding additional photometric errors introduced via transformation relations. Atmospheric extinction and detector temperature variations were corrected for \citep{clausen2001}. However, based on numerous bright standard stars observed each night, standard Johnson $V$ and standard Str\"omgren indices $(b-y, m_1, c_1$ and $\beta$) for NY\,Hya were determined at all orbital phases. See Table \ref{nyhya_photometric_data} for details. We note only minuscule colour differences between the two components during primary and secondary eclipses indicating near-equal \Teff s.

Original time stamps (HJD) of all SAT observations refer to the midpoint of the integration time interval. We have transformed the HJD times in the UTC time standard to BJD times in the TDB time standard using the online time conversion\footnote{\url{http://astroutils.astronomy.ohio-state.edu/time/}} provided by \citet{eastman2010}. This was necessary in order to have a consistent time standard as the SAT eclipse minimum timings were later combined with space-based photometry.

\subsection{Space-based {\it TESS} observations}
\label{space-basedTESSobservations}

NY\,Hya was observed by the {\it TESS} (Transiting Exoplanet Survey Satellite) space telescope \citep{TESS,TESS2,TESS3}. NY\,Hya was observed at a 2 min cadence in Sector 8 with Camera 1 and CCD 1. The photometric data were retrieved from MAST\footnote{\url{https://archive.stsci.edu/}} (Mikulski Archive for Space Telescopes) using {\sc lightkurve} \citep{Lightkurve,lightkurve2019}. Since NY\,Hya is a binary system the SAP (Simple Aperture Photometry) light curve was extracted. We discarded photometric measurements with the following data quality flags: `attitude tweak', `safe mode', `coarse point', `earth point', `desat', and `manual exclude' (see {\it TESS} science data products description\footnote{\url{https://archive.stsci.edu/missions/tess/doc/EXP-TESS-ARC-ICD-TM-0014.pdf}}). The SAP light curve is produced by {\it TESS} SPOC (Science Processing Operations Center) \citep{Jenkins2016} designed for producing optimal aperture photometry. 

\begin{figure}
\includegraphics[width=0.48\textwidth,angle=0]{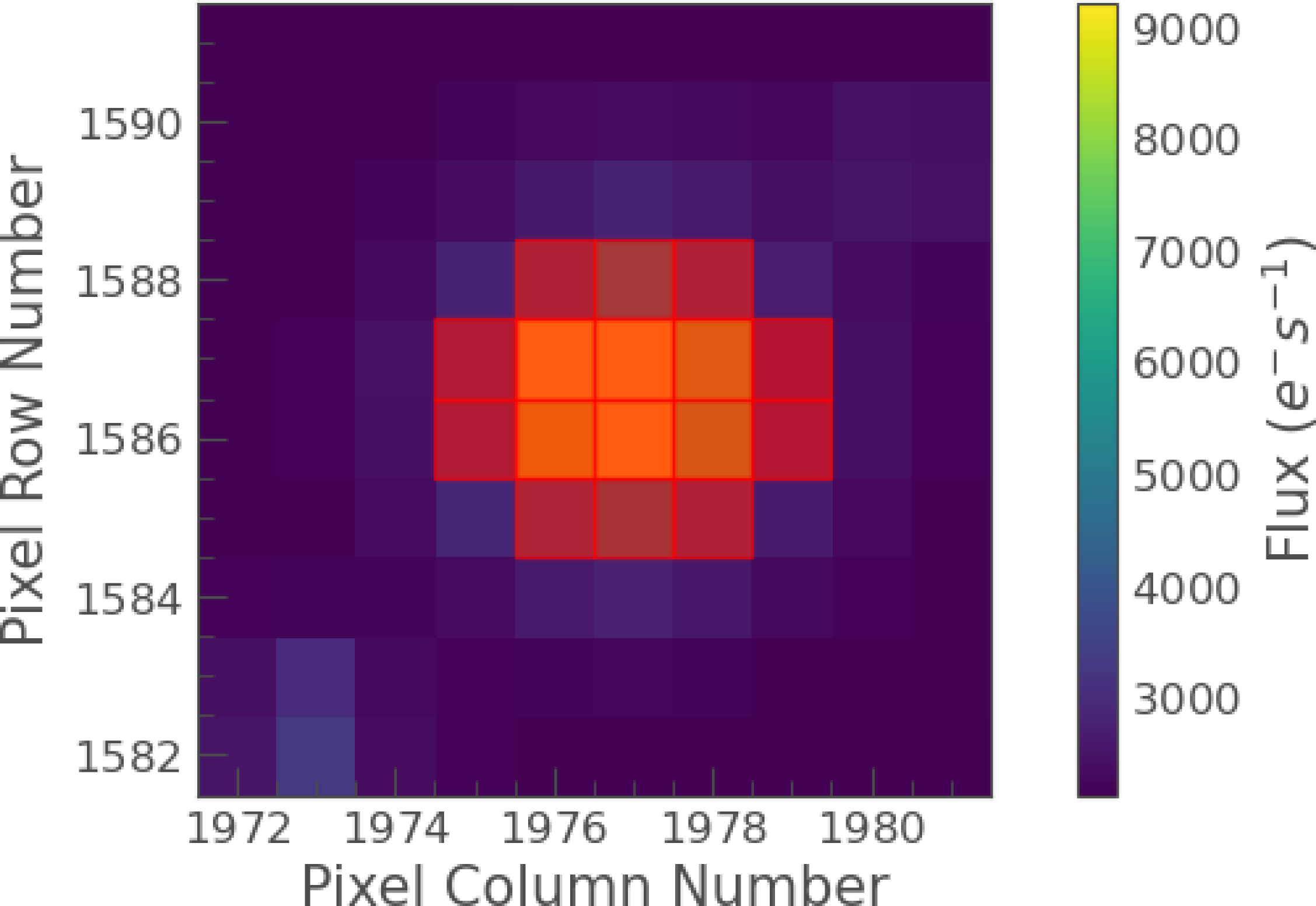}
\includegraphics[width=0.48\textwidth,angle=0]{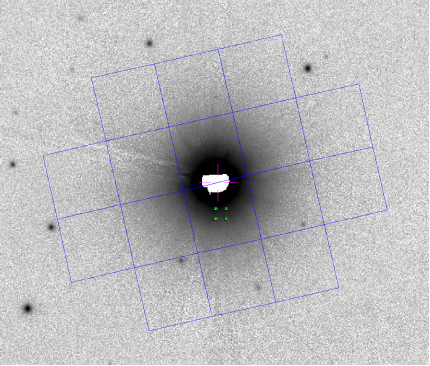}
\caption{Photometric images of NY,HYA from two different instruments. \emph{Top panel}: Camera 1 \& CCD 1 target pixel file image of the first {\it TESS} cadence for NY\,Hya. The 
{\it TESS} target aperture mask used to generate the SAP light curve is highlighted in red. \emph{Bottom panel}: SAP aperture superimposed on a {\it PanSTARRS} image. North is up and east
is left. Each box is  $21 \times 21$ arcsec in size. Three faint companions are positioned at around 22 arcsec from NY\,Hya.}
\label{fig:1}
\end{figure}

Fig.\,\ref{fig:1} shows a pixel-mosaic image of the first cadence for NY\,Hya along with the target aperture mask. The data release notes\footnote{\url{https://archive.stsci.edu/missions/tess/doc/tess_drn/tess_sector_08_drn10_v02.pdf}} for sector 8 describe that an instrument anomaly began at UTC 2019-02-17 05:48:35 which ceased all data and telemetry collection. Normal operations resumed at 2019-02-20 12:02:38. Data collection was paused for 1.19 days during perigee passage while downloading data. The `release notes' for Sector 8 also describe how Camera 1 was affected by scattered light from the Earth and Moon. The background pixels were primarily affected at the end of the two orbits by Earth and towards the start of orbit 24 by the Moon.

\begin{figure*}
\includegraphics[width=0.98\textwidth,angle=0]{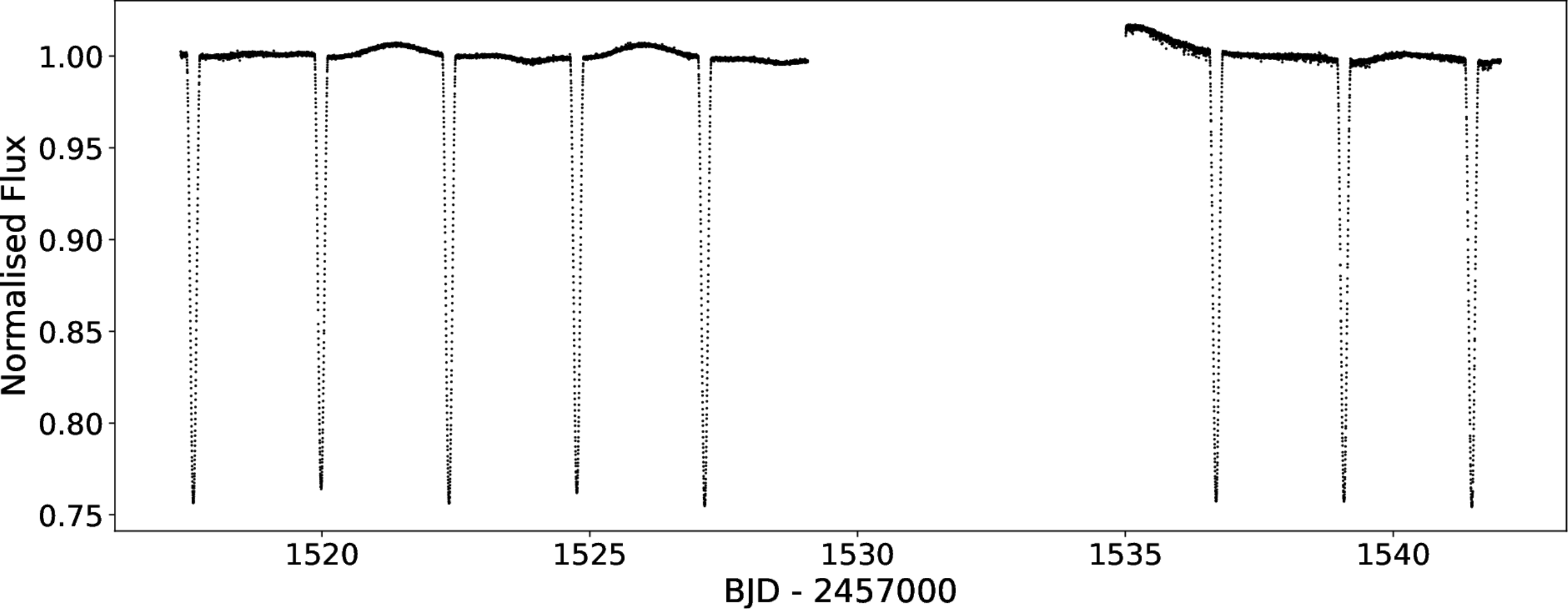}
\caption{Extracted {\it TESS} SAP light curve of NY\,Hya from Sector 8 with a sampling of 2 min. The
first eclipse is a primary eclipse. Five primary and three secondary eclipses were observed. Some
out-of-eclipse variation is clearly detected. The increase in flux at BJD 2,458,536 changes depending on the parameters used in the {\sc lightkurve} detrending process. We therefore believe that this is systematic noise.}
\label{fig:3}
\end{figure*}

Because the available data affected by the scattered light do not contain an eclipse we masked the data between BJD 2,458,530.0 and 2,458,536.3, resulting in a 6.3 day data gap. We show the resulting SAP light curve in Fig.~\ref{fig:3}.

Due to {\it TESS} having a large pixel scale (21 arcsec/pixel), it is important to check for faint nearby stars which may contaminate the point spread function (PSF) of the target star. To do this, we follow the procedure given by \citet{Swayne2020}. We overlaid the {\it TESS} target aperture mask over an image of NY\,Hya from the {\it PanSTARRS} image server \citep{Flewelling2016}, which indicated that three faint objects (lower three pixels) are within the {\it TESS} aperture mask used by SPOC. We cross-referenced these objects with the {\it TESS} input catalogue \citep{Stassun2018,Stassun2019}. The brightest two were found to have a {\it TESS} magnitude of 16 and 19 (NY\,Hya has a {\it TESS} magnitude of 8) which translates to 1600 and 25000 times fainter than NY\,Hya providing a total contribution of flux of approximately 0.01\% and hence these sources are negligible. The faintest of the three is not in the {\it TESS} input catalogue, indicating that it is fainter than {\it TESS} magnitude 19.

The {\it Gaia} Data Release 2 (DR2) \citep{GAIA2018} indicates the presence of a nearby star ({\it Gaia} 5746104876937814912) that is 6 arcsec from NY\,Hya and has since been confirmed to have the same parallax as NY\,Hya, from the {\it Gaia} Early Data Release 3 \citep[EDR3][]{GAIA2021}. When compared with the PanSTARRS image (Fig.~\ref{fig:1}) the companion lies within the saturated pixels of NY\,Hya and therefore, lies within the TESS target aperture. Consequently, we crossed-referenced the companion star with the TESS input catalogue and found the companion (TIC 876368206) to have a TESS magnitude of 16.5, indicating a negligible ($\approx$0.01\%) contribution to the measured flux.

\subsection{EMCCD lucky-imaging observations}
\label{emccdlucky-imagingobservations}

We further obtained high-resolution imaging data with the purpose to detect nearby companions that could 
be the source of observed photometric signals or the dilution of such signals. NY\,Hya was observed on the
beginning night of May 21, 2019 and May 23, 2021 with the two-colour instrument (TCI; $v$ and $z$ bands) attached 
to the Danish 1.54m telescope at the ESO/La Silla Observatory, Chile. For optimal detection, all (May 21, 2019) images 
were obtained at airmass 1.10 with a seeing of <1 arcsec. The observations were conducted as part of the 2019/2021 MiNDSTEp\footnote{\url{http://mindstep-science.org}} campaign. Each TCI consists of a
$512 \times 512$ pixel electron-multiplying (Andor, iXon+897) CCD capable of imaging simultaneously in two
colours. The field of view is about $45 \times 45$ arcsec yielding a pixel scale of approximately 0.09
arcsec/pixel. A detailed description of the instrument and the lucky-imaging reduction pipeline is given in
\citet{Skottfelt2015}.

The observations and data reduction were carried out using the methods outlined as follows, which resemble the methods described by \citet{Evans2016}. The exposure times of the targets in the high-resolution imaging observations were selected to detect (at 5-$\sigma$) a star up to 5\,mag fainter than the target in $V$. Fainter stars can be safely ignored as their contamination to the eclipse depth would be $<0.1$\,$m$mag, which is less than the RMS noise (0.382\,$m$mag) of the TESS data presented in this work. NY\,Hya was observed for 120\,s at a frame rate of 10\,Hz, allowing for near-diffraction limited images to be constructed by combining only frames with the least atmospheric distortion. The raw data were reduced by a custom pipeline that performs bias and flat frame corrections, removal of cosmic rays, determination of the quality of each frame and frame re-centring. The end product consists of ten sets of stacked frames ordered by quality.

\begin{figure}
\includegraphics[width=0.48\textwidth,angle=0]{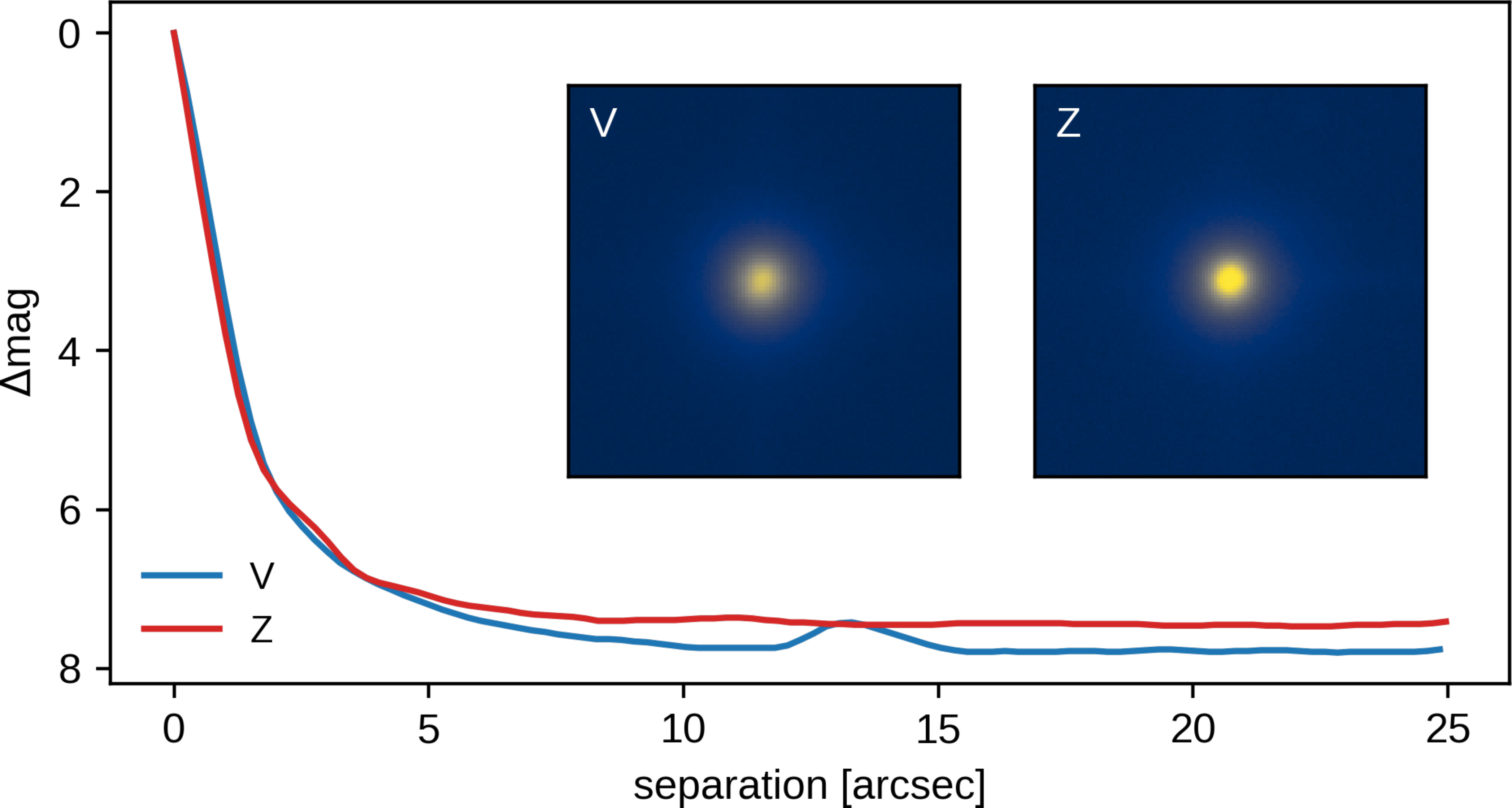}
\caption{$5\sigma$ detection contrast-curves ($\Delta{\rm mag}$) from the TCI images using the top
20\% stacked frames. Inset images are $10 \times 10$ arcsec. The full width at half maximum (seeing) was measured to be 0.97 arcsec in $v$ and 0.85 arcsec in $z$.}
\label{fig:nyhya_contrastcurves}
\end{figure}

We computed $5\sigma$ contrast curves from image statistics for each camera from selecting the first 20\% best images (i.e. stacked the first 240 frames) and summed flux contributions from each pixel in concentric rings centred on the star's location as determined by a two dimensional Gaussian fit to the point spread function and extending radially away from the star. We used a cubic spline to smooth the contrast in delta magnitudes as a function of separation, assuming a minimum of $5\sigma$ for the detection of a point source. The resulting contrast curves are shown in Fig.~\ref{fig:nyhya_contrastcurves}. We detect a small brightness increase of $\Delta{\rm mag} = 7.5$ at a separation of about 14 arcsec in the $v$-band camera. At this point, we are not sure whether this is instrumental or caused by the faint nearby companions as shown in Fig.~\ref{fig:1}.

The nearby companion at 6 arcsec that was detected in the {\it Gaia} DR2 and EDR3 was not detected in the presented high-resolution imaging data, which is expected considering the companion is approximately 8.5 magnitudes fainter than NY\,Hya in the TESS bandpass which has a similar central wavelength to the $z$-band camera.

\subsection{Spectroscopic observations}
\label{spectroscopicobservations}

\citet[his Table 2.4]{ribas1999} presented and analysed 22 unpublished RV (CORAVEL scanner, decommissioned) measurements ($\sigma_{RV} = 0.5$ km/s per measurement). \cite{ribas1999} found NY\,Hya to be a double-lined spectroscopic binary and derived preliminary spectroscopic elements. He found two components (NY\,Hya A and B), near-identical in minimum mass, orbiting each other in a circular orbit with a systemic RV of around 40 km/s. The spectroscopic orbital period agrees with the \citet{clausen2001} value ($0.015\sigma$). Unfortunately, a complete simultaneous photometric and spectroscopic analysis of NY\,Hya was never presented in the literature. Furthermore, a check in the Detached Eclipsing Binary Catalogue (DEBCat;\footnote{\url{https://www.astro.keele.ac.uk/jkt/debcat/}} \citealt{debcat}) revealed no information on NY\,Hya.

A total of 35 high-resolution spectra were obtained of NY\,Hya using the ESO 1.52m telescope and the Fibre-fed Extended Range Optical Spectrograph (FEROS;\footnote{ \url{https://www.eso.org/sci/facilities/lasilla/instruments/feros.html},\url{https://www.eso.org/public/teles-instr/lasilla/mpg22/feros/}, \url{http://www.mpia.de/FEROS/about\_feros.html}, \url{http://www.ls.eso.org/lasilla/Telescopes/2p2T/E1p5M/FEROS/index.html}}, \cite{kaufer1999}) at the ESO/La Silla observatory, Chile. The resolving power of FEROS is $\lambda/\Delta\lambda = 48000$ (constant velocity offset of 2.2 km/s/pix) covering an effective wavelength range of 3600--9200 \si{\angstrom} starting from the Balmer discontinuity and spanning 39 \'echelle orders. The object-sky (OS) observing mode was used for the two fibre apertures for optimal background subtraction. The entrance aperture of each fibre is 2.6 arcsec. The minimum and maximum standard error for the wavelength calibration (all nights) was 0.007 \si{\angstrom} and 0.019 \si{\angstrom}, respectively.

Observations were carried out as part of the FEROS commissioning and guaranteed time period (commissioning II \& 62.H-0319/GT I) between November 23, 1998 and January 21, 1999. All spectra obtained during the two commissioning periods are public and were retrieved from the FEROS spectroscopic database hosted at the Landessternwarte Heidelberg.\footnote{\url{https://www.lsw.uni-heidelberg.de/projects/instrumentation/Feros/ferosDB/search.html}} During observations, the spectrograph was located in a temperature-controlled vacuum chamber. Typically, calibration frames were obtained at the beginning, middle and end of the night resulting in an average to account for thermal drift and environmental changes. Wavelength scales were established on a nightly basis from Thorium-Argon (Th-Ar) exposures. Pre-normalisation fluxes are relative fluxes (object/flat). In Table \ref{specdata} in appendix \ref{appendix1} we list details of the nightly observed FEROS spectra.

Raw spectral data reduction was performed with the ESO 
{\sc midas}\footnote{\url{https://www.eso.org/sci/software/esomidas/}} package. The spectra are available as standard reduced (i.e. standard, not optimal), order-extracted, flat-fielded,
bias-subtracted (from between orders), wavelength calibrated (barycentric rather than heliocentric correction to the wavelength scale is applied at the re-binning stage of the \'echelle orders) one-dimensional flux tabulated
spectra. The instrumental blaze function was removed to first order using the flat exposure. Removal of
scattered light was applied. Exposure times\footnote{exposure time is the half shutter open time as opposed to the time when half the detected photons have arrived.} were either 420s, 600s or 900s with the vast majority (32) of spectra taken with 600s resulting in a signal-to-noise $(S/N)$ ratio in the range 174 - 261. We had to discard two spectra entirely (\#2266 at the orbital phase 0.31 and \#2912 at 0.27) due to low quality and wavelength calibration problems in the data. Since we have several high-quality spectra in this orbital phase range (four others between phases 0.24 and 0.35) the coverage of orbital phase is still good.

\subsection{Identities of the two stars}
\label{Identitiesofthetwostars}

The standard definitions are that the primary eclipse is deeper than the secondary eclipse, and that the primary star is eclipsed at primary eclipse. We follow these conventions here, but with caveats. The two stars are almost identical, and starspots affect the light curve shape, so choosing which is the primary is not trivial. We inspected the three TESS sectors and found that one type of minimum is clearly deeper than the other in two cases; we label this the primary eclipse. In the third TESS sector starspot activity causes the primary and secondary eclipses to be of almost identical depth. The greater scatter in the $uvby$ light curves means the eclipse depths are not significantly different in those data.

The orbital ephemeris given below (Eq.~\ref{eq_ephemeris2} in Section~\ref{timesofminimumlightandnewephemeris}) defines our identification of the stars. Orbital phase zero in this ephemeris corresponds to the primary eclipse and is where the primary star is eclipsed by the secondary star. We refer to the primary as star~A in the following analysis, and its companion as star~B. These identifications are consistent with the ephemeris given by \citet{clausen2001}. We ultimately find that the masses, radii and temperatures of the two stars are so similar as to be identical to within their uncertainties.

\section{Radial velocity and light ratio determination}
\label{spectralanalysis}

\subsection{Continuum normalisation}
\label{continuumnormalisaton}

All FEROS spectra were normalised to the continuum level iteratively and via interactive comparisons with a synthetic spectrum computed using the {\sc synth3} code \citep{kochukhov2007} which is based on {\sc atlas}/{\sc synthe} program suite of \cite{kurucz1993}, the {\sc synth} and {\sc synthmag} codes of \cite{piskunov1992} and the {\sc sme} program written by \cite{valenti1996}. We made use of the Vienna Atomic Line Database ({\sc vald}) \citep{piskunov1995} for line identification and initial estimates for the fundamental atmospheric parameters for both components ($T_{\rm eff}$ = 5700 K, $\log g = 4.15$ (cgs), [Fe/H] = 0.0 dex). The $T_{\rm eff}$, $\log g$, and [Fe/H] were obtained from Str\"omgren photometry and a preliminary SED analysis (see Sect.~\ref{spectralenergydistributionmodeling}) to obtain composite spectra for different orbital phases matching that of the observations for continuum normalisation. We then compared these spectra with the FEROS spectra visually by using the {\sc binmag6} code \citep{kochukov2018} and continuum-normalised our observational spectra.

We examined four different wavelength regions centred on the Str\"omgren ($vby$) and {\it TESS} passbands to normalise the observed spectra. Wavelength ranges of these segments correspond to the Full Width at Half Maximum (FWHM) of the passband response curves \citep{rodrigo2012,rodrigo2013}. We made use of interactive {\sc idl} widgets {\sc xregister\_1d.pro} and {\sc line\_norm.pro} in the {\sc fuse}\footnote{\url{https://archive.stsci.edu/fuse/analysis/idl_tools.html}} package, respectively, for cosmic-ray removal and continuum normalisation. 

We measured the signal-to-noise $(S/N)$ with relevant tools in the {\sc guiapps} package of the {\sc iraf} package\footnote{{\sc iraf} is distributed by the National Optical Astronomy Observatories, which is operated by the Association of Universities for Research in Astronomy, Inc. (AURA) under cooperative agreement with the National Science Foundation.} from different segments of the continuum and averaged them. We double-checked our measurements by also employing the `SNR Estimator' in the current version of the {\sc ispec}\footnote{\url{https://www.blancocuaresma.com/s/iSpec}} software package \citep{ispec1,ispec2}.

\subsection{Radial velocities}
\label{radialvelocities}

RVs were measured using $i$) one-dimensional cross-correlation functions (CCFs, \citet{tonrydavis1979}), $ii)$ Rucinski's broadening functions \citep{rucinski1999}, and $iii)$ the two-dimensional cross-correlation technique (hereafter {\sc todcor}) \citep{mazehzucker1994} in $160-250\,\si{\angstrom}$ wide spectral windows. Similar results were obtained from all three methods lending confidence in the accuracy of our RV measurements.

In this work, we chose to present and adopt measurements obtained from {\sc todcor} since this technique provides the additional option to measure a light ratio (luminosity or flux ratio) for selected spectral windows. Estimates of the spectral light ratios will be beneficial in the analysis of light curves (see Sect.~\ref{tessphotometry}) as discussed later. The {\sc todcor} technique aims to correlate an observed binary spectrum against a combination of two template spectra with all possible RV shifts to determine the unknown wavelength displacements in the spectra of each individual component at any orbital phase. Therefore, the correlation is a two-dimensional function of the velocity shifts of the two templates. The position of the correlation maxima corresponds to the RVs of individual components. {\sc todcor} has the advantage over the traditional one-dimensional cross correlation in the determination of the correlation maxima even when blends of the correlation peaks are inevitable and for the case when one companion is significantly fainter than the other.

We applied the {\sc todcor} technique to wavelength regions in close agreement with the Str\"omgren and {\it TESS} filter\footnote{\url{http://svo2.cab.inta-csic.es/theory/fps/}} transmission profiles \citep{rodrigo2012,rodrigo2013,rodrigo2020} ($v: 4030-4190\,\si{\angstrom}, b: 4570 - 4770\,\si{\angstrom}, y: 5360-5600\,\si{\angstrom}$ and $7850-8100\,\si{\angstrom}$ for the {\it TESS} filter). We made use of synthetic spectra  as templates for each of the stars that were generated for spectral normalisation. We experimented with templates accounting for the rotational broadening as well, but in line with other CCF-based techniques relying on delta-functions, we ignored stellar rotation during the synthesis of our templates to reduce the noise in the measurements due to line blending caused by the orbital motion exacerbated by rotational broadening. Assuming an a priori small difference between the \Teff's of components in the computation of their synthetic spectra would help us follow which star is which at any given orbital phase, and their comparisons with the observed spectra without affecting the RV measurements. Hence, we chose a hotter template for one of the stars (star A) by 50 K ($T_{\rm eff, A}$ = 5750 K and $T_{\rm eff, B}$ = 5700 K, while fixing $v \sin i = 0.0\,\rm km/s$, $\log g$ = 4.16 (cgs) and [Fe/H] = 0.0 dex for both components). Estimates for RV uncertainties were obtained as the FWHM of Gaussian fits to the two-dimensional correlation functions with centres at the RV for which the two-dimensional correlation function is algorithmically maximised.

We had to discard the {\it TESS} wavelength region ($7850-8100\,\si{\angstrom}$) dominated by telluric lines.  We therefore determined RVs from the Str\"omgren passbands and adopted velocities from a weighted mean of these TODCOR measurements, which we provide in Table \ref{tab:todcor} in appendix \ref{appendix2} together with their uncertainties.

The spectral light ratio ($l_{\rm B} / l_{\rm A}$) is computed from eq.\,A4 in \citet{mazehzucker1994}. For each spectrum, we determined an estimate of light ratio uncertainties from the scatter of ten light ratio measurements in close proximity to the {\sc todcor} maxima. In Table \ref{table:lightratios}, we provide a list of final measurements as adopted in this work. We excluded the light ratio measurements at orbital phases near primary and secondary eclipses. We provide and adopt the weighted mean of spectroscopic light ratio measurements and uncertainties for each of the wavelength region in Table \ref{table:lightratios}. The $u$ passband is partially outside the FEROS wavelength coverage, hence it is not suitable for light ratio estimations.

\begin{table} 
\caption{Spectroscopic light ratio ($l_{\rm B} / l_{\rm A}$) measurements from FEROS data.}
\centering
\begin{tabular}{cccc}
\hline
Passband    &  $\lambda_{\rm int}$ [\si{\angstrom}] & $\lambda_{\rm cen}$ [\si{\angstrom}] & Sp. $l_{\rm B} / l_{\rm A}$ \\
\hline
\hline
Str\"om-$u$ &  3350 - 3650   & 3500  & n/a \\
Str\"om-$v$ &  4030 - 4190   & 4110  & $1.021 \pm 0.064$ \\
Str\"om-$b$ &  4570 - 4770   & 4670  & $1.007 \pm 0.069$ \\
Str\"om-$y$ &  5360 - 5600   & 5480  & $1.004 \pm 0.042$ \\
{\it TESS}  &  7850 - 8100   & 7975  & $1.149 \pm 0.035$ \\
\hline
\end{tabular}
\label{table:lightratios} 
\tablefoot{Measurements were carried out with {\sc todcor} in the Str\"omgren and {\it TESS} passbands. The $u$ passband is partially outside the FEROS wavelength coverage, and hence not determined. In general, the blue end of FEROS is noise-dominated. The $\lambda_{\rm cen}$ and $\lambda_{\rm int}$ refer to the central wavelength and wavelength interval of the filter as obtained from the VOSA filter service (\url{http://svo2.cab.inta-csic.es/theory/fps/}).}
\end{table}

We verified the {\sc todcor} light ratio measurements on the basis of computing the relative strengths of a few 
isolated absorption lines. Light ratio measurements derived from the ratio of equivalent widths (EWs) is a well established method \citep{petrie1939}. We made use of three relatively isolated neutral iron lines in the wavelength window $5360-5600\,\si{\angstrom}$ (Str\"omgren $y$ band) for this purpose and only used spectra at orbital phases (0.07, 0.93, and 0.94) to avoid line blending as much as possible. We then measured the EWs of the selected lines absorbed by each of the components from the Gaussian fits with the {\sc ispec} package for spectroscopic measurements. We averaged all the line strength ratios, weighted by the goodness of the Gaussian fits from all three lines recorded at all three orbital phases (nine measurements in total) and determined a light ratio value of $l_{\rm B} / l_{\rm A} = 1.063 \pm 0.021$, which is in close agreement ($1.3\sigma$) with the estimate obtained from {\sc todcor} ($1.004 \pm 0.042$). However, we chose to adopt the light ratios as determined from {\sc todcor}, which is readily available for wavelength regions we need. They are also not affected by line blending.

\section{Times of minimum light and new ephemeris}
\label{timesofminimumlightandnewephemeris}

The times of minimum light from SAT and {\it TESS} data allow a precise determination of the eclipse ephemeris due to an extended temporal base-line coverage. For both data sets, we applied the \citet{kwee1956} method (KvW) for the computation of eclipse times. Realistic timing uncertainties were obtained via Monte-Carlo bootstrapping (with replacement) based on 100,000 bootstrap trials. Reducing this number by half produced identical results. The 16th, 50th and 84th percentiles were determined from the resulting distribution providing a median and $\pm 1\sigma$ uncertainty where the maximum of the lower and upper percentile was chosen. A total of five primary and four secondary eclipses were determined from SAT data from $y$ band data only due to a higher photometric precision. Table \ref{times_of_minima_sat} and \ref{times_of_minima_tess} provide an overview of measured times of minimum for SAT, and {\it TESS} data, respectively.

A new ephemeris based on a weighted least-squares fit was calculated based on primary and  secondary eclipses. We chose the first primary eclipse of {\it TESS} as the reference eclipse ($T_0$) in an attempt to avoid parameter correlation between $T_0$ and the orbital period $P$. The scatter around the best-fit line could either be of astrophysical nature or the timing uncertainties as obtained from the KvW method (+ bootstrapping) are underestimated. We have no evidence for a third body that might introduce eclipse-timing variations. Therefore, we chose to re-scale (inflate) the KvW timing uncertainties with the reduced $\sqrt{\chi^2}$ in order to obtain a reduced $\chi^2 = 1.0$. This assumes that we trust a linear ephemeris with no additional astrophysics that could cause timing variations. We found this new ephemeris to be
\begin{equation}
T_{\rm min}(E) = 2,458,517.59882(12) + 4^{\rm d}.7740549(41) \times E
\label{eq_ephemeris2}
,\end{equation}
\noindent
where $E$ is the orbital epoch. The root cause for underestimated timing uncertainties is not fully understood but may be stellar surface activity. This will be discussed in a later section.

\begin{table}[hbt!]
\caption{Times of primary (P) and secondary (S) minima of NY\,Hya from SAT data only (combined 97/98 \& 98/99 season).}
\label{times_of_minima_sat}
\centering                         
\begin{tabular}{cccc}      
\hline
$T_{\rm min}$ & $\sigma$ & Type & O$-$C\tablefootmark{b}\\
BJD(TDB)-2,400,000.0 & (days) & P/S & (days) \\
\hline
\hline
50833.75770\tablefootmark{~}     & 0.00026       & S     & ~0.00024\\
50845.69227\tablefootmark{~}     & 0.00031       & P     & -0.00032\\
50857.62812\tablefootmark{a}     & 0.00023       & S     & ~0.00039\\
50876.72498\tablefootmark{~}     & 0.00051       & S     & ~0.00103\\
\hline
51160.77705\tablefootmark{~}     & 0.00018       & P     & -0.0032\\
51203.74803\tablefootmark{~}     & 0.00013       & P     & ~0.0013\\
51227.61577\tablefootmark{~}     & 0.00045       & P     & -0.0012\\
51234.77826\tablefootmark{~}     & 0.00032       & S     & ~0.00019\\
51246.71363\tablefootmark{~}     & 0.00018       & P     & ~0.00042\\
\hline                                   
\end{tabular}
\tablefoot{Light minima were determined as described in Section~\ref{Identitiesofthetwostars}). Corresponding uncertainties are not scaled. \tablefoottext{a}{Not published in \citet{clausen2001}}.~\tablefoottext{b}{Based on the ephemeris in Eq.~\ref{eq_ephemeris2}.}}
\end{table}

\begin{table}[hbt!]
\caption{Times of minimum of NY\,Hya determined from {\it TESS} SAP short-cadence data.}
\label{times_of_minima_tess}
\centering                         
\begin{tabular}{cccc}      
\hline
$T_{\rm min}$ & $\sigma$ & Type & O$-$C\tablefootmark{a}\\
BJD(TDB)-2,400,000.0 & (days) & P/S & (days) \\
\hline
\hline
58517.599300 & 0.000050 & P & ~0.00048 \\
58519.985500 & 0.000040 & S & -0.00035 \\
58522.373050 & 0.000040 & P & ~0.00017 \\
58524.759320 & 0.000040 & S & -0.00059 \\
58527.146730 & 0.000040 & P & -0.00020 \\
58536.695020 & 0.000040 & P & -0.00002 \\
58539.082630 & 0.000040 & S & ~0.00056 \\
58541.469390 & 0.000060 & P & ~0.00029 \\
\hline                                   
\end{tabular}
\tablefoot{We retained 30 points on each side of the minimum flux. Some eclipses had data gaps within the 30 point span causing a slight increase in the timing precision. Uncertainties are not scaled.\tablefoottext{a}{Based on the ephemeris in Eq.~\ref{eq_ephemeris2}.}}
\end{table}

\section{Light- and RV-curve analysis}
\label{Light-andRVcurveanalysis}

Before we embark on a fully quantitative modelling of spectroscopic and photometric data we note the following. From a visual inspection of the Str{\"o}mgren and {\it TESS} light curves, the two components of NY\,Hya are similar to each 
other in size: the two eclipses are almost identical in duration and depth. The stars appear to be grazing each other during conjunctions resulting in partial eclipses.\footnote{For two similar stars, the amount of light lost during a total eclipse must be roughly a factor of 0.5. This corresponds to a change of $-2.5\log (1/2) \simeq 0.75$ mag from the flux out of the eclipses. For NY\,Hya, the eclipses are only about 0.30--0.40 mag deep.} Furthermore, no interval of flat totality is seen pointing towards a relatively low orbital inclination. Colour variations during primary and secondary eclipse are small, indicating near-identical atmospheric properties of the components. 
Reflection effects are likely negligible. The long orbital period implies this system to be well-detached and hence little deformation in stellar shape is expected. Out-of-eclipse brightness changes during both SAT seasons are present. This can be seen more  clearly from the continuous {\it TESS} data. Changes in brightness during ingress--egress phases within a season are also observed. Subtle changes in the depth of light minima are present. Seasonal and inter-seasonal variability is likely due to atmospheric surface activity in the form of spots. Finally, the two eclipses appear at phase 0.0 and 0.5 indicating a near-circular orbit.

\begin{figure}
\centering
\includegraphics[width=0.47\textwidth]{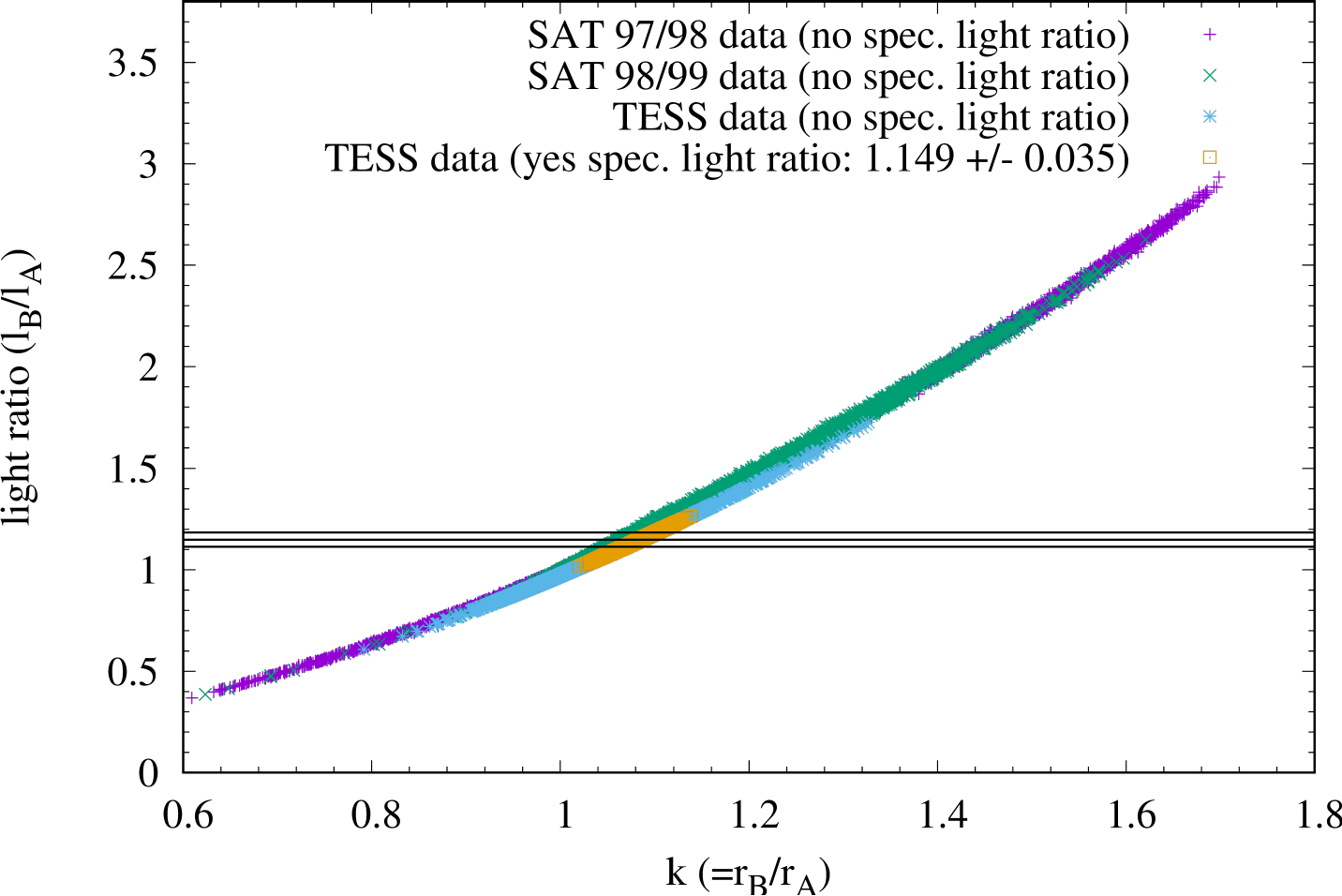}
\caption{Strong parameter correlation \citep{torres2000} between light ratio $l_B/l_A$ and radius ratio for {\it TESS} and SAT data. The horizontal lines show the spectroscopic light ratio for the {\it TESS} passband and its uncertainty.}
\label{fig:LRATExperiments_UpdatedPlot1}
\end{figure}

\subsection{Str\"omgren photometry}


We modelled the SAT data with the purpose of deriving passband-dependent light ratios. These can be used to determine Str\"omgren colours for each binary component. The modelling of SAT data utilised the {\sc jktebop}\footnote{\url{http://www.astro.keele.ac.uk/jkt/codes/jktebop.html} code, which implements the IAU 2015 Resolution B3 \citep{IAU2015-Res-B3} for Solar values.} (Ver. 41) code \citep{southworth2004,southworth2007,southworth2013}. The code is an extension of the {\sc ebop} code \citep{etzel1975,etzel1981,popperetzel1981} and is based on the Nelson-Davis-Etzel model \citep[NDE;][]{nelsondavis1972,etzel1981}. The two components are modelled as biaxial spheroids for the calculation of
the reflection and ellipsoidal effects, and as spheres during the eclipse phases. The validity of applying this model to NY\,Hya is warranted: the two components turn out to have fractional radii much smaller than 0.3 \citep{northzahn2004} and the oblateness of both components was found to be smaller than 0.04 \citep{popperetzel1981} indicating NY\,Hya to be a detached binary system. A best-fit model is determined via a Levenberg-Marquardt (LM) least-squares minimisation \citep{press1992}. For NY\,Hya, no prior information on light curve parameters exists and hence we apply a frequentist statistical data modelling approach.

To investigate whether or not the two SAT data sets (97/98 and 98/99) can be merged we considered each $y$ data set separately and determined best-fit parameters and realistic uncertainties. We considered only seven basic parameters: the reference epoch of primary eclipse $T_0$, orbital period $P$, sum of fractional radii $r_A + r_B$ where $r_A=R_A/a$ and $r_B=R_B/a$ with $a$ measuring the orbital semi-major axis and $R$ denoting the stellar radius, the ratio of radii $k=r_B/r_A$, the orbital inclination $i$, the central surface brightness ratio $(J=J_B/J_A)$ and a constant light scale (sfact) parameter (dependent on the choice of comparison/check stars) of the out-of-eclipse baseline magnitude. In this work, the latter parameter is treated as a nuisance parameter and we marginalise over it to account for its possible effect on all other parameters. The NDE model surface integration ring size was set to 5 degrees (a test with 1 degree resulted in no difference). The photometric mass ratio was set to force the two stars to be spherical.

To constrain the ratio of fractional radii ($k$) we added a spectroscopic light ratio as measured from FEROS in the $y$ band (see Sect.~\ref{tessphotometry} and Fig.~\ref{fig:LRATExperiments_UpdatedPlot1}). We quantitatively investigated the parameter consistency between the two data sets and found differences in the range $1.1\sigma$ ($r_{A}+r_{B}$) to $2.0\sigma$ ($J = J_B/J_A$). The largest contribution to parameter uncertainties originates from the 97/98 data due to a larger number of incomplete eclipses. We take this as evidence to justify a merging of the two data sets in each pass-band for a final analysis.

In addition to the seven basic light curve model parameters, additional parameters and their statistical significance were systematically tested for. RV data were not included at this stage. We tested for: one- and two-parameter LD laws (with coefficients fixed / free), orbital eccentricity ($e\cos\omega$ and $e\sin\omega)$, and third-light ($l_3$). Model selection is based on a standard hypothesis test invoking the Fisher-Snedecor $\mathcal{F}$-statistic \citep{lucysweeney1971} and is akin to a analysis-of-variance (ANOVA) test. At each testing stage, when including a new parameter, we tested for the rejection of the null-hypothesis at a 0.1 per cent significance level. This was done for each additional parameter sequentially. The parameter that passed the hypothesis test while simultaneously resulting in a maximum improvement ($\Delta\chi^2$) to the model fit, was then included consecutively replacing the previous model with an extended model. The testing sequence was then repeated for all remaining parameters.

We found that the data support the detection of limb-darkening. The next best improvement in the $\chi^2$ statistic was found by including a quadratic limb-darkening law. We do not see any statistical evidence for the SAT data to support a detection of the $e\cos\omega$ eccentricity term either after the $\mathcal{F}$-test based model comparison.

An additional consideration examined whether spots during eclipses could be accounted for with the application of a polynomial trend removal (invoking the poly option in {\sc jktebop}), we calculated $k$ with and without detrending while considering parameter uncertainties from TASK8 and TASK9. We found differences at the $0.70\sigma$ (TASK8) and $1.1\sigma$ (TASK9) levels. We conclude that imposing any attempt of detrending the eclipses does not improve the measurement of $k$.

\begin{figure*}
\centering
\includegraphics[width=0.48\textwidth]{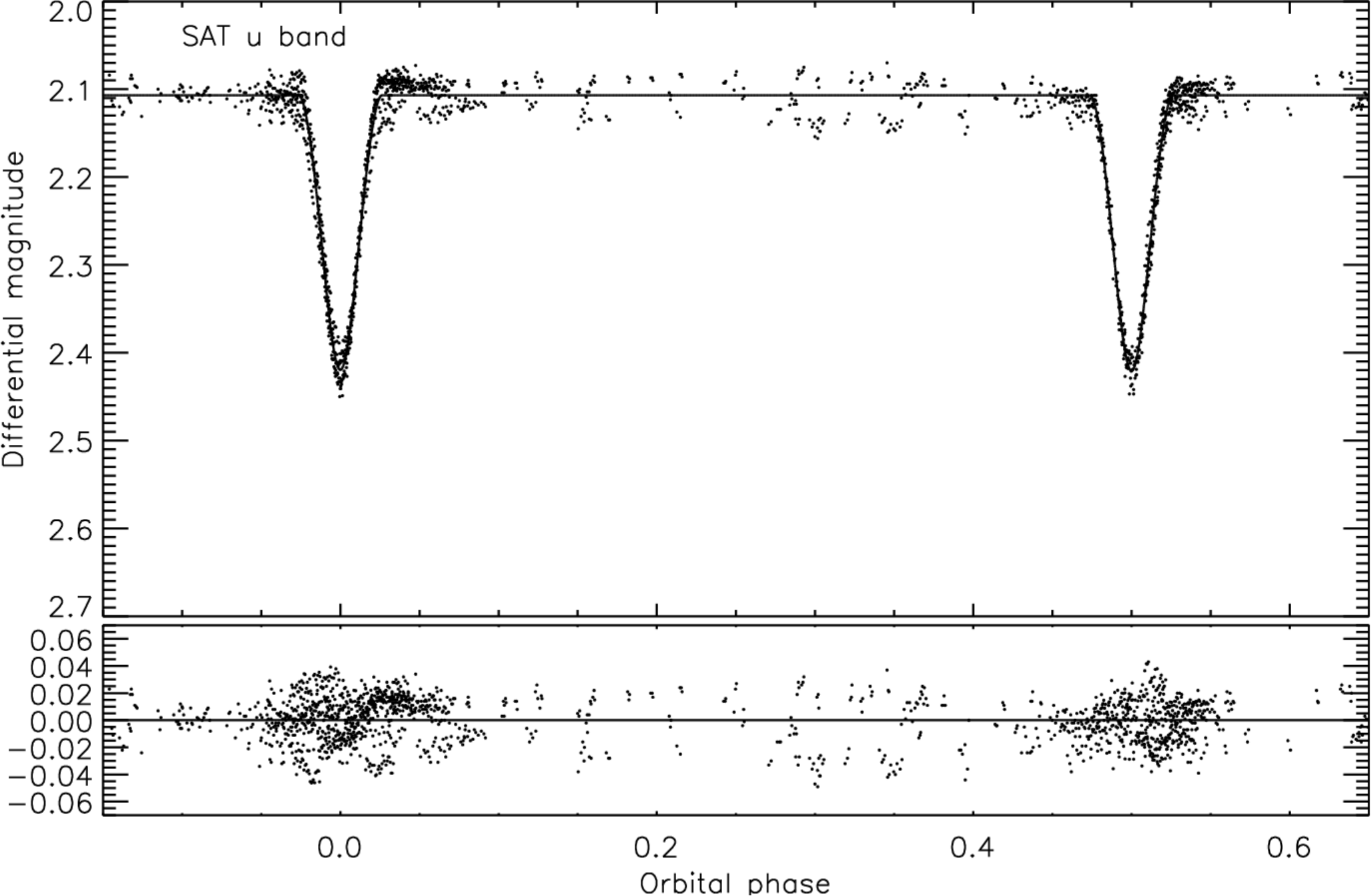}
\includegraphics[width=0.48\textwidth]{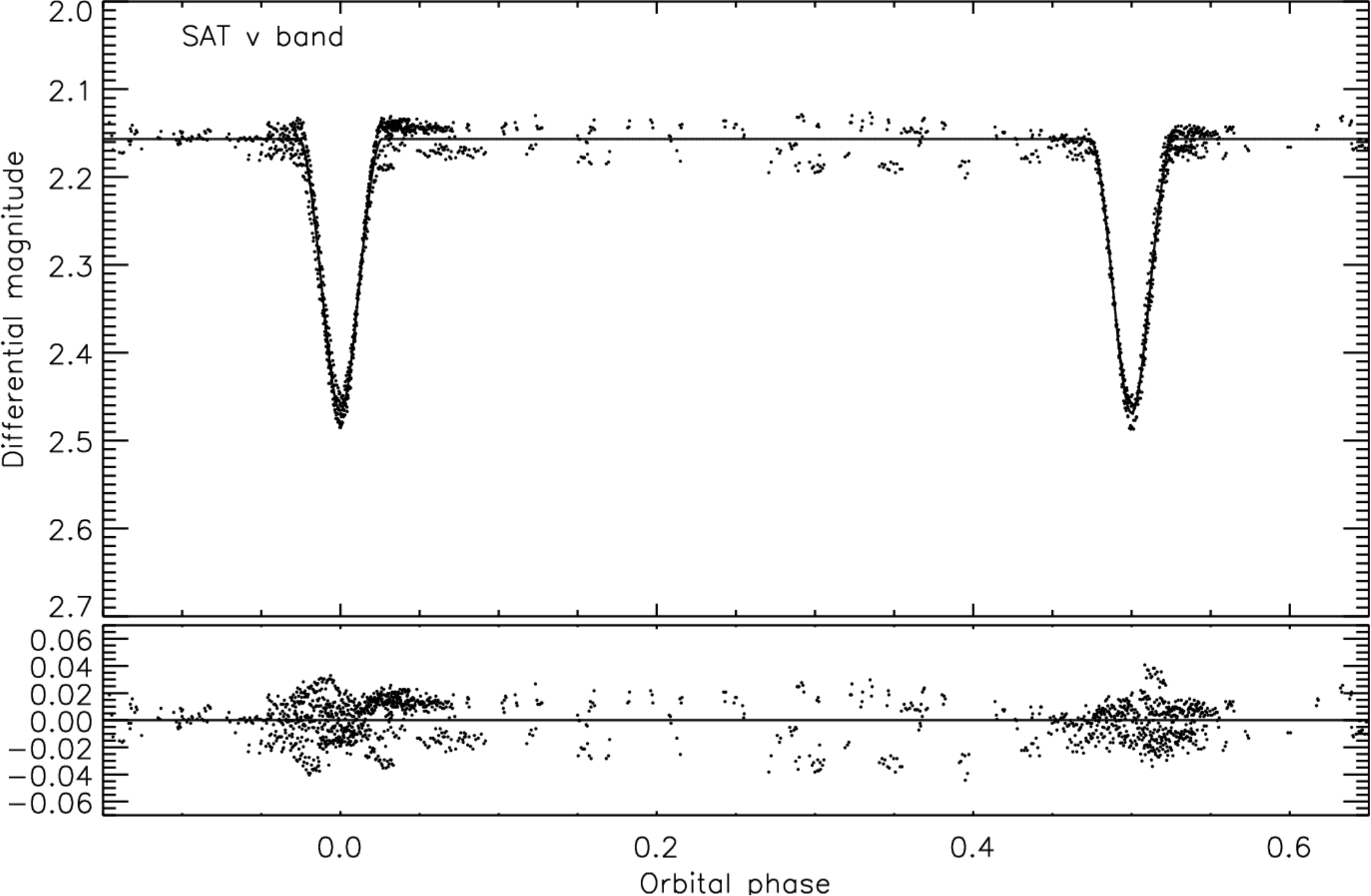}
\includegraphics[width=0.48\textwidth]{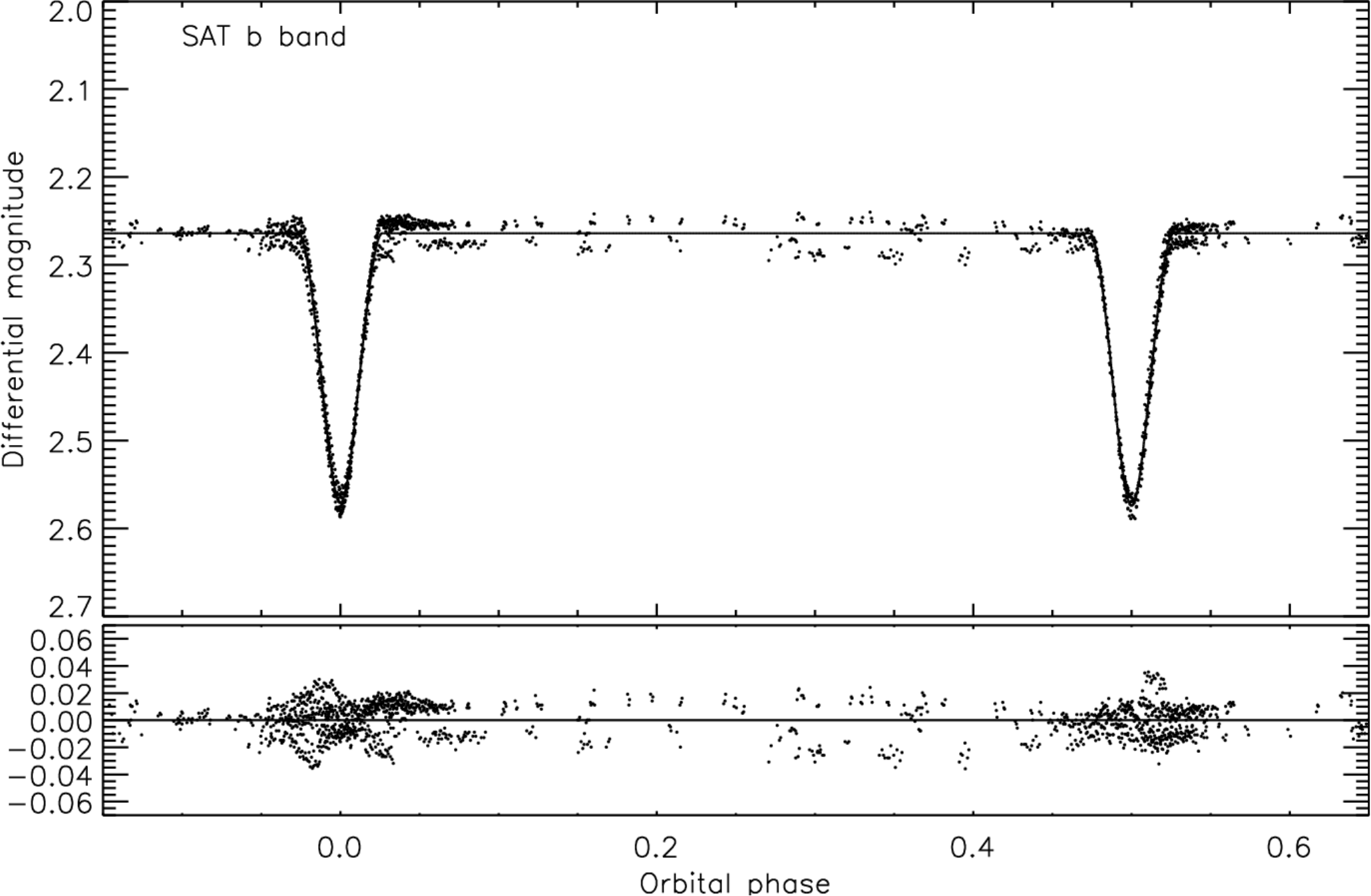}
\includegraphics[width=0.48\textwidth]{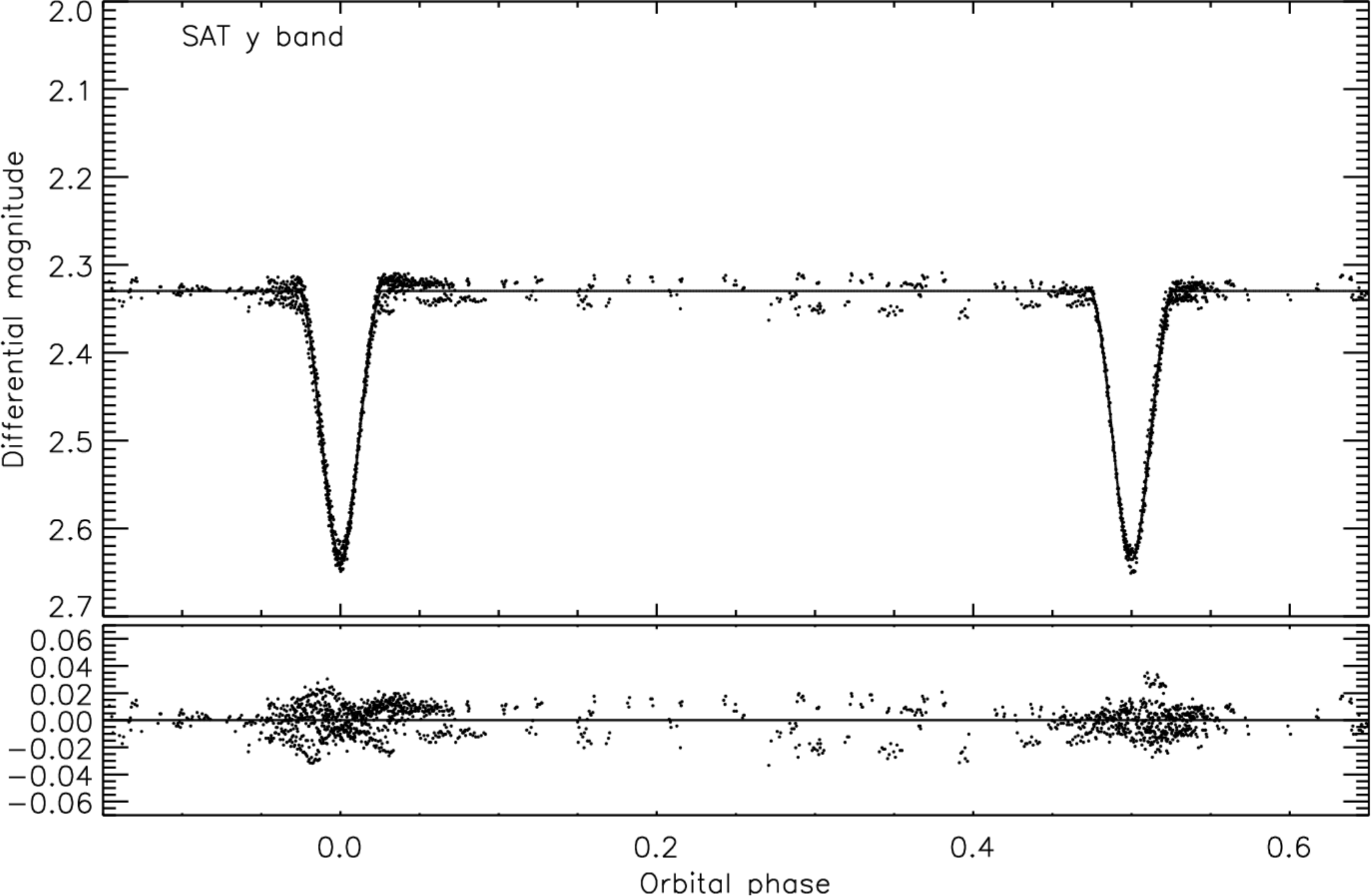}
\caption{Str{\"o}mgren (SAT) $uvby$ light curves of NY\,Hya. We show data from  the 97/98 and the 98/99
season. The solid line shows the best-fit model obtained from {\sc jktebop}. The residuals are shown below
each model fit.}
\label{fig:SATuvbyplots}
\end{figure*}

We decide to only retain a quadratic limb-darkening law with coefficients obtained from the {\sc limbdark}\footnote{\url{https://github.com/john-livingston/limbdark}} code. Uncertainties were obtained from proper error propagation by fixing $k-\delta k$, $k$ and $k + \delta k$ in turn. We scaled data errors to force the reduced $\chi^2 = 1.0$. Final SAT light curve parameters are presented in Table \ref{tab:sat_data}. We note that the best-fit LD coefficients are not to be trusted and are poorly constrained by the data. Plots of best-fit models in each SAT band are shown in Fig.~ \ref{fig:SATuvbyplots}.

The code allows the simultaneous fitting of RV data \citep{southworth2013}. The five parameters are the semi-amplitudes $K_{\rm A}, K_{\rm B}$, eccentricity $e$, argument of pericentre $\omega$ and the systemic velocity $\gamma$. Instead of fitting for $e$ and $\omega$ individually, the code fits for $e\cos\omega$ and $e\sin\omega$ to break a strong correlation between $e$ and $\omega$ \citep{pavlovski2009}. In the code, the parameters $e\cos\omega$, $e\sin\omega$, $T_0$ and $P$ are constrained simultaneously by photometric and spectroscopic data. The inclusion of RV data can result in a significant improvement in the photometric measurement of $e\sin\omega$ \citep{wilson1979}. As described in \cite{southworth2013}, {\sc jktebop} allows the two stars to have different systemic velocities, $\gamma_A$ and $\gamma_B$, in order to detect any possible systematic effects in RVs due to discrepancies between the FEROS spectra and a chosen template spectrum. We fitted for $\gamma_A$ and $\gamma_B$ and found a negligible difference at the $0.28\sigma$ level. This result is consistent with and as expected for two near-identical stars. Therefore, we fitted for a single systemic velocity. Finally, we did not perform a $\mathcal{F}$-test on RV parameters and assume that the above model is a correct description of spectroscopic data.

\begin{table*}
\caption{Best-fit photometric parameters for SAT $uvby$ bands.}
\label{tab:sat_data}
\centering
\begin{tabular}{lcccc}
\hline
parameter                                   & $u$       &       $v$     &       $b$     &       $y$     \\
\hline
\hline
$k$                                         & fixed & fixed & fixed & $0.999 \pm 0.020$ \\
$r_A + r_B$                                 & $0.180\pm0.032\tablefootmark{b}$  & $0.182\pm0.032\tablefootmark{b}$  & $0.181\pm0.022\tablefootmark{b}$ & $0.180\pm0.013\tablefootmark{b}$ \\
$i$ $(^\circ)$                              & $85.6\pm1.3\tablefootmark{b}$ & $85.5\pm1.3\tablefootmark{b}$ & $85.58\pm0.89\tablefootmark{b}$ & $85.61\pm0.53\tablefootmark{b}$ \\
$J_B/J_A$                                   & $1.003\pm0.043\tablefootmark{b}$ & $1.006\pm0.041\tablefootmark{b}$ & $1.009\pm0.034\tablefootmark{b}$ & $1.004\pm0.014\tablefootmark{b}$ \\
quad $u_A = u_B\tablefootmark{c}$           & $0.9\pm1.8\tablefootmark{b}$ & $0.7\pm1.7\tablefootmark{b}$ & $0.6\pm1.2\tablefootmark{b}$ & $0.62\pm0.64\tablefootmark{b}$ \\
quad $\nu_A = \nu_B\tablefootmark{c}$       & $-0.1\pm2.4\tablefootmark{b}$ & $0.26\pm2.2\tablefootmark{b}$ & $0.3\pm1.7\tablefootmark{b}$ & $-0.01\pm0.94\tablefootmark{b}$ \\
\hline
$r_A$                                       & $0.090\pm0.016\tablefootmark{b}$  & $0.091\pm0.016\tablefootmark{b}$ & $0.090\pm0.011\tablefootmark{b}$ & $0.0898\pm0.0067\tablefootmark{b}$ \\
$r_B$                                       & $0.090\pm0.016\tablefootmark{b}$  & $0.091\pm0.016\tablefootmark{b}$  & $0.090\pm0.011\tablefootmark{b}$ & $0.0898\pm0.0065\tablefootmark{b}$ \\
$l_B/l_A$                                   & $1.001\pm0.061\tablefootmark{b}$ & $1.004\pm0.059\tablefootmark{b}$ & $1.007\pm0.054\tablefootmark{b}$ & $1.002\pm0.042\tablefootmark{a}$ \\
RMS ($m$mag)                                & 17.0  & 15.0  & 13.0 & 11.0 \\
\hline
\end{tabular}
\tablefoot{The ratio of fractional radii $k$ for $y$ data was constrained with an empirical spectroscopic light ratio ($l_B/l_A = 1.004 \pm 0.042$). Consistency checks: Differences in photometric and spectroscopic light ratios: $u~({\rm n/a})$, $v~(0.20\sigma)$, $b~(0.0\sigma),$ and $y~ (0.03\sigma)$. We also note the decrease in RMS scatter consistent with nightly RMS scatter for SAT observations, which further justifies the use of $y$ band data to constrain $k$. The two limb-darkening coefficients were freely varying for all four bands and associated uncertainties were found to be large producing unphysical results. The best-fit $J_B/J_A$ values are all larger than unity for all four bands, and   contradict  the results using {\it TESS} data. A potential path of reconciliation is found by realising the existence of a strong parameter correlation between the limb-darkening and surface brightness ratio possibly throwing off the solutions. Considering $y$ data we therefore carried out tests by investigating fixed and varying combinations of the limb-darkening parameters and found no differences.}
\tablefoot{\tablefoottext{a}{from MC.} \tablefoottext{b}{from RP-PB.}\tablefoottext{c}{Not to be trusted. For some runs the total limb-darkening was found to be unphysical.}}
\end{table*}

\subsection{{\it TESS} photometry}
\label{tessphotometry}


We also analyzed the {\it TESS} short-cadence data (13420 data points) since this is the most precise data potentially providing  tighter constraints on model parameters as compared to the SAT data. We included all times of minimum light from the SAT 97/98 and 98/99 eclipses as an additional observational constraint on the ephemeris. { The largest difference in residuals between SAT data and the ephemeris was found to be $1.2\sigma$.} {\it TESS} SC data posits no danger for light curve smearing effects \citep{southworth2011}. {\it TESS} flux units and associated uncertainties were converted to magnitudes. The out-of-eclipse baseline magnitude was detrended and normalised with a cubic spline function. Finally, we iteratively removed outliers that are present in the {\it TESS} photometry by rejecting data points lying greater than $3\sigma$ from the best fit. We tested for any differences (considering $r_A$) for a $4\sigma$ clipping and found none.

\begin{table}
\centering
\caption{Best-fit parameters to the {\it TESS} and FEROS data for \emph{fixed} values of $k$ as obtained from SAT $y$ band data with proper error propagation.}
\label{tab:tessfinal}
\begin{tabular}{l c}
\hline
Parameter                                       & Value \\
\hline
\hline
$T_0$ ($\rm BJD_{\rm TDB}$-2,458,000.0)         & $517.598957 \pm 0.000081$\tablefootmark{b}\\
$P$ (days)                                      & $4.7740563 \pm 0.0000012$\tablefootmark{b}\\
$r_{\rm A}+r_{\rm B}$                           & $0.17789 \pm 0.00085$\tablefootmark{b}\\
$i$ ($^\circ$)                                  & $85.631 \pm 0.035$\tablefootmark{b}\\
$J=J_{\rm B}/J_{\rm A}$                         & $0.9732 \pm 0.0021$\tablefootmark{b}\\
sqrt $u_{\rm A} = u_{\rm B}$                    & $0.201 \pm 0.085$\tablefootmark{b}\\
sqrt $\nu_{\rm A} = \nu_{\rm B}$                & $0.30 \pm 0.14$\tablefootmark{b}\\
$K_{\rm A}$ $(\rm{km/s})$                       & $83.81 \pm 0.31$\tablefootmark{a} \\
$K_{\rm B}$ $(\rm{km/s})$                       & $83.29 \pm 0.29$\tablefootmark{a}\\
$e\cos\omega$                                   & $-0.000106 \pm 0.000046$\tablefootmark{b}\\
$\gamma_{\rm A} = \gamma_{\rm B}$ $(\rm{km/s})$ & $40.79 \pm 0.16$\tablefootmark{b}\\
\hline
$r_{\rm A}$                                     & $0.08899 \pm 0.00090$\tablefootmark{b}\\
$r_{\rm B}$                                     & $0.0889 \pm 0.0011$\tablefootmark{b}\\
photom. $l_B/l_A$ (from $k$)                    & $1.029 \pm 0.041$\tablefootmark{a,b}\\
$e\sin\omega$                                   & $0.0$ (fixed)\\
$e$                                             & $0.000106 \pm 0.000046$\tablefootmark{b}\\
$\omega$ ($^\circ$)                             & $180.0$ \\
pri. $T_{14}$ (days)                            & $0.2461 \pm 0.0026$ \\
sec. $T_{14}$ (days)                            & $0.2461 \pm 0.0026$ \\
LC RMS ($m$mag)                                 & 0.86 \\
RV RMS A,B (km/s)                               & 0.64, 0.52 \\
\hline
\end{tabular}
\tablefoot{The lower part displays derived (calculated) with fixed quantities indicated. The eclipse duration $T_{14}$ was calculated from \cite{winn2010} and depends on the stellar radius among other parameters. The reduced $\chi^2$ for both data sets is $\approx 1.0$. Uncertainties are $1\sigma$ (68.3\%) confidence intervals chosen to be the larger from MC or RP-PB simulations. The eccentricity assumes $e\sin\omega = 0.0$. The period agrees with the ephemeris period in Eq. \ref{eq_ephemeris2} at a $0.30\sigma$ level.}
\tablefoot{\tablefoottext{a}{from MC.} \tablefoottext{b}{from RP-PB.}}
\end{table}

We investigated the effect of various LD laws (linear, logarithmic, quadratic and square-root) on the fractional radii ($r_A, r_B$). We find no evidence for any systematic errors with a maximum difference at the $0.02\sigma$ level. Third light was not detected at a significant level within the {\it TESS} aperture mask. This result is consistent with the high-resolution Lucky-Imaging observations as well as non-detection of third-light in FEROS spectra indicating the non-detection of stellar companions. As a result, we found that accounting for the limb darkening through the square-root law and the addition of geometric parameter $e\cos\omega$ improved the success of the fits for a 10-parameter model statistically significantly. The non-zero eccentricity cannot be explained by the light-travel time effect because the mass ratio for the two components is very close to unity. Also, no apsidal motion is expected for a circular orbit. The RMS scatter for this 10-parameter model was found to be significantly larger than the mean value of photometric error returned by the {\sc lightkurve} pipeline. We therefore conclude that the original {\it TESS} error bars are underestimated and likely account only for Poisson counting statistics per point. Although we tested with the geometric parameter $e\sin\omega$ from the pool of parameters, {\it TESS} data do not support the inclusion of $e\sin\omega$ (or any other parameter that was tested for in the sequence) beyond what was found. Finally, we note that the $\mathcal{F}$-test is consistent with the general finding that $e\sin\omega$ is less constrained than $e\cos\omega$ \citep{kallrathmilone2009,southworth2013}. Surprisingly, this is also the case for the high-precision {\it TESS} data presented in this work.

For partial eclipses, the ratio of the radii becomes strongly correlated with orbital inclination and surface brightness ratio which implies that the light ratio becomes strongly correlated as well with the ratio of radii. We therefore added a spectroscopic light ratio from FEROS in the {\it TESS} passband ($l_B/l_A = 1.149 \pm 0.035$) as an additional constraint on the light curve model \citep{andersen1990,torres2000,southworth2007}. In Fig.~\ref{fig:LRATExperiments_UpdatedPlot1}, we show the results of including and not including a spectroscopic light ratio for {\it TESS} data. The difference in photometric and spectroscopic light ratio was found at a $0.41\sigma$ level. However, final parameters have been obtained by adopting and fixing $k \pm \delta k$ from the SAT $y$ band constrained with a spectroscopic light ratio which is much more accurately determined due to the narrower passband compared to the {\it TESS} filter. The final modelling included times of minimum light from {\it TESS} as an additional observational constraint on the ephemeris.

\begin{figure*}
\centering
\includegraphics[width=1.0\textwidth]{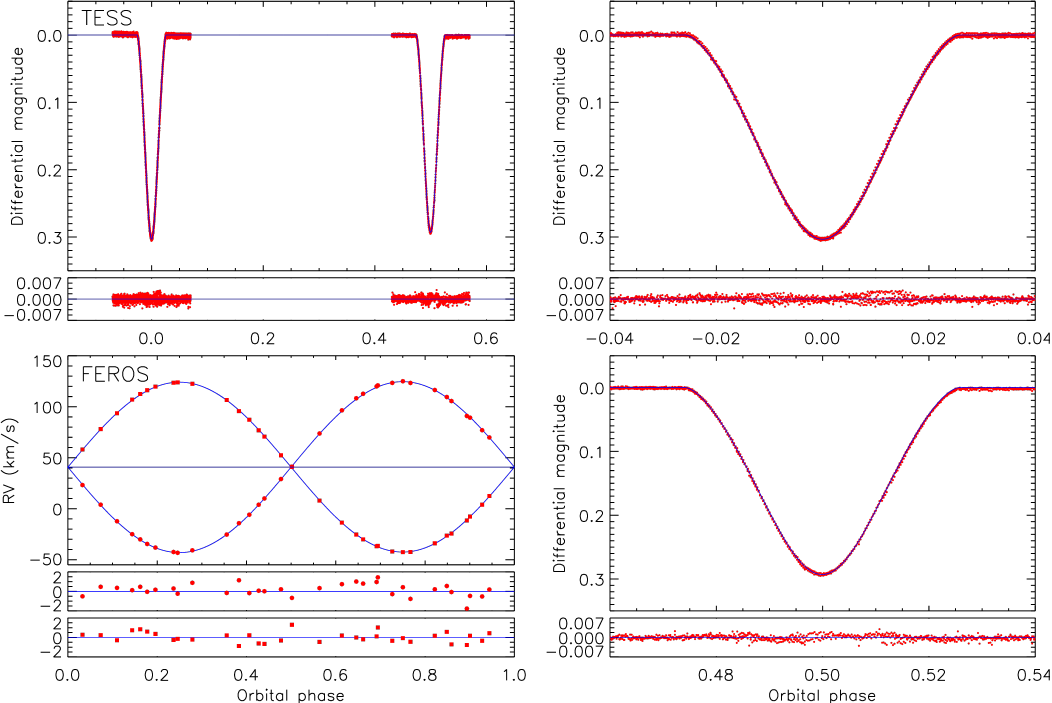}
\caption{Result of simultaneous {\sc jktebop} fitting of NY\,Hya. Observational data are shown in red and best-fit models as solid lines. No error bars are plotted. \emph{Left column}: {\it TESS} light curve (top) and FEROS RV curves (bottom). \emph{Right column}: Details of the primary (top) and secondary (bottom) eclipse clearly showing the presence of time-correlated red noise in the residuals.}
\label{fig:plotLCRV_jkt}
\end{figure*}

\begin{figure*}
\centering
\includegraphics[width=1.0\textwidth]{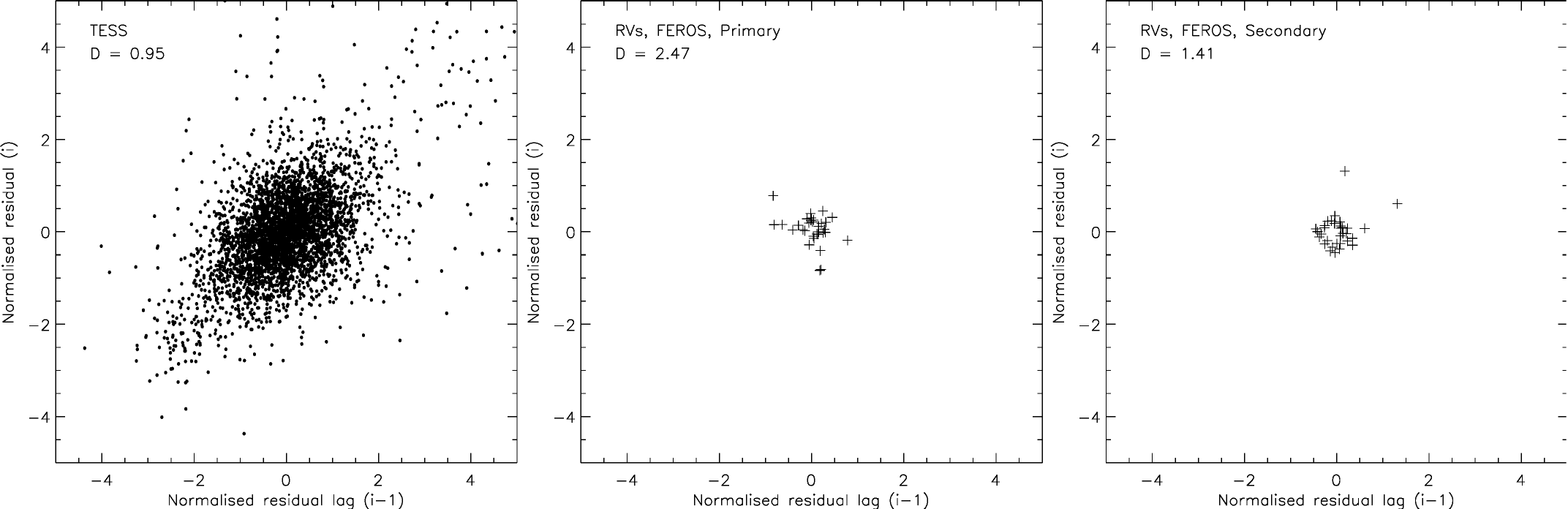}
\caption{Durbin-Watson (lag) plots for the three different data sets: {\it TESS}, RVs of star~A, and RVs of star~B. A $\mathcal{D} = 2.0$ indicates no auto-correlation with data randomly distributed following a standard normal distribution. For the  {\it TESS} data and the  RV data of star~B (based on two outliers) we see a positive auto-correlation, while for the  RVs of star~A we see a negative auto-correlation.}
\label{fig:DWPlots}
\end{figure*}

\subsection*{Final solution and parameter uncertainties}
We fitted clipped {\it TESS} data with RVs simultaneously. Although {\it TESS} is a space-based observing platform we cannot rule out instrument-related correlated noise (pixel-to-pixel variations, temperature fluctuations, etc) or correlated noise of astrophysical nature (star spots) that are not accounted for in the modelling process. We therefore inferred parameter uncertainties from running the classic bootstrap (TASK7), Monte Carlo (TASK8, MC) as well as the Residual-Permutation Prayer-Bead (TASK9, RP-PB) algorithms within {\sc jktebop}. As a test we ran 5000 and 10000 trials for TASK7 and TASK8 and found no difference.

For the {\it TESS} data, we found that the nominal photometric errors as returned by {\sc lightkurve} were underestimated by a factor of roughly 1.6 based on the inherent data scatter. Therefore, for the final uncertainty estimation the input errors were scaled by $\sqrt{\chi^2_{\nu}}$ to force the reduced $\chi^2 = 1.0$. The scaling relies on several assumptions \citep{andrae2010}: $i)$ errors follow a Gaussian distribution, $ii)$ the model is linear in all parameters and $iii)$ the applied model is a \emph{correct} description of the data. In Fig.\,\ref{fig:plotLCRV_jkt}, we show the best-fit model and Table \ref{tab:tessfinal} lists the final best-fit parameters. We did not detect any skewness in the resulting parameter parent distributions for each algorithm and thus report a symmetric $1\sigma$ uncertainty. Parameter uncertainty covariances and parameter parent distributions for light curve and RV data are shown in Figs. \ref{fig:corner_lc_task8} and \ref{fig:corner_rv_task8} in the appendix. We found the PB-RP algorithm to yield the largest uncertainty for most parameters and refer to Table \ref{tab:tessfinal} for details. In general, we adopt and report the largest uncertainty for a conservative estimate and refer to \citet{turner2016} for a discussion on realistic uncertainties.

Finally, we performed a quantitative auto-correlation test on the residuals of photometric and RV data. This test aims to detect any non-random structure or systematics in the residuals indicating a deviation from a standard normal distribution. For this, we calculated the Durbin-Watson ($\mathcal{D}$) statistic \citep{hugheshase2010} for each data set and found $\mathcal{D}=0.95, 2.47$ and 1.41 for the {\it TESS}, RV (star~A) and RV (star~B) data sets, respectively. We show the resulting DW lag plots in Fig.~\ref{fig:DWPlots}. Relatively, the auto-correlation in the {\it TESS} data is stronger compared to the two spectroscopic data sets. This result quantitatively confirms the presence of systematic noise in the {\it TESS} data most prominently seen during primary and secondary eclipses.

\section{Physical properites}
\label{physicalproperties} 

\subsection{Spectral disentangling}
\label{spectraldisentangling}

In order to disentangle our spectra and obtain a single spectrum for each of the components, we first performed a preliminary fit to our RV measurements with {\sc todcor} by making use of the {\sc idl} code 
{\sc rvfit.pro} \citep{rvfit2015}, and derived a preliminary set of orbital parameters for the system, which turned out to be in good agreement with the results from {\sc jktebop} presented in Table~\ref{tab:tessfinal} and hence support our findings from the simultaneous modelling of light and radial velocity curves. We made use of the {\sc fdbinary} code \citep[Ver. 3.0]{ilijic2004} to disentangle our continuum-normalised, composite spectra of NY\,Hya. We converted the wavelength axis of the spectra to logarithmic units and sampled each spectrum equidistantly by making use of cubic spline interpolation. We assigned the orbital phases computed from the linear ephemeris in Eq.~(\ref{eq_ephemeris2}) to each of the processed input spectra. We made use of four of the orbital elements (eccentricity, the longitude of the periastron and semi-amplitudes) which we derived from our preliminary RV fit. Since we did not detect a phase shift or apsidal motion in the system, we have fixed the relevant parameters to zero. We set the number of optimisation runs to 1000 and the number of iterations to 10000.

Spectral disentangling is performed in Fourier space using the Keplerian RVs based on five orbital parameters by {\sc fdbinary}. Light ratios form an optional parameter set. We experimented using the light ratios we derived from {\sc todcor} as well as adjusting them (fixing its value to one for each input spectrum). We had very similar results in both attempts in terms of the quality (expressed with the $S/N$) and depths of the spectral lines. The code outputs the disentangled spectra as well as the residuals $(O-C)$ and the RVs. We compared the RVs computed by {\sc fdbinary} with that from other independent methods and assessed its results as consistent with the others. 

\subsection{Stellar atmospheric parameters}
\label{stellaratmosphericparameters}

The SED analysis, described in more detail in Section \ref{spectralenergydistributionmodeling}, provides an estimate for atmospheric parameters of the system. Therefore, we decided to make use of the synthetic spectrum fitting technique with these estimates \citep{valentifischer2005}, better suited to estimate the \Teff of stars with heavy line blending, which becomes an important problem in the spectral analysis of stars on the cooler side of 5500 K. Since other methods heavily rely on EWs of individual lines and their ratios, they are not deemed to be optimum to extract stellar atmospheric parameters \citep{tsantaki2013}.

We used the latest version of {\sc ispec} \citep{ispec2} in order to derive the stellar atmospheric parameters of each component. We first corrected both of our disentangled spectra for the systemic velocity. We made use of the rough estimates for T$_{\rm eff}$ (5550 and 5500 K, for star A and B respectively), $\log g$ (4.15 for both components), and [M/H] (solar abundance) as initial parameters. We experimented with different amounts of line broadening to determine an initial value for the projected rotational velocity $(v \sin i)$ by employing the template spectra we used in the measurement of the RVs with {\sc todcor} for the purpose and determined a value of 15 km/s. Since we expect the components to be tidally locked we assumed the same rotational velocity for both of the components. We used the {\sc synth3} code \citep{kochukhov2007} for computation, {\sc atlas9} \citep{kurucz2003} for the stellar atmosphere model, solar abundances from \citet{asplund2009}, and a modified version of the {\sc vald} line list in the wavelength ranges of our spectra \citep{piskunov1995}.

\begin{figure}
\centering
\includegraphics[angle=0,width=0.47\textwidth]{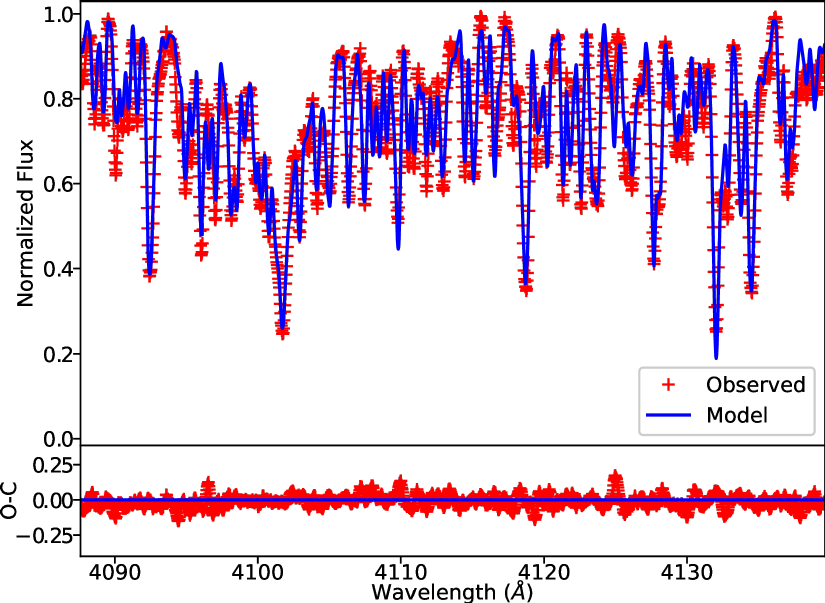}
\includegraphics[angle=0,width=0.47\textwidth]{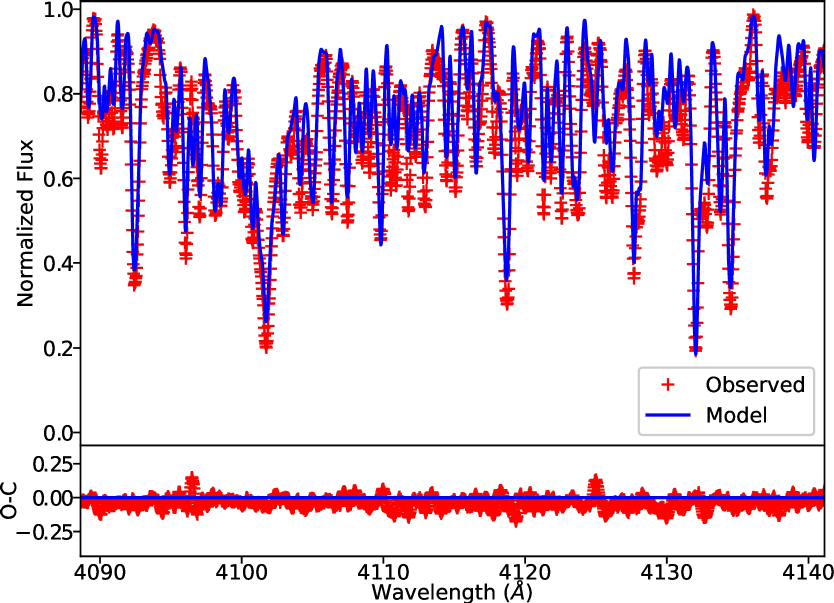}
\caption{Model and disentangled observed FEROS spectrum in the 4090--4140 \si{\angstrom} region. 
\emph{Top panel}: Disentangled spectrum of star A (plus symbols) and its best model. \emph{Bottom panel}: Same content, but for star B.}
\label{fig:model_spectrum_4090_4140}
\end{figure}

\begin{figure}
\centering
\includegraphics[angle=0,width=0.47\textwidth]{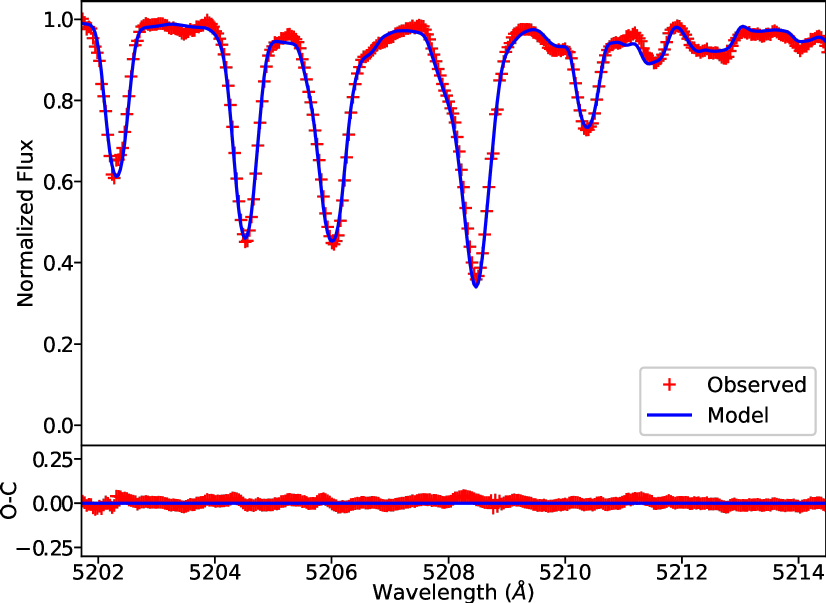}
\includegraphics[angle=0,width=0.47\textwidth]{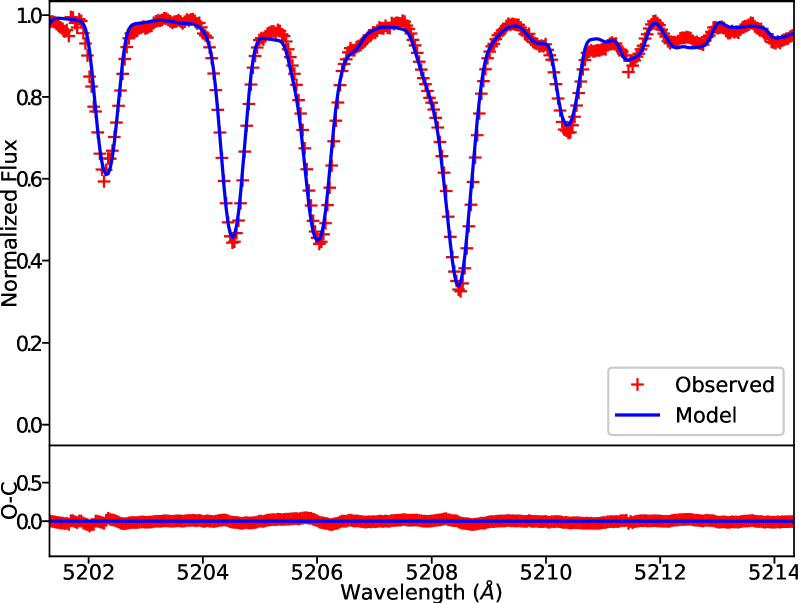}
\caption{Same as Fig.~\ref{fig:model_spectrum_4090_4140}, but   for the 5202--5214 \si{\angstrom} region for star A (top panel) and star B (bottom panel).}
\label{fig:model_spectrum_5202_5214}
\end{figure}

We combined all disentangled spectra from four different wavelength regions to achieve a global fit consistent with all of them for both of the spectra separately. We then selected the entire combined spectra from four different wavelength regions with {\sc ispec} to fit with the least-squares minimisation method. {\sc ispec} also filters out the lines from its analysis for which a Gaussian fit fails. We then ran the code and had our first fits, which agreed mostly with the observed spectra. However, we had very large error bars in all parameters and significantly lower value ($\log g_A \sim 3.95$ cgs) compared to that we have from the first light curve modelling attempts ($\log g_A = 4.183$, $\log g_B = 4.188$). Surface gravity is the least dominant factor in shaping the spectra of cool dwarfs. Therefore, it leads to significant degeneracies in the fitting procedure and large error bars in all the parameters. Since its effect is only marginal in synthetic spectrum fitting, it will be more adequate to determine $\log g$ from light- and RV curve analysis. Therefore, we fixed the surface gravity at $\log g_A = 4.2059$ for star A and  $\log g = 4.210$ for star B from their measured masses and radii. We found that fixing the parameters to these values significantly decreased the fitting uncertainties on other parameters. We provide the values of the atmospheric parameters for each of the components derived from modelling of the disentangled spectra in Table \ref{tab:atmospheric_parameters}. In Figs.\ \ref{fig:model_spectrum_4090_4140} and \ref{fig:model_spectrum_5202_5214}, we also provide two spectra for each of the components (star A at the top, star B at the bottom) from two different wavelength regions to illustrate the success of our models from the synthetic spectrum fitting.

\begin{table}
\caption{Atmospheric parameters of the components in NY\,Hya.}
\label{tab:atmospheric_parameters}
\centering
\begin{tabular}{lcc}  
\hline
Parameter & Star A & Star B \\
\hline
\hline
$T_{\rm eff}$ (K)    & $5595\pm61$     & $5607\pm61$ \\
$\log g$ (cgs)       & $4.2059$ (fixed) & $4.210$ (fixed) \\
${\rm [M/H]}$ (dex)  & $0.00$ (fixed)  & $0.00\pm0.05$ \\
${\rm [Fe/H]}$ (dex) & $-0.01\pm0.06$  & $0.02\pm0.07$ \\
$v\sin i$ (km/s)     & $15.77\pm1.34$  & $15.71\pm1.35$ \\
$v_{\rm mic}$ (km/s) & $1.05$          & $1.05$ \\
$v_{\rm mac}$ (km/s) & $3.62$          & $3.66$ \\
\hline
\end{tabular}
\end{table}

\begin{figure*}
\centering
\includegraphics[width=0.48\textwidth,angle=0]{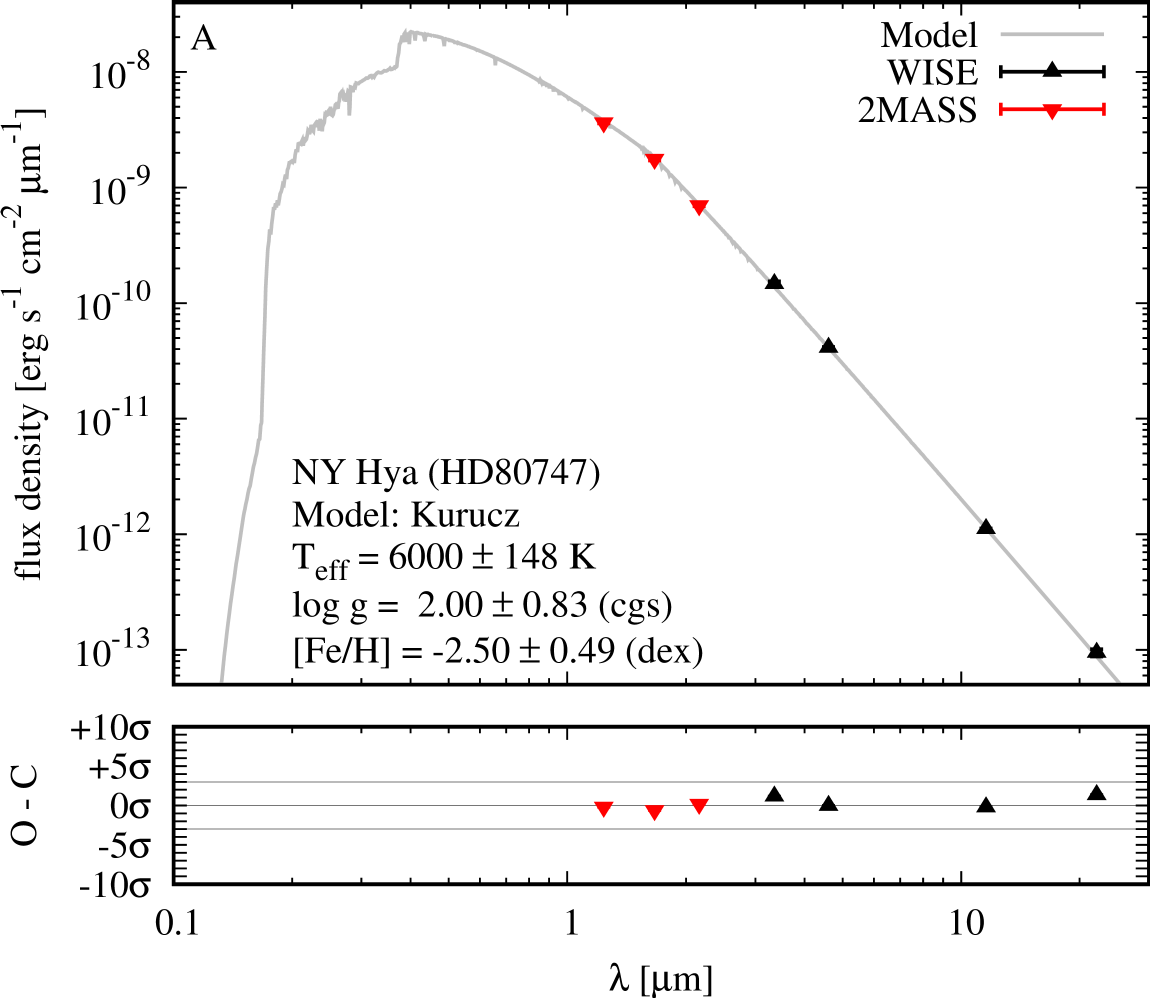}
\includegraphics[width=0.48\textwidth,angle=0]{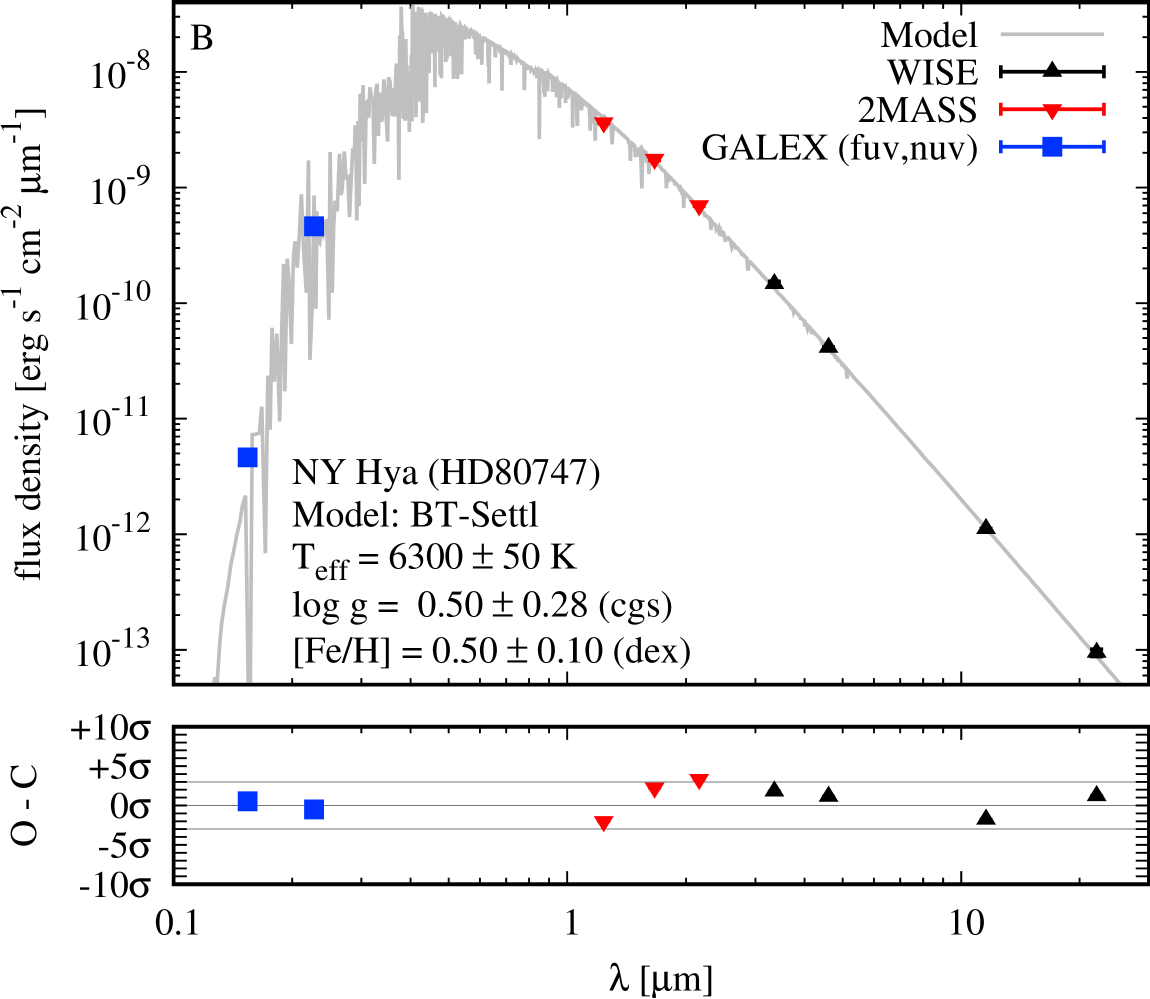}
\includegraphics[width=0.48\textwidth,angle=0]{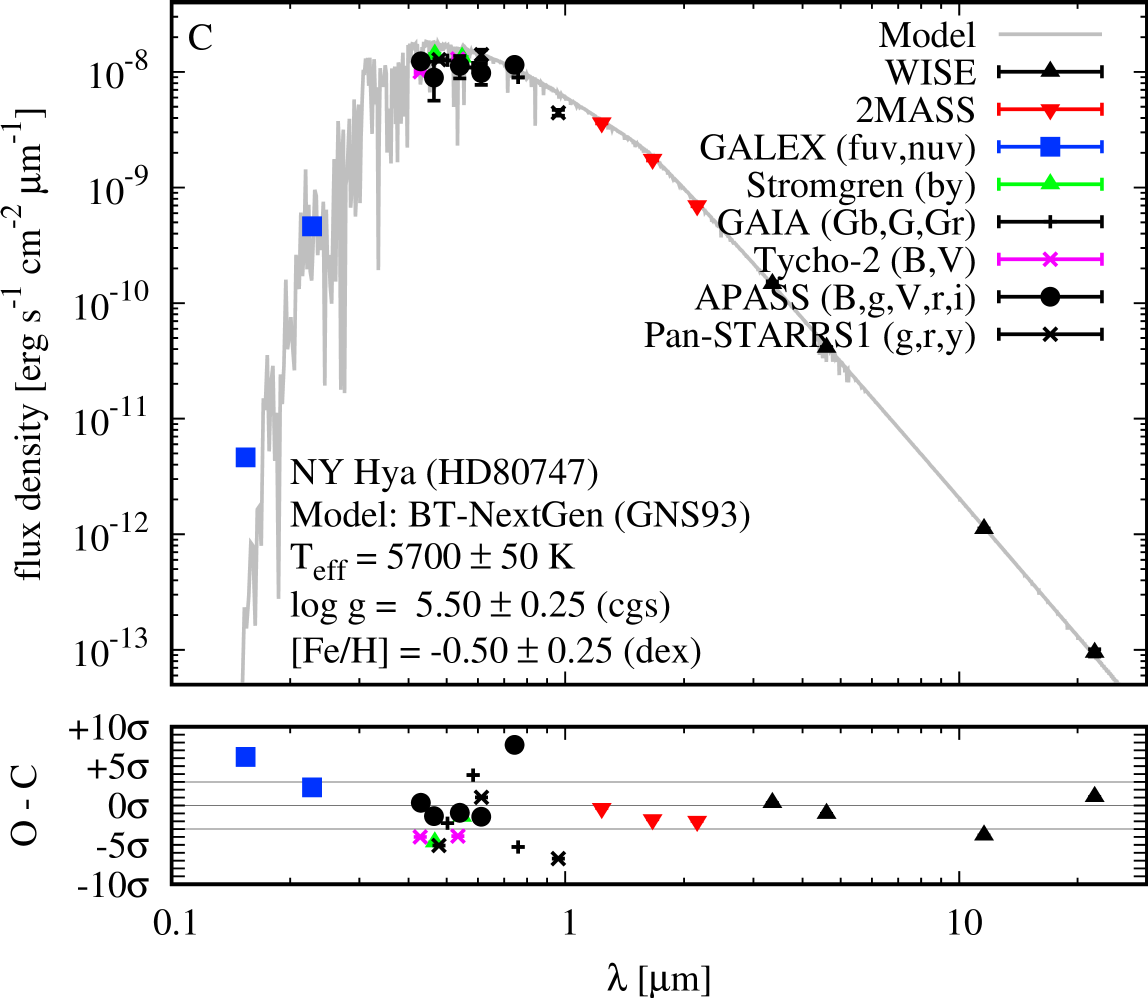}
\includegraphics[width=0.48\textwidth,angle=0]{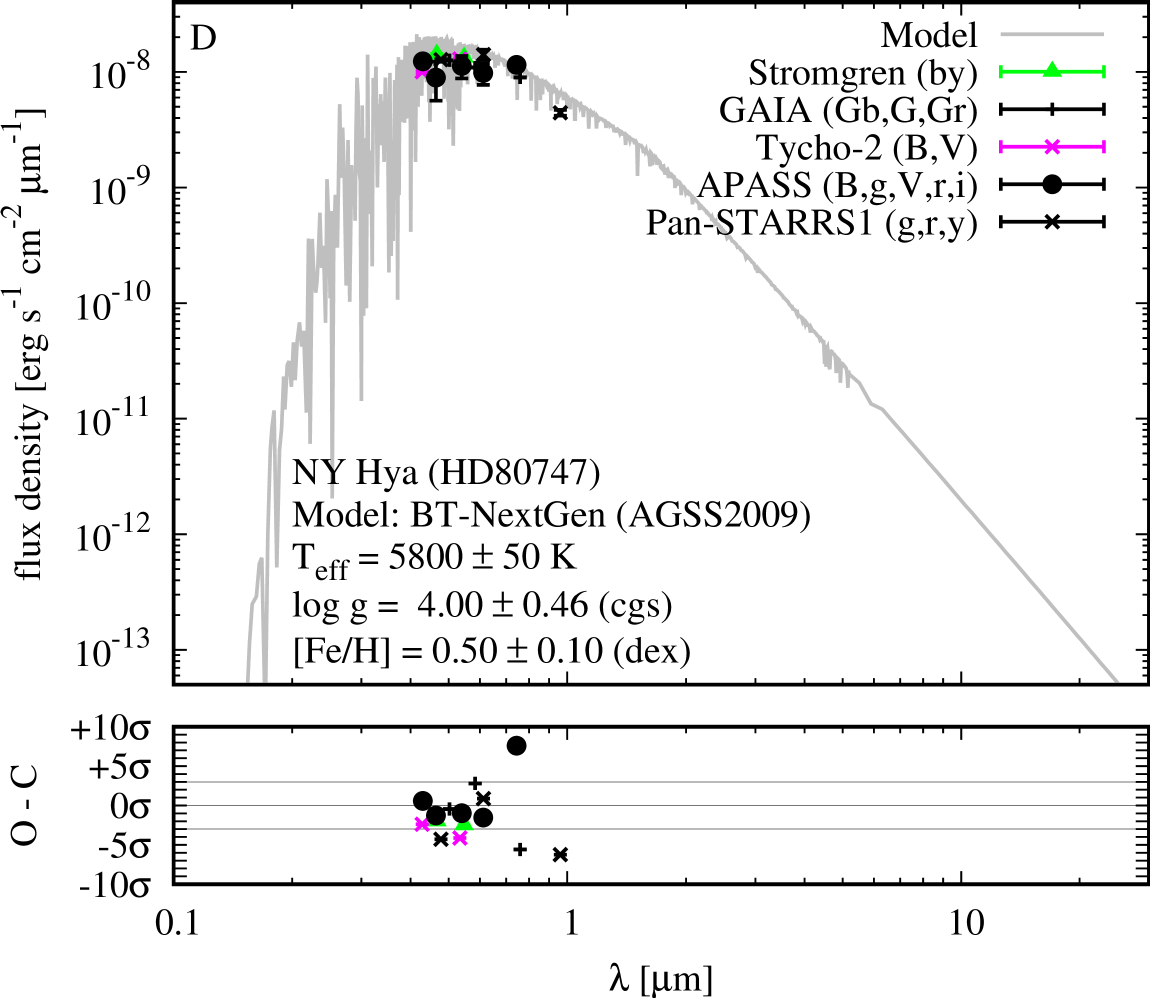}
\caption{VOSA SED model fits (corrected for extinction) for various data sets of flux measurements for NY\,Hya. The horizontal lines in the residual plots mark $\pm 3\sigma$ deviation from the model. \emph{Panel A}: WISE and 2MASS flux measurements $(\chi^2_{\nu} = 1.2)$. \emph{Panel B}: WISE, 2MASS, and GALEX fluxes $(\chi^2_{\nu} = 6.0)$. \emph{Panel C}: Almost all available archive data with <$10\sigma$ deviations $(\chi^2_{\nu} = 15.3)$ (see Table \ref{tab:photo_info} for details). \emph{Panel D}: Only retaining flux measurements around 7000 \si{\angstrom} $(\chi^2_{\nu} = 17.6)$. The APASS Sloan $i$ measurement   deviates  at a $8-9\sigma$ level (see text for details). Some error bars are smaller than the symbol size. In all cases, no infrared excess was detected eventually pointing towards a circumstellar disc and/or a cool companion. We also note that data for each model has been thinned to reduce data size.}
\label{fig:sed}
\end{figure*}

We attempted to estimate an age for the system by employing different techniques for the purpose. We found neither core-emission signals in the H$_{\alpha}$ line nor the signs of lithium absorption in the spectra recorded at convenient orbital phases for the job. We also employed the chromospheric activity indicators $S_{HK}$ and $\log R^{'}_{HK}$ determined as 0.37 and -4.5, respectively by \cite{isaacsonfisher2010} from three low S/N ($\simeq 35$ each) spectra they acquired at the orbital phases 0.65, 0.86 and 0.28 as a result of which they reported an age of 0.47 Gyr for NY\,Hya using calibration relations from \citet{mamajekhillenbrand2008}. These measurements should only be regarded as an upper limit because any flux velocity-shifted out of the passband will cause this number to increase to more-active levels when the binarity is ignored. Applying the empirical X-ray activity based on the fractional X-ray luminosity ($R_X = L_X/L_{\rm Bol}$) vs age relation from \cite{mamajekhillenbrand2008}, we derived an age of 0.19 Gyr. However, we defer the age determination to a comprehensive analysis of the system's evolutionary history.

\subsection{Interstellar extinction and reddening}
\label{interstellarextinctionandreddening}

Correcting for extinction is important to infer an accurate \Teff\ estimate from 
broad- and intermediate-band photometry predominantly in the blue wavelength region. Following
\citet{mann2015} and \citet{aumerbinney2009}, the effect of extinction/reddening is essentially zero for
stars within 70 pc due to a low-density cavity defining the `Local Bubble' in the solar neighbourhood. This finding is consistent with \citet{reis2011} and \citet{bessel1998} suggesting that reddening may be ignored for stars within 100 pc \citep{perry1982}. Therefore, towards the direction of NY\,Hya, galactic coordinates $(l,b)=(239^{\circ},29^{\circ})$, a small and possibly negligible amount of extinction is expected at the measured distance of around 106 pc.

\begin{table*}[hbt!]
\caption{Extinction-corrected Johnson $V$ magnitude (from extinction law) and de-reddened Str\"omgren colour indices for NY\,Hya and (constant) check stars.}
\label{dereddened}
\centering                         
\begin{tabular}{ccccc}      
\hline
Object & $V_0$ (mag) & $(b-y)_0$ (mag) & $m_0$ (mag) & $c_0$ (mag) \\
\hline                                   
\hline                                   
NY\,Hya   & $8.564 \pm 0.057$ & $0.439 \pm 0.019$ & $0.247 \pm 0.0093$ & $0.3491 \pm 0.0078$ \\
HD\,82074 & $6.250 \pm 0.044$ & $0.503 \pm 0.011$ & $0.2872 \pm 0.0069$ & $0.3059 \pm 0.0063$ \\
HD\,80446 & $9.138 \pm 0.048$ & $0.380 \pm 0.012$ & $0.1995 \pm 0.0088$ & $0.3137 \pm 0.0092$ \\
HD\,80633 & $7.385 \pm 0.047$ & $0.306 \pm 0.012$ & $0.1622 \pm 0.0079$ & $0.4239 \pm 0.0073$ \\
\hline
\end{tabular}
\tablefoot{NY\,Hya is considered as a single star. The values for NY\,Hya are valid at orbital phase 0.25. Data for the comparison--check stars were obtained from {\sc stilism} with the reddening $E(B-V)$ evaluated at their respective distances (all three are within 78 pc) (see text for details).}
\end{table*}

The reddening $E(B-V)$ can be estimated from maps of galactic dust, based on which colour indices can be de-reddened and an interstellar extinction value ($A_V$) can be derived. We made use of the dust maps provided by \citet{sfd1998}, \cite{schlaflyfinkbeiner2011}, \citet{green2014, green2018}, \citet{stilism1} and \citet{stilism2} to estimate the reddening value. In that regard, as pointed out by the referee, the \citet{sfd1998} and \cite{schlaflyfinkbeiner2011} dust maps\footnote{NASA/IPAC Infrared Science Archive; \url{https://irsa.ipac.caltech.edu/applications/DUST/}} estimate the {\it total} reddening towards the direction of NY\,Hya to infinity. When making use of the NASA/IPAC service no distance information is provided. This approach would overestimate the reddening for NY\,Hya. Reddening/extinction at the distance of NY\,Hya is given by the {\sc argonaut}\footnote{\url{http://argonaut.skymaps.info}} \citep{green2014, green2018} and {\sc stilism}\footnote{\url{https://stilism.obspm.fr}} \citep{stilism1,stilism2} tools. Both projects yielded a reddening consistent with zero. From {\sc argonaut} we found $E(B-V) = 0.01_{-0.01}^{+0.02}$ mag. From {\sc stilism} we found $E(B-V) = 0.0030 \pm 0.015$ mag. Formally, the resulting (conservative) weighted mean for the reddening at NY\,Hya is $E(B-V) = 0.0065 \pm 0.025$ mag (consistent with zero) and was adopted in subsequent calculations. Reddening values for HD\,82074, HD\,80446 and HD\,80633 were found using {\sc stilism}.

We then de-reddened the SAT Str\"omgren colour indices using the reddening relations given by \citet{crawford1975} and \citet{crawford1976}. The colour excess for the $(b-y)$ colour is given as $E(b-y) = 0.74 E(B-V)$ \citep{crawford1975}. The remaining (updated) relations are found from \citet{crawford1976} which were also applied by \citet{casagrande2011} and are valid for stars covering a wide range of spectral types including late-type F-, G- and M-dwarfs. In particular, the relations $E(m_1) = (-0.3333 \pm 0.0051) E(b-y)$, $E(c_1) = (0.1871 \pm 0.0071) E(b-y)$ were implemented in the {\sc idl}\footnote{\url{https://www.l3harrisgeospatial.com}} routine 
{\sc deredd.pro} and applied in \citet{southworth2004}. In Table \ref{dereddened}, we quote the reddening-free colour indices. Uncertainties were obtained from Monte Carlo simulations assuming Gaussian errors.

We calculated the interstellar extinction using the extinction laws $A_V = (3.14 \pm 0.10) E(B-V)$ \cite{schultz1975,fitzpatrick1999} and $A_V = (4.273 \pm 0.012) E(b-y)$ \citep{crawford1976}. We find very good agreements between the various methods at a $0.05\sigma$ level. In Table \ref{dereddened}, we quote the extinction-free Johnson $V$ magnitudes for NY\,Hya and comparison/check stars. We quote the value obtained from the extinction law. Uncertainties were also obtained from Monte Carlo simulations assuming Gaussian errors.

In addition, the colour excess $E$ can be calculated via intrinsic colour calibration relations based on Str\"omgren $uvby-\beta$ photometry. We checked the range of validity of two calibration relations \citep{SN1989, KS2010} for the measured colour indices and $\beta$ value measured for NY\,Hya, both of which are valid for late F- to G-dwarfs. We find good agreement $(<1\sigma)$ between the values we found based on both calibrations and also with our previous estimates presented in Table \ref{dereddened}. In conclusion, the colour excess estimates indicate a very small amount of reddening with $E(b-y)$ nearly being consistent with zero. This implies that interstellar absorption/scattering is small at the direction and distance of NY\,Hya.

\section{Spectral energy distribution modelling}
\label{spectralenergydistributionmodeling}

We compiled archive broad-band (reddened) photometry covering the wavelength range from far-ultraviolet to mid-infrared to determine a SED model for NY\,Hya as a single star. The near-identical nature of the two stars justifies this and provides a unique opportunity to determine a relatively accurate estimate of atmospheric properties. The \Teff\ is most reliable since the broad-band photometry aims at measuring the flux continuum and therefore probes various slopes of the SED depending on wavelength. We assume that the archive photometry data regards NY\,Hya as a single star. The resulting \Teff\ will be a mean value which will be close to the true values for each star. The derived flux will be overestimated by a factor of two. We first present some details of photometric sky surveys relevant for NY\,Hya.

\subsection{Broad-band photometry}
\label{broadbandphotometry}

We obtained broad-band magnitudes of NY\,Hya from different photometric archives. Str\"omgren $uvby$ magnitudes can be reconstructed from Str\"omgren colour indices $(b-y), m_1, c_1$ and from assuming Johnson $V$ equals Str\"omgren $y$. This assumption is valid since the two filter transmission profiles are near-identical (with $V$ more broader than $y$) and generally there are no strong features in the $V$ bandpass. Furthermore, the transformation from the instrumental Str\"omgren $y$ ($y_{\rm instr}$) to the standard Johnson $V$, based on a large number of bright standard stars, is of the form $V = A + B(b-y)_{\rm st} + y_{\rm instr}$ \citep{crawford1970}, where the coefficient $B$ is around 0.02. Since $(b-y)_{\rm st}$ varies from about 0.000 for an A0V star to about 0.500 for a K1V star, the difference between Johnson $V$ and $y_{\rm instr}$ for most of the here relevant part of the main-sequence stars is between 0.000 and 0.010 mag (apart from the arbitrary zero point $A$). On nearly all 63 nights, the number of bright standard stars was sufficient to ensure a correct transformation of $y_{\rm instr}$ to Johnson $V$.

As noted earlier, \cite{paunzen2015} provides Str\"omgren data for NY\,Hya in his catalogue, but we deem those measurements not trustworthy. Standard Str\"omgren magnitudes are given as

\begin{eqnarray}
    y   &=& V\, ({\rm assumed}),\\
    b   &=& (b-y) + y, \\
    v   &=& m_1 + 2(b-y) + y, \\
    u   &=& c_1 + 2m_1 + 3(b-y) + y,
\end{eqnarray}
\noindent
where the associated $1\sigma$ uncertainties are found from standard error propagation, assuming no correlations between uncertainties and independent variables and are given as
\begin{eqnarray}
    \sigma_{y} &=& \sigma_{V}\, ({\rm assumed}), \\
    \sigma_{b} &=& \sqrt{\sigma_{b-y}^2 + \sigma_{y}^2}, \\
    \sigma_{v} &=& \sqrt{\sigma_{m_1}^2 + 4\sigma_{b-y}^2 + \sigma_{y}^2}, \\
    \sigma_{u} &=& \sqrt{\sigma_{c_1}^2 + 4\sigma_{m_1}^2 + 9\sigma_{b-y}^2 + \sigma_{y}^2}.
\end{eqnarray}
\noindent
We have implemented the above equations and numerically propagated errors via Monte Carlo simulations and can confirm the validity of the expressions in uncertainties. We used the $V$ band estimate and Str\"omgren colour indices at 
phase 0.25 (cf. Table~\ref{nyhya_photometric_data}). However, we find that the $u$ and $v$ band estimates are systematically deviating $>10\sigma$ in all our SED models. We have therefore not included those points.

We also decided to discard the DENIS\footnote{http://cds.u-strasbg.fr/denis.html} (Deep Near Infrared Survey of the Southern Sky, \citet{denis}) $I, J, K_S$ photometry, since the DENIS Johnson $J$ ($1.25\, \mu m$) and $K_S$ ($2.16\, \mu m$) magnitudes are {\it nearly} identical to the 2MASS \citep{2MASS} photometry. Given the higher 2MASS photometric precision (indicated by good survey quality flags), we chose to only retain the 2MASS data. 2MASS $JHK_S$ observations were taken at epoch JD 2,451,199.8029. This corresponds to an orbital phase of 0.17 for NY\,Hya and hence reflects an out-of-eclipse brightness in each passband. We also ignored the SkyMapper data recorded in similar passbands ($(u,v,g,r,i,z)$) to other all-sky surveys because they are based on fewer images, even though they are recorded during out-of-eclipse phases.

We have retrieved raw data from {\it APASS} (American Association of Variable Star Observers Photometric All Sky Survey, \citet{apass}) via the URAT1 (US Naval Observatory Robotic Astrometric Telescope, \citet{urat1}) catalogue. While the original APASS (DR9) archive does not provide a Sloan $i$-band magnitude and measurement uncertainties in all five APASS passbands, the URAT1 catalogue does. However, the data reported in the URAT1 catalogue does not match the APASS (DR9) catalogue listings. We therefore obtained unpublished photometry, which were acquired at times outside eclipses, following a conversation with A.\ Henden, and calculated their average values and standard deviations for each filter. Comparing APASS $V_J$ with Str\"omgren $y$ and Tycho-2 $V_T$, we find good agreement and use them in the SED modelling as provided in Table \ref{tab:photo_info}.

All {\it PanSTARRS} \citep{panstarrs1, panstarrs2} data have been included in our analysis except for the $i_{P1}$ point which we found to deviate by $>10\sigma$ for all our SED models and hence seem unreliable.

The recently published {\it Gaia} DR2 data ($G_{BP}, G, G_{RP}$, \citealt{GAIA2018}) were included in our SED analysis too. We converted the calibrated flux into magnitudes using the revised zero-points for each pass-band. Uncertainties were propagated via Monte Carlo simulations. The resulting distribution in magnitude-space was found to be near-Gaussian. The median value and 68\% confidence interval ($\pm 1\sigma$) was determined to infer uncertainties in the three magnitudes.

\begin{table}
    \caption{Compiled archive photometric measurements of NY\,Hya.}

    \centering
    \begin{tabular}{l c c r}
    \hline
    
    \multicolumn{3}{c}{Photometry} & SED \\
    \hline
    \hline

    $fuv$ (mag)          & $19.989      \pm 0.128$ & GALEX & yes\\
    $nuv$ (mag)          & $14.146      \pm 0.006$ & GALEX & yes\\
    
    $g_{P1}$ (mag)          & $8.9321 \pm 0.0140$ & Pan-STARRS1 & yes \\
    $r_{P1}$ (mag)          & $8.2799 \pm 0.1136$ & Pan-STARRS1 & yes \\
    $i_{P1}$ (mag)          & $8.1992 \pm 0.0010$ & Pan-STARRS1 & no \\
    $y_{P1}$ (mag)          & $8.5613 \pm 0.0650$ & Pan-STARRS1 & yes \\
    
    $G_{\rm BP}$ (mag)      & $8.764  \pm 0.0031$ & {\it Gaia} (DR2) & yes \\
    $G$      (mag)          & $8.396  \pm 0.0013$ & {\it Gaia} (DR2) & yes \\
    $G_{\rm RP}$ (mag)      & $7.898  \pm 0.0046$ & {\it Gaia} (DR2) & yes \\
    
    $B_T$    (mag)          & $9.543   \pm 0.026$  & Tycho-2 & yes \\
    $V_T$    (mag)          & $8.723   \pm 0.018$  & Tycho-2 & yes \\
    
    

    $B_J$  (mag)            & $9.30 \pm 0.12$ & APASS & yes \\
    $g'$   (mag)            & $9.38 \pm 0.40$ & APASS & yes \\
    $V_J$  (mag)            & $8.80 \pm 0.24$ & APASS & yes \\
    $r'$   (mag)            & $8.68 \pm 0.23$ & APASS & yes \\
    $i'$   (mag)            & $8.08 \pm 0.02$ & APASS & yes \\

    $u$ (mag) & $10.756 \pm 0.025$ & this work & no \\
    $v$ (mag) & $ 9.717 \pm 0.019$ & this work & no \\
    $b$ (mag) & $ 9.028 \pm 0.017$ & this work & yes \\
    $y$ (mag) & $ 8.584 \pm 0.016$ & this work & yes \\

    $J$     (mag)           & $7.330   \pm 0.020$  & 2MASS & yes \\
    $H$     (mag)           & $7.016   \pm 0.033$  & 2MASS & yes \\
    $K_S$   (mag)           & $6.961   \pm 0.023$  & 2MASS & yes \\
    
    $W1$     (mag)          & $6.868   \pm 0.0580$ & WISE & yes \\
    $W2$     (mag)          & $6.924   \pm 0.019$  & WISE & yes \\
    $W3$     (mag)          & $6.923   \pm 0.016$  & WISE & yes \\
    $W4$     (mag)          & $6.828   \pm 0.078$  & WISE & yes \\

\hline
\end{tabular}
\label{tab:photo_info}
\tablefoot{Notes on various missions and surveys: GALEX (Galaxy Evolution Explorer, \citet{GALEX}) measures near-UV ($nuv$) and far-UV ($fuv$) fluxes. Tycho-2 \citep{hog2000} measures $B_T$ and $V_T$ fluxes. {\it Gaia} (Global Astrometric Interferometer for Astrophysics, \citet{GAIA2018}). {\it PanSTARRS} (Panoramic Survey Telescope and Rapid Response System, \citet{panstarrs1,panstarrs2}). WISE (Wide-field Infrared Survey Explorer, \citet{wise}). Data points that were included in Fig.\,\ref{fig:sed}C are indicated with `yes'.}
\end{table}

\subsection{SED modelling}
\label{SEDmodeling}

We utilised the Virtual Observatory SED Analysis ({\sc vosa}\footnote{\url{http://svo2.cab.inta-csic.es/theory/vosa/}} Ver.\ 6.0) online tool \citep{bayo2008}. {\sc vosa} derives stellar properties using theoretical atmosphere models from which synthetic fluxes are calculated to fit observations of NY\,Hya. All archival photometry has been compiled in Table \ref{tab:photo_info}. Astrophysical constants follow the IAU 2015 Resolution B3 recommendation for Solar values \citep{IAU2015-Res-B3}. All archival photometry is corrected for extinction using $R = 3.14$ and $E(B-V) = 0.0065$ mag (see  the previous section),  thus avoiding a possible underestimation of the \Teff. We also provided a {\it Gaia} DR2 parallax to {\sc vosa} allowing a semi-empirical estimate of the total flux emitted and derived stellar radius.


The five SED model parameters are \Teff, surface gravity $(\log g)$, metallicity ([m/H]), extinction $(A_V)$ (by providing $R = 3.14$ and $E(B-V) = 0.0065$ mag) and a flux density proportionality factor $(M_d)$ to scale the theoretical spectrum to observed fluxes. The surface gravity and metallicity are generally poorly constrained from broad-band photometry and are therefore the least accurate quantities. The reduced $\chi^2$ statistic ($\chi^2_{\nu}$) is used to assess the quality of the fit. As a rule of thumb (C.\ Rodriges, private communication) reasonable accurate SED models have $\chi^2_{\nu}<50$. Errors were found from a Monte Carlo bootstrapping algorithm and are mainly limited by the parameter grid mesh for a given atmosphere model. We searched for best-fit models from the following atmosphere models: {\sc kurucz} \citep{kurucz2003}, {\sc coelho} \citep{coelho2014} and a suite of {\sc phoenix}\footnote{\url{https://phoenix.ens-lyon.fr}} \citep{allard1994, allard2012} models: {\sc bt-nextgen-gns93}, {\sc bt-nextgen-agss2009}, {\sc bt-settl} {\sc bt-cond}, {\sc bt-dusty}: \citep{allard2012}, {\sc bt-settl-cifist} \citep{btsettlcifist}, {\sc nextgen} \citep{nextgen}. No parameter limits were chosen a priori. 

{ In Table \ref{tab:sed_models} we list the first few best-fit models in order of increasing $\chi^2_{\nu}$;  the first model is shown in Fig.~\ref{fig:sed}C. We note that SED models with an outdated solar composition \citep{grevessenoels1993,grevessesauval1998} tend to systematically yield lower \Teff\ by about 200~K compared to atmosphere models adopting more recent \citep{grevesse2007,asplund2009,caffau2009} solar abundances. Considering the formal uncertainties in $T_{\rm eff}$, this difference between the competing models is significant.} In Fig.~\ref{fig:sed}, we show four different SED models considering various data sets. Formally, the best-fit model yields a temperature of $T_{\rm eff} = 5700$ K, which agrees with the {\it Gaia} \citep{GAIA2018} mean estimate to within $1.7\sigma$ although the SED uncertainties of 50 K is generally judged to be likely too optimistic by community members. To obtain an idea of a realistic uncertainty, we refer to the {\sc vosa} documentation where the modelling performance was investigated by comparing model estimates with high-quality \Teff\ estimates of FGK stars from the ELODIE V3.1 \citep{prugniel2007} catalogue. The catalogue was cross-matched with the {\it Gaia} EDR3 catalogue retaining stars with a parallax $> 10$ $m$as to avoid extinction. For a statistically large sample and considering the {\sc bt-settl-agss2007} and {\sc kurucz} models, mean differences in the range 6--50 K were determined. From this, we conclude that the uncertainties in $T_{\rm eff}$ in Table \ref{tab:sed_models} seem to be realistic given that NY\,Hya has a parallax of $\pi = 9.4$ $m$as.

\begin{table*}
    \caption{Results of the first few best-fit (extinction corrected) SED models for the data in Table \ref{tab:photo_info}.}
    \centering
    \begin{tabular}{ccccccc}
    \hline
SED model             & $\chi^2_{\nu}$ & $T_{\rm eff}$ (K) & $\log g\,\rm (cgs)$ & [m/H] (dex) & $F_{\rm tot}$ ($10^{-9} \rm erg/\rm cm^{2}/\rm s$) & $L_{\rm bol}~({\rm L}_{\odot})$ \\ 
\hline
\hline
\multicolumn{7}{c}{Solar abundances from \cite{grevessenoels1993} or \cite{grevessesauval1998}} \\
\hline
1 & 15.3      & $5700 \pm 50$    & $5.50 \pm 0.25$      & $-0.50 \pm 0.25$ & 
$10.6232 \pm 0.0024$ & $3.728 \pm 0.050$ \\ 
8 & 20.7      & $5600 \pm 100$    & $5.00 \pm 0.18$      & $-0.50 \pm 0.25$ & 
$10.6481 \pm 0.0057$ & $3.741 \pm 0.051$ \\ 
10 & 21.2      & $5600 \pm 100$    & $4.00 \pm 0.25$      & $-0.50 \pm 0.20$ & 
$10.6482 \pm 0.0057$ & $3.742 \pm 0.051$ \\ 

11 & 22.5      & $5600 \pm 100$    & $4.50 \pm 0.20$      & $-0.50 \pm 0.25$ & 
$10.6632 \pm 0.0053$ & $3.748 \pm 0.051$ \\ 

12 & 23.4      & $5600 \pm 50$    & $4.00 \pm 0.25$      & $-0.50 \pm 0.25$ & 
$10.7421 \pm 0.0022$ & $3.775 \pm 0.050$ \\ 

13 & 23.7      & $5600 \pm 50$    & $3.00 \pm 0.25$      & $-0.50 \pm 0.25$ & 
$10.7233 \pm 0.0022$ & $3.769 \pm 0.050$ \\ 

14 & 24.2      & $5750 \pm 125$    & $3.00 \pm 0.25$      & $0.00 \pm 0.18$ & 
$10.5041 \pm 0.0020$ & $3.690 \pm 0.049$ \\ 

15 & 25.2      & $5750 \pm 125$    & $4.50 \pm 0.25$      & $0.000 \pm 0.075$ & 
$10.6022 \pm 0.0057$ & $3.728 \pm 0.051$ \\ 

\hline

\multicolumn{7}{c}{Solar abundances from \cite{grevesse2007} or \cite{asplund2009} or \cite{caffau2009}} \\

\hline

2 & 19.3      & $5800 \pm 50$    & $4.00 \pm 0.25$      & $0.0 \pm 0.20$ & 
$10.6244 \pm 0.0023$ & $3.706 \pm 0.049$ \\ 

3 & 19.4      & $5800 \pm 50$    & $4.00 \pm 0.25$      & $0.0 \pm 0.20$ & 
$10.5432 \pm 0.0023$ & $3.705 \pm 0.049$ \\ 

4 & 19.4      & $5800 \pm 50$    & $4.00 \pm 0.25$      & $0.0 \pm 0.20$ & 
$10.5467 \pm 0.0024$ & $3.705 \pm 0.049$ \\ 

5 & 19.4      & $5800 \pm 50$    & $5.50 \pm 0.26$      & $0.0 \pm 0.0$ & 
$10.567 \pm 0.022$ & $3.711 \pm 0.056$ \\ 

6 & 19.4      & $5800 \pm 50$    & $4.00 \pm 0.25$      & $0.0 \pm 0.20$ & 
$10.5432 \pm 0.0024$ & $3.704 \pm 0.049$ \\ 

7 & 20.2      & $5700 \pm 50$    & $4.00 \pm 0.26$      & $0.0 \pm 0.0$ & 
$10.632 \pm 0.022$ & $3.737 \pm 0.057$ \\ 

9 & 21.0      & $5800 \pm 50$    & $4.00 \pm 0.25$      & $0.0 \pm 0.20$ & 
$10.5463 \pm 0.0023$ & $3.705 \pm 0.049$ \\ 


\hline
\end{tabular}
\label{tab:sed_models}
\tablefoot{The first model is shown in Fig.\,\ref{fig:sed} panel C. From $T_{\rm eff}$ and $L_{\rm bol}$ we calculate the stellar radius to be $R = 1.93 \pm 0.12\,{\rm R}_{\odot}$ resulting in a mean of $0.97 \pm 0.06\,{\rm R}_{\odot}$ for each component in NY\,Hya. Some uncertainties are too optimistic. Models: {\sc bt-nextgen-gns93} (1,11,12,13), {\sc kurucz} (14), {\sc bt-nextgen-agss2009} (2,9), {\sc bt-dusty} (3), {\sc bt-cond} (4), {\sc bt-settl-agss2009} (6), {\sc bt-settl-cifist} (5,7), {\sc nextgen} (8,10,11), and {\sc coelho} (15) with $[\alpha/{\rm Fe}] = 0.4$ dex. Formal weighted means for old estimates of solar abundances: $T_{\rm eff} = 5636 \pm 25 $ K, $\logg = 4.286 \pm 0.081$ (cgs), ${\rm [m/H]} = -0.167 \pm 0.056$ dex and for recent solar abundances: $T_{\rm eff} = 5788 \pm 18 $ K, $\logg = 4.202 \pm 0.096$ (cgs), ${\rm [m/H]} = 0.0 \pm 0.0$ dex. For a reference to each model, see text for details.}
\end{table*}

\subsection{Physical properties and photometric distance}
\label{physicalpropertiesandphotometricdistance}

Final physical properties of NY\,Hya were calculated using {\sc jktabsdim}\footnote{\url{https://www.astro.keele.ac.uk/jkt/codes/jktabsdim.html}} \citep[Ver. 15]{southworth2005a} from the measured values $P, r_A, r_B, i, e, K_A$ and $K_B$. Consistency is ensured by implementing the IAU 2012/2015 Resolution B2/B3 \citep{IAU2015-Res-B3} for Solar values. Proper error propagation is performed and we list final values in Table~\ref{tab:physprop}.

\begin{table}
%
%
\centering
\caption{Physical properties of the two components in NY\,Hya as derived from {\it TESS} photometry and FEROS spectral data.}
\label{tab:physprop}
\begin{tabular}{l c c}
\hline
Parameter                            & star A                              & star B \\
\hline
\hline
Mass $({\rm M}_{\odot})$             & $1.1605 \pm 0.0090$ & $1.1678 \pm 0.0096$ \\ 
Semi-major axis $({\rm R}_{\odot})$  & \multicolumn{2}{c}{$15.814 \pm 0.040$} \\ 
Radius $({\rm R}_{\odot})$           & $1.407 \pm 0.015$   & $1.406 \pm 0.017$ \\ 
$\log g$ $(\rm cm/s^2)$              & $4.2059 \pm 0.0089$ & $4.210 \pm 0.011$ \\ 
$v_{\rm synch}$ $(\rm km/s)$         & $14.91 \pm 0.16$    & $14.90 \pm 0.18$ \\ 
$T_{\rm eff}$ $(\rm K)$              & $5595 \pm 61$       & $5607 \pm 61$ \\ 
$\log(L_{\rm Bol}/{\rm L}_{\odot})$  & $0.243 \pm 0.021$   & $0.246 \pm 0.022$ \\ 
$V$ (mag)                            & $9.247 \pm 0.043$   & $9.244 \pm 0.043$ \\ 
$(b-y)_0$ (mag)                      & $0.425 \pm 0.026$   & $0.420\pm 0.025$\\ 
$m_{1,0}$ (mag)                      & $0.249 \pm 0.047$   & $0.256\pm 0.047$ \\ 
$c_{1,0}$ (mag)                      & $0.346 \pm 0.054$   & $0.346\pm 0.054$ \\ 
$M_{\rm Bol}$\tablefootmark{a} (mag) & $4.133\pm 0.053$    & $4.126 \pm 0.054$ \\ 
$M_{V}$\tablefootmark{b} (mag)       & $4.241\pm 0.063$    & $4.231 \pm 0.064$ \\ 
$M_{V}$\tablefootmark{b} (mag)       & \multicolumn{2}{c}{$3.483 \pm 0.045$} \\ 
$M_{V}$\tablefootmark{c} (mag)       & \multicolumn{2}{c}{$3.365 \pm 0.022$} \\ 
$K_S$ dist. (pc)                     & \multicolumn{2}{c}{$104.1 \pm 1.5$} \\ 
{\it Gaia} DR2 dist. (pc)            & \multicolumn{2}{c}{$106.10 \pm 0.70$} \\
\hline
\end{tabular}
\tablefoot{The systemic absolute $V$-band magnitude agrees with the fully empirical measurement obtained from the {\it Gaia} parallax to within $1.9\sigma$. All magnitudes are extinction--reddening corrected.}
\tablefoot{\tablefoottext{a}{Using $M_{\rm Bol,\odot} = 4.74$ (derived from IAU 2015 Res. B3).} \tablefoottext{b}{Derived using bolometric corrections from \citet{girardi2002} and adopting a weighted mean of ${\rm [Fe/H]} = 0.003\pm 0.046$ dex.}\tablefoottext{c}{Fully empirical: derived from Str\"omgren photometry and {\it Gaia} parallax.}}
\end{table}

Tidal interactions in binary star systems will have effects on both stellar rotation and orbital (mainly eccentricity) evolution \citep{torres2010}. For the measured orbital period and mass-ratio the turbulent dissipation and radiative damping formalism by \citet{zahn1977,zahn1989} predicts a ($e$-folding) synchronisation timescale of 5.2 Myr and a ($e$-folding) timescale for orbit circularisation of 1.3 Gyr. We found the orbit of NY\,Hya to be near-circular. However, an indication of system age is obtained only if the initial orbital eccentricity is known. A circular orbit does not necessarily imply an old system. As an example, the V615 Per system has a circular orbit even though the period is relatively long (13.7~d) and the system is only $\approx 20$ Myr old \citep{southworth2004a}.

From the measured radii and orbital period, we find the stellar equatorial rotation velocities to { be $14.91 \pm 0.16$ km/s and $14.90 \pm 0.18$ km/s for star~A and star~B}, respectively. The spectroscopic rotational velocities were found to be $15.77 \pm 1.34$ km/s and $15.71 \pm 1.35$ km/s and agree with the synchronous speeds to within $0.64\sigma$. From this, we conclude that the system is tidally locked in a 1:1 spin-orbit resonance. This finding is in agreement with the period analysis of the {\it TESS} flux variation out of the eclipses from which we independently concluded that the system is rotationally synchronised.

Finally, we carried out a consistency check and calculated a semi-empirical distance from photometry and \Teff. The photometric distance to NY\,Hya is computed from the surface brightness relations \citep{southworth2005a} adopting the empirical calibrations from \citet{kervella2004} valid for dwarf stars between spectral types A0 and M2. We adopt the spectroscopic $T_{\rm eff}$ measurements presented in Sect.~\ref{stellaratmosphericparameters} along with the 2MASS \citep{2MASS} standard $JHK_S$ magnitudes and the Str\"omgren (SAT) derived Johnson $V$ magnitude. We adopted the interstellar reddening $E(B-V) = 0.0065 \pm 0.025$ mag from Sect.~\ref{interstellarextinctionandreddening} to align the distances found in the various passbands. Having experimented with different passband magnitudes, which disagreed with each other slightly but provided photometric distances between 100 and 104 pc, we report a final photometric distance of $104.1 \pm 1.5$ pc from the 2MASS $K_S$ band \citep{kervella2004} since this measurement is affected least from uncertainties in interstellar reddening effects. This $K_S$-band distance agrees with the astrometric {\it Gaia} DR2 measurement ($106.10 \pm 0.70$ pc) to within $1.3\sigma$.

\subsection{Atmospheric properties from empirical relations}

The SAT data provide an opportunity for consistency checks on the atmospheric properties from empirical calibration relations based on Str\"omgren colours. For a short review, we refer to \citet{arnadottir2010}. We have made use of the {\sc uvbysplit}\footnote{\url{https://www.astro.keele.ac.uk/jkt/codes.html}} code with proper error propagation. The light ratios in the $uvby$ passbands allow the derivation of Str\"omgren colours for each component star. Several calibration relations for $T_{\rm eff}, \log g$ and [Fe/H] exist. In all cases, we checked the validity of each calibration relation. Uncertainties in the colour indices were propagated via Monte Carlo simulations assuming Gaussian errors. The uncertainty from the calibration relation was added in quadrature to obtain the final uncertainty. This assumes no correlations between uncertainties. Important to note is that all calibration relations are based on FGK dwarfs (more or less). The work in \citet{arnadottir2010} offers some methodology to distinguish between dwarf and (sub-)giant stars. Based on the measured surface gravities, the two components in NY\,Hya are likely dwarf stars still on the main-sequence.

\citet{Olsen1984} provides a calibration relation (eq. 14) for the metallicity based on standard relations for
de-reddened colour indices \citep{arnadottir2010, torres2014}. We find ${\rm [Fe/H]_A} = 0.17 \pm 0.18$ dex and ${\rm
[Fe/H]_B} = 0.15 \pm 0.18$ dex. These estimates, albeit low precision, are consistent with solar metallicity. Metallicity estimates with high fractional uncertainties are obtained from \citet{casagrande2011} and
\cite{holmberg2007} using calibration relations with de-reddened Str\"omgren colour indices. We find ${\rm [Fe/H]_A} =
0.06 \pm 0.39$ dex and ${\rm [Fe/H]_B} = 0.14 \pm 0.38$ dex \citep{holmberg2007} and ${\rm [Fe/H]_A} = 0.17 \pm 0.40$ dex
and ${\rm [Fe/H]_B} = 0.23 \pm 0.39$ dex \citep{casagrande2011}. Again, these estimates are consistent with a solar metallicity as was also determined from disentangled FEROS spectra.

\Teff s are obtained from various sources and some relations \citep{holmberg2007,casagrande2011} depend on metallicity. For solar metallicity, we find $T_{{\rm eff},{\rm A}} = 5561 \pm 196$ K and $T_{{\rm eff},{\rm B}} = 5591 \pm 193$ K \citep[eq. 9]{alonso1996}, $T_{{\rm eff},{\rm A}} = 5637 \pm 179$ K and $T_{{\rm eff},{\rm B}} = 5669 \pm 172$ K \citep[eq. 2, p. 522 considering $0.33 < (b-y) < 0.50$]{holmberg2007}, $T_{{\rm eff},{\rm A}} = 5672 \pm 181$ K and $T_{{\rm eff},{\rm B}} = 5705 \pm 175$ K \citep{casagrande2011} where we have added 20 K in quadrature as an additional uncertainty in the zero-point of the temperature scale (see their table 4). For all these relations, we find an increase/decrease in $T_{\rm eff}$ of 30 - 40 K when [Fe/H] is increased/decreased by 0.10 dex. An unpublished $T_{\rm eff}-(b-y)_0$ calibration relation is obtained from Mamajek (2014).\footnote{\url{https://figshare.com/articles/figure/Stromgren\_b\_y\_colour\_Versus\_Effective\_Temperature\_for\_AFGK_Main\_Sequence\_Stars/949735/1}} This relation is independent of metallicity. However, the median and 68.3\% confidence interval for the metallicity of stars in the sample was found to be $0.0 \pm 0.2$ dex. We find $T_{{\rm eff},{\rm A}} = 5648 \pm 197$ K and $T_{{\rm eff},{\rm B}} = 5675 \pm 194$ K. Finally, \citet{alonso1996} provide a calibration relation in the infrared based on TCS $J-K$ colour and thus less affected by interstellar reddening. We find a mean temperature of $\langle T_{\rm eff}\rangle = 5673 \pm 235$ K. In general, we find good agreement with measurements obtained from FEROS data.

Several survey studies included NY\,Hya in their spectroscopic monitoring programs providing estimates of atmospheric properties. However, all these studies assumed NY\,Hya to be a single star. Measurements of stellar metallicity, surface gravity, and rotational velocity are all highly likely to be in error. The only quantity with some accuracy from these surveys is the \Teff. As will later be argued, the two components in NY\,Hya are almost identical. Hence various measurements of $T_{\rm eff}$ would represent a mean value not far from the actual value for each star. The works (no uncertainties were quoted) in \cite{nordstroem2004,holmberg2007} and \cite{holmberg2009} report the same 
$T_{\rm eff} = 5458$ K and \cite{robinson2007,mcdonald2012} report 5678 K and 5719 K, respectively. We find a literature mean value of 5618 K for NY\,Hya as a single star. Assuming an uncertainty of 100 K this estimate is in good agreement ($1.3\sigma$) with a detailed spectroscopic analysis presented later in this work and also agrees $(0.23\sigma)$ well with the {\it Gaia} \citep{GAIA2018} estimate of $5647 \pm 76$~K. \citet{gaspar+2016} reports a stellar metallicity of 
[Fe/H] = $0.16 \pm 0.06$ dex, but this measurement does not account for binarity as well. Analysing double stars as single stars can introduce a significant bias in the metallicity estimate.


\begin{figure*}
\centering
\includegraphics[angle=0,width=0.48\textwidth]{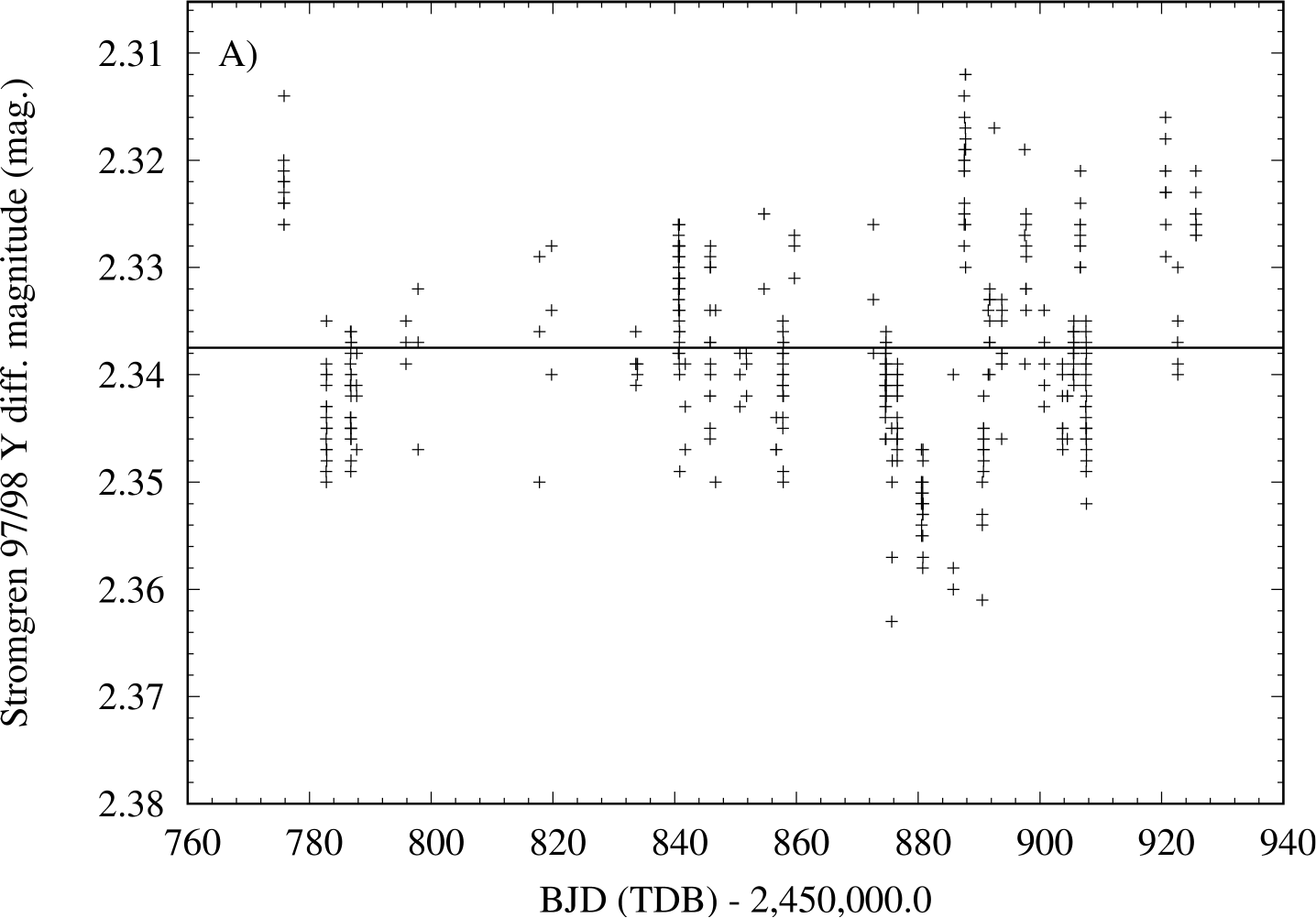}
\includegraphics[angle=0,width=0.48\textwidth]{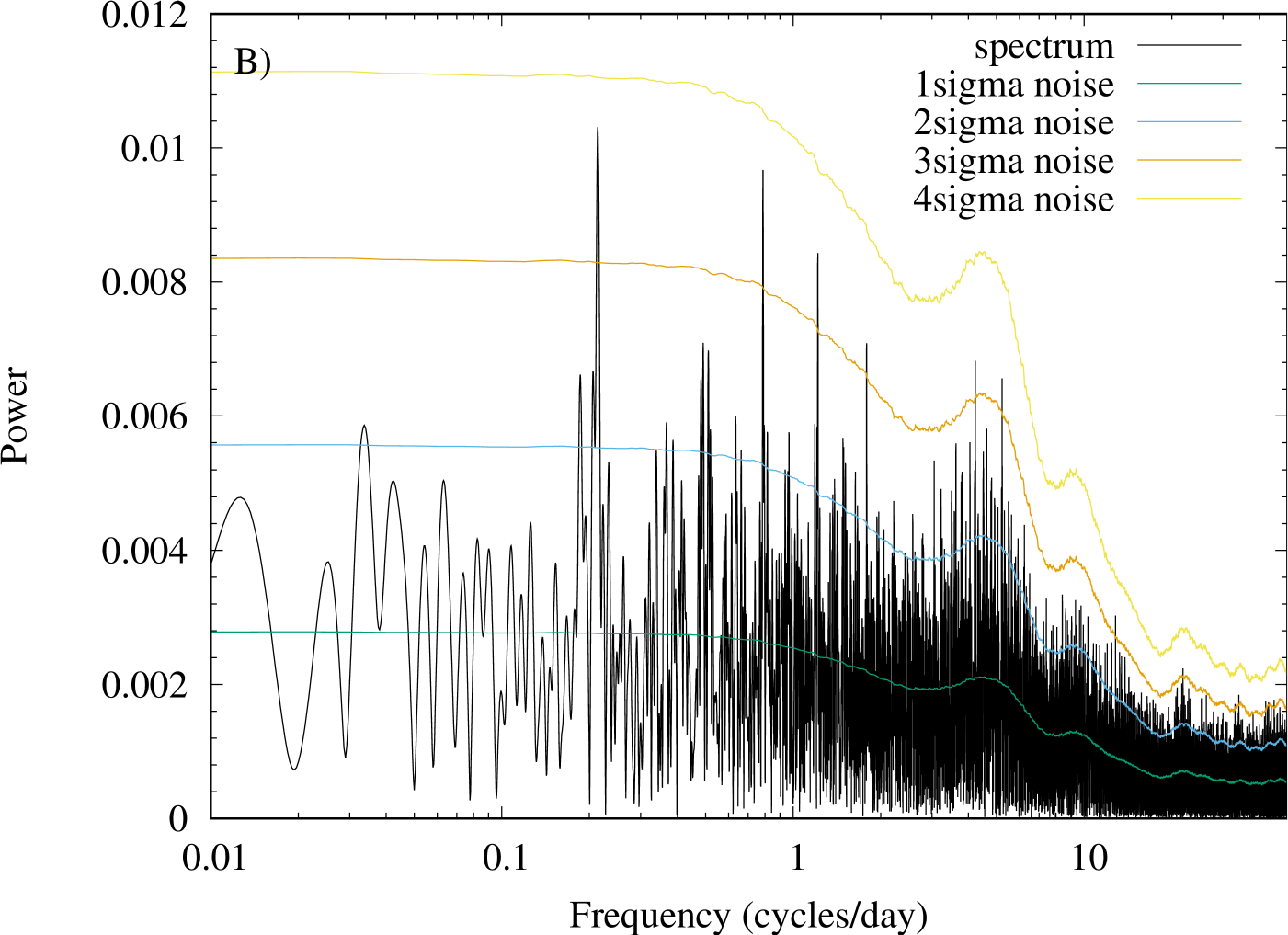}
\includegraphics[angle=0,width=0.48\textwidth]{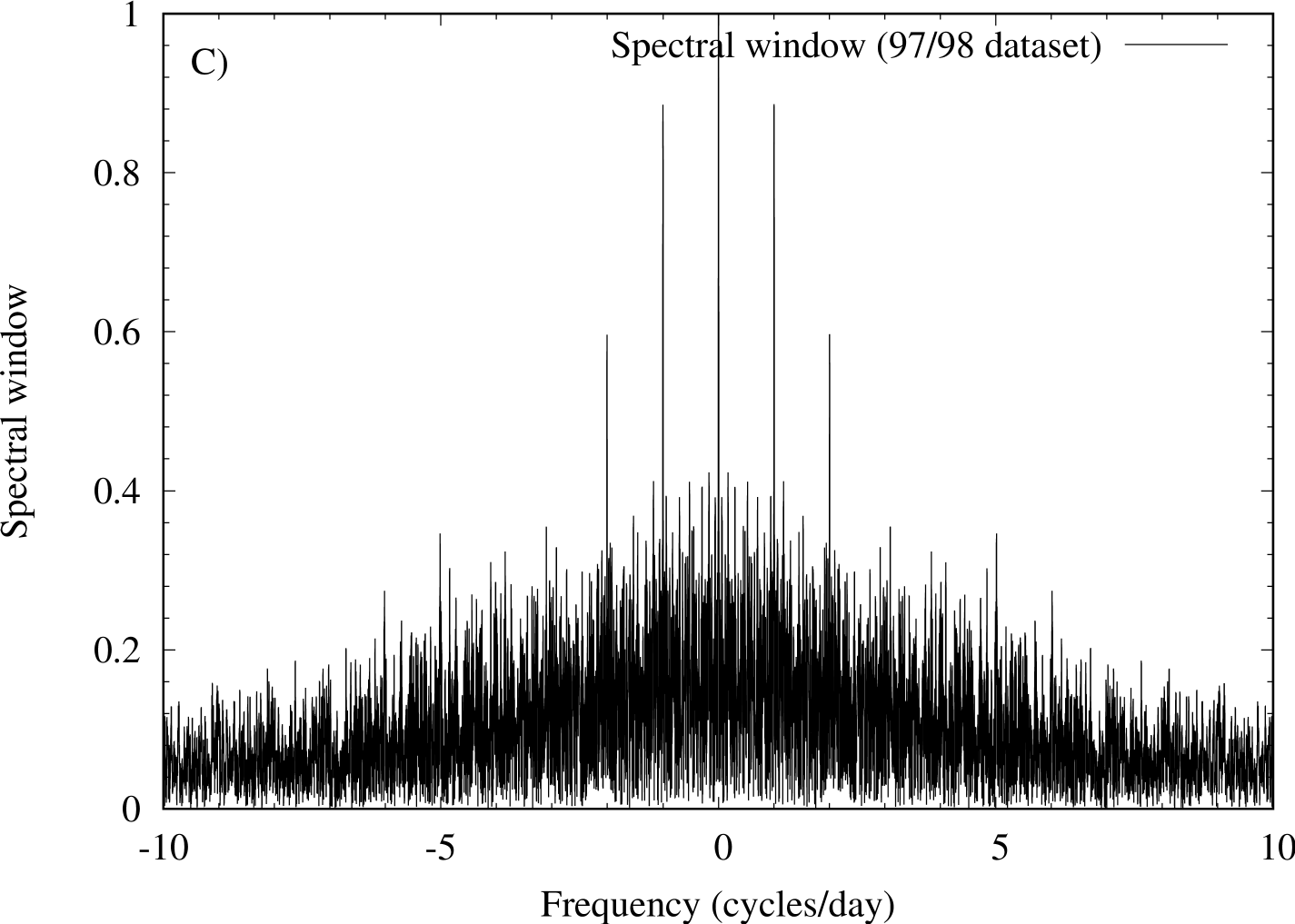}
\includegraphics[angle=0,width=0.48\textwidth]{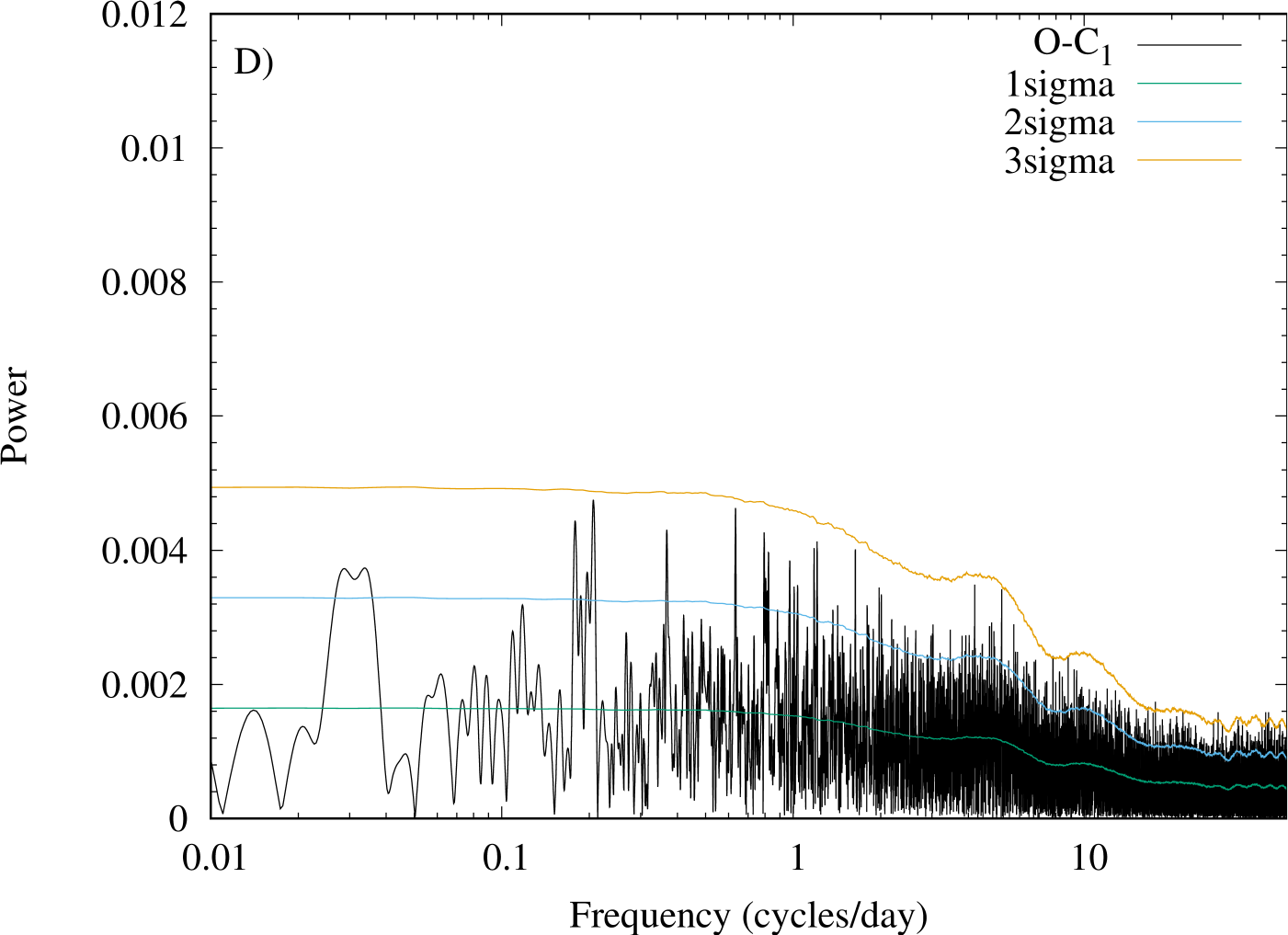}
\caption{Results of the period analysis for the SAT 97/98 $y$-band data set. \emph{Panel A}: 404 $y$
out-of-eclipse photometry showing magnitude vs time where the primary and secondary eclipses have been
removed. The horizontal line is the mean flux out of the eclipses as determined by {\sc period04} and was subtracted
prior to the computation of the Fourier spectrum. \emph{Panel B}: Power spectrum and $1,2,$ and $3\sigma$ noise levels. A frequency at $f_1 = 0.2138$ c/d is detected corresponding to a period of 4.68 days. Frequencies near 1 c/d are likely daily/nightly alias frequencies. \emph{Panel C}: Power spectrum 
of the spectral (data gap structure) window. \emph{Panel D}: Power spectrum after removing the
$f_1$ frequency component with the highest power. Panels B and D have the same scale along the
y-axis.}
\label{fig:PeriodAnal9798}
\end{figure*}

\section{Stellar activity}
\label{stellaractivity}

We performed a period analysis on the three data sets, (SAT 97/98, SAT 98/99 and {\it TESS}). We applied
{\sc period04}\footnote{\url{https://www.univie.ac.at/tops/Period04/}} \citep{period04}. This package implements the Discrete Fourier Transform (DFT) algorithm \citep{press1992} and is suitable
for the detection of periods in unevenly sampled time series data. In all three data sets, we removed all eclipses from the original data.

For the SAT 97/98 season, we show the results in Fig.~\ref{fig:PeriodAnal9798} based on the Str\"omgren $y$-band light curve. The first panel shows the out-of-eclipse variability when all eclipses are removed. The second panel shows the spectral power for frequencies up to 50 cycles per day (c/d). No significant spectral power is found beyond this limit. Because of irregular
sampling the Nyquist frequency of $\nu_{\rm Ny} \approx 123$ c/d was computed iteratively as the expectation value by means of a histogram analysis, based on recursive nested intervals (see p. 11 in \citealt{LenzMScThesis2005}). 

In Fig.~\ref{fig:PeriodAnal9798}B the power spectrum of the recorded data shows several significant frequencies. We quantify the
significance of a given frequency by calculating the $1\sigma$, $2\sigma$ and $3\sigma$ noise levels as
implemented within {\sc period04}. We used the recommended (see p.\ 14 in \citealt{LenzMScThesis2005})
stepping rate of $1/20 T$ where $T$ is the observational baseline time period (approximately 150 days). We
determined a frequency with maximum power at $f_1 = 0.21381 \pm 0.00018$ c/d corresponding to a period of
around 4.68 days. The best-fit amplitude and phase is of no interest and was omitted here. The
uncertainty in frequency was obtained using the {\sc period04} built-in bootstrapping error estimation
function. We chose 5000 Monte Carlo simulations. This result indicates that at least one component in 
NY\,Hya is contributing to a periodic modulation of the flux at the orbital phases out of the eclipses. To determine whether the remaining
detected frequencies are real or spurious we pre-whitened the original power spectrum, as a result of which we ended up with one that resembles the noise level of the original power spectrum indicating that no further frequencies are to be found within the original data set.

The power spectrum of the out-of-eclipse fluxes for the SAT 98/99 data contains more information since the observation 
cadence was increased. We applied the same Fourier analysis methodology as
for the 97/98 data set. The mean Nyquist frequency $\nu_{\rm Ny}$ was found to be around 199 c/d. A
frequency of $f_1 = 0.20414 \pm 0.00031$ c/d was found at the $3\sigma$ noise level. This corresponds to a
period of around 4.89 days and differs from the SAT 97/98 $f_1$ frequency by $27\sigma$. This result
indicates that the out-of-eclipse variability is not constant suggesting a complex variation. The pre-whitened power spectrum still shows a number of frequencies although at a smaller significance (around $2.5\sigma$) level. The highest frequency
was found at $f_2 = 0.42824 \pm 0.00018$ c/d and corresponds to a period of around 2.35 days. This period
is close to half the orbital period of NY\,Hya. This result suggests that the SAT 97/98 data set now contains a
flux modulation out of eclipses at a period comparable to 1/1 and 1/2 of the binary period. The decrease in power
for the $f_2$ frequency corresponds roughly to around $0.5\sigma$ and hence the significance of $f_1$ and
$f_2$ are somewhat comparable. Pre-whitening the original power spectrum with both $f_1$ and $f_2$ results
in a power spectrum with no significant (beyond $2\sigma$) frequency content (between 0.1 and 5 c/d).

\begin{figure}
\centering
\includegraphics[angle=0,width=0.47\textwidth]{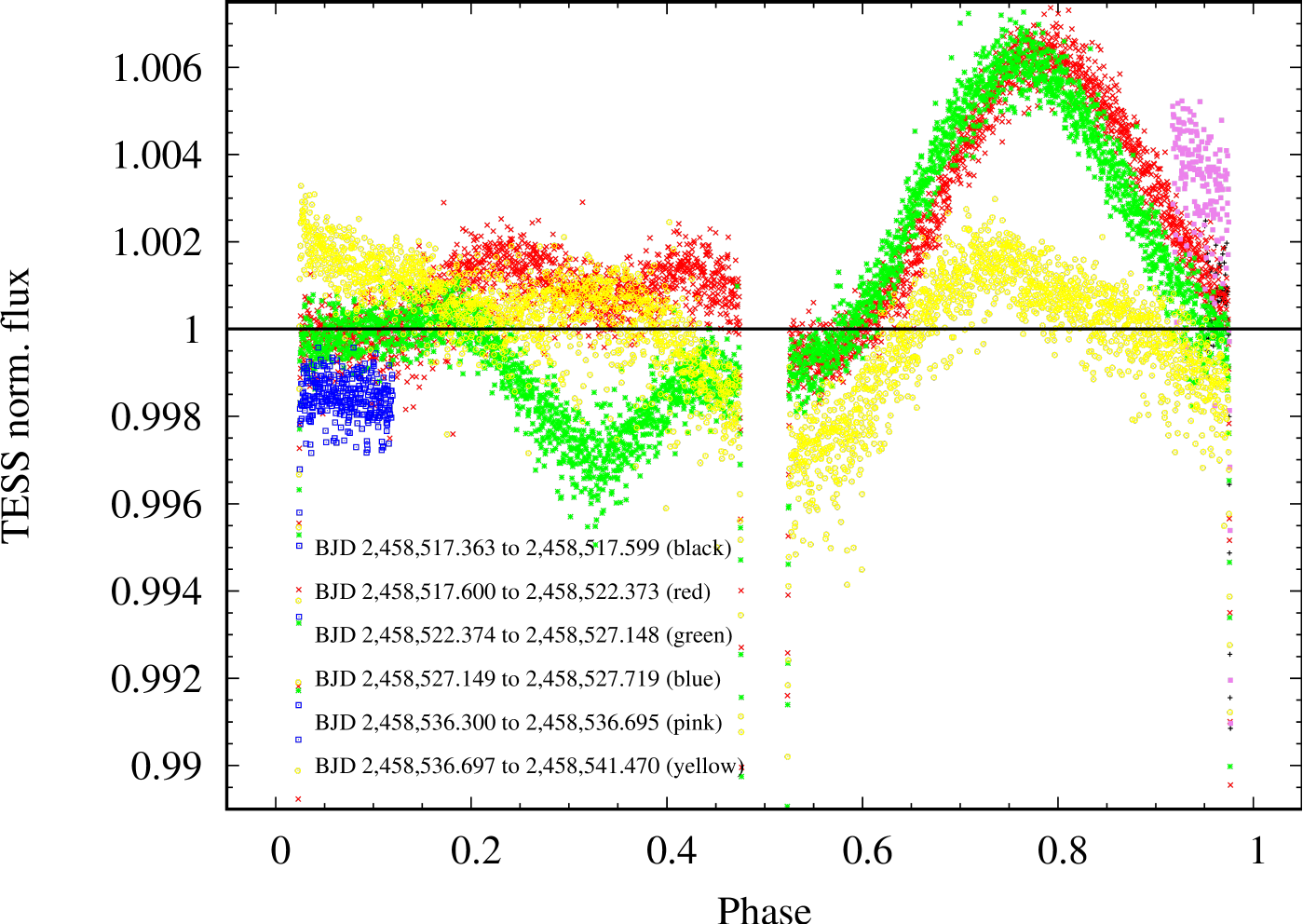}
\caption{Normalised {\it TESS} baseline flux (with eclipses) phases folded with the orbital period of NY\,Hya. The different colours encode successive times in the order black, red, green, blue, pink and yellow.}
\label{fig:TESS-NYHya}
\end{figure}

\begin{figure*}
\centering
\includegraphics[width=1.0\textwidth]{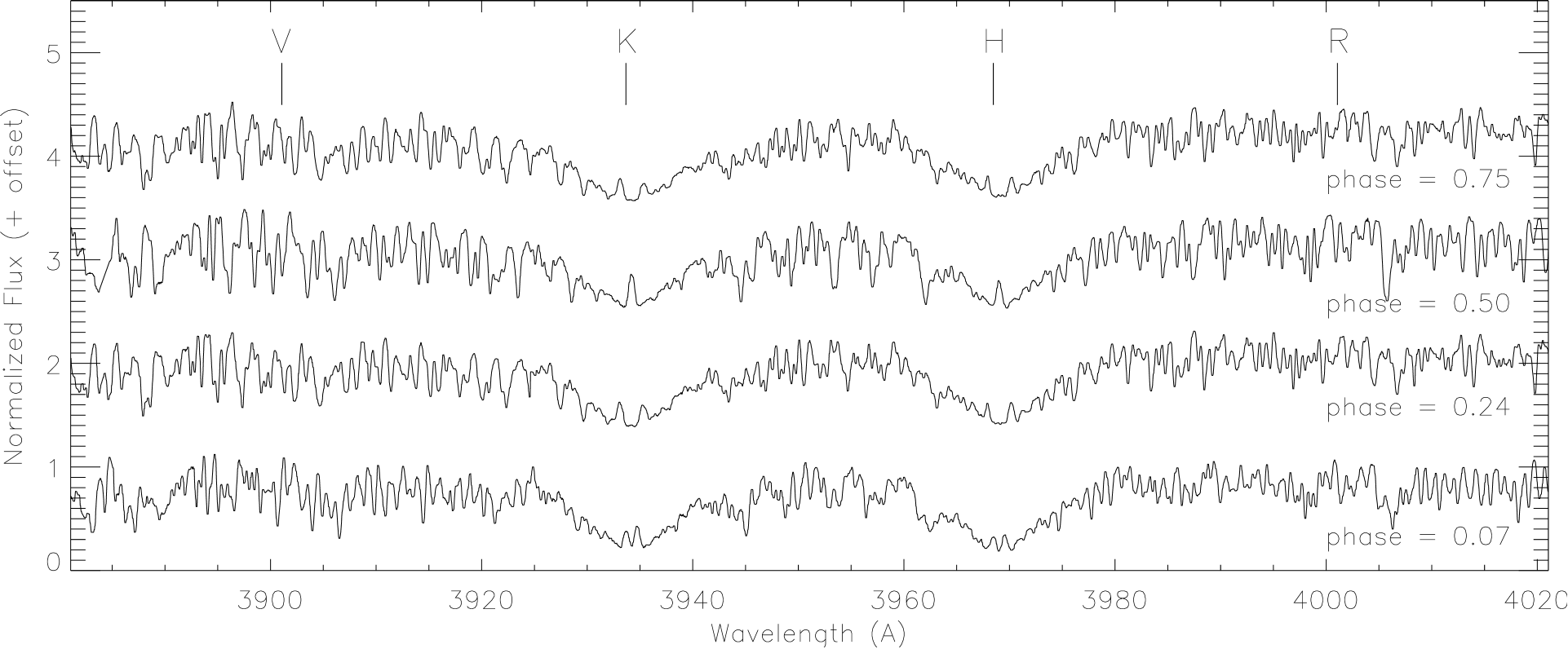}
\caption{Wavelength region around the Ca II K\&H lines (3933.66~\si{\angstrom} and 3968.47~\si{\angstrom}) at
different orbital phases. Core emission lines are visible for both components demonstrating chromospheric
activity. The outermost vertical lines mark the location of the V and R continuum bands at
3901.07~\si{\angstrom} and 4001.70~\si{\angstrom}.}
\label{fig:CaII}
\end{figure*}

The {\it TESS} data offer the potential for the most detailed period analysis due to the relatively 
high sampling cadence and photometric precision. We first phase folded the complete {\it TESS} data with
the orbital period of NY\,Hya (4.77 days). The result is shown in Fig.~\ref{fig:TESS-NYHya}. The
observations clearly show photospheric flux changes at phases around 0.30--0.35 and 0.80--0.85 and hence are separated in time by approximately half an orbital period. The flux changes at phase 0.30--0.35 seem to change from orbit to orbit while the flux change at phase 0.80--0.85 are more or less constant. If these changes are caused by photospheric spot evolution, then the {\it TESS} data provides strong evidence for the presence of at least two spot features (during the {\it TESS} sector 8 time window) on either one or both components. Qualitatively, the spot at phase 0.30--0.35
changes in both flux (low temperature) and longitude/latitude position as the phase of minimum flux is changing. The spot feature at phase 0.80--0.85 is clearly larger spanning a longer phase period and also
significantly hotter compared to the previous spot feature. We also see {\it TESS} systematics as discussed
earlier, but their presence does not alter the spot evolution interpretation. Assuming synchronous rotation (see Sect.~\ref{physicalpropertiesandphotometricdistance} for an argument to render this assumption true), the flux feature at phase 0.30--0.35 could be a spot located on the inward-facing hemisphere of one component or a spot located on the outward-facing hemisphere of the other component rotating into the line of sight of the observer after primary eclipse at phase 0.0. A similar chain of reasoning applies for the flux feature at phase 0.80--0.85. Without additional evidence it is impossible to decide whether both features belong to a single component.

We then computed a power spectrum from {\sc period04} using the complete flux data with {\it TESS} eclipses removed. The mean Nyquist frequency $\nu_{\rm Ny}$ was found to be around 359 c/d. As before, we have computed $1\sigma, 2\sigma$ and $3\sigma$ noise levels. As a `sanity check', we have also computed a power spectrum using a {\sc python} implementation of the {\sc lombscargle} algorithm (part of the {\sc
astropy}\footnote{\url{https://docs.astropy.org/en/stable/timeseries/lombscargle.html}} package). The results were identical. We omitted the calculation of the False-Alarm-Probabilities (FAPs) from the {\sc lombscargle} algorithm as FAPs become less reliable for time series data with long data gaps \citep{reegen2007}. Following successive pre-whitening stages, the first five frequencies were found at $f_1 = 0.2029 \pm 0.0014$, 
$f_2 = 0.071 \pm 0.011$, $f_3 = 0.4180 \pm 0.0012$, $f_4 = 0.1055 \pm 0.013$ and 
$f_5 = 0.64 \pm 0.31$ c/d corresponding to periods of around 4.9, 14.1, 2.4, 9.5 and 1.6 days.

We conclude this section as follows. From the out-of-eclipse variability in {\it TESS} light curves, we have identified two clear increases in brightness after the 2nd and 4th eclipses. Assuming that these are the same feature seen twice, we get a rotational 
period of about 4.7 days. This was also inferred quantitatively from a period analysis where we detected a 4.7 and 4.9 day period when considering SAT 97/98 and {\it TESS} out-of-eclipse data, respectively. This is approximately the orbital period, 
but has been estimated from the stellar rotational modulation and not from orbital eclipses. We can interpret this as being indicative of at least one of the stars showing star-spot activity at the time of the {\it TESS} observations and that its rotation period is the same as the orbital period, i.e. the star has synchronous rotation. The synchronisation between the rotation of the components and the orbital period points towards a not very young system, with an age of at least few million  years (see Sect.~\ref{physicalpropertiesandphotometricdistance}). Unless the stellar rotational velocities $(v\sin i)$ are significantly different (which they are not) it is not possible to work out which of the stars shows this spot activity. Evidence for photospheric activity via star-spot(s) is supported from spectroscopic observations. Several FEROS spectra exhibit clear Ca II K \& H core emission lines (see Fig.~\ref{fig:CaII}) for each component which usually indicates photospheric activity.

From catalogue data, NY\,Hya appears to be active. This is supported by photometric data \citep{clausen2001} which shows out-of-eclipse variations in the Str{\"o}mgren bands with an approximate 0.04 magnitude variability. 

Smoking-gun evidence for stellar activity is found via the detection of significant X-ray emission with {\it ROSAT} and {\it XMM-Newton}. \cite{szczygiel2008} used the {\it ROSAT} X-ray fluxes to determine a hardness ratio of 0.48 and estimate the system's bolometric and X-ray luminosities. They found $\log ( L_{\rm X} / L_{\rm Bol} ) = -4.26$ indicating a strong X-ray emission often associated with stellar youth as a first interpretation. They estimate a Rossby number of $\log (P_{\rm rot} / \tau_{\rm conv}) = -0.51$ assuming NY\,Hya to be a single star. Stellar activity was also reported by \cite{isaacsonfisher2010} and \citet{pace2013} from spectroscopic observations measuring the core-emission component in the Ca II K and H lines. They reported $\log R^{'}_{\rm HK} = -4.463$ and $S_{\rm max} = 0.375$, respectively, indicating a high level of chromospheric activity. However, these values should be treated with some caution since the binary nature of NY\,Hya might have shifted the Ca II lines away from their respective spectral windows used to define the $R^{'}_{\rm HK}$ index. No observational time stamps were provided by the authors to determine an orbital phase.

\section{New {\it Gaia} astrometry and stellar kinematics}
\label{newGAIAastrometryandstellarkinematics}

From the {\it Gaia} satellite \citep[DR2]{GAIA2018}, the stellar parallax of NY\,Hya was measured to be 
$\pi = 9.429 \pm 0.062$ $m$as (fractional uncertainty of $\sigma_{\pi}/\pi \approx 0.0066$). From 
$d = 1/\pi$ and Monte Carlo error propagation, the median distance to NY\,Hya is found to be $d = 106.10 
\pm 0.70$ pc. The new {\it Gaia} parallax measurement is a significant improvement on the revised {\it Hipparcos} \citep{hog2000} measurement of $\pi = 12.22 \pm 1.16$ $m$as \citep{vanleeuwen2007}.

NY\,Hya has {\it Gaia} DR2 proper motion in RA and Dec measured to be $\mu_{\alpha}\cos\delta = -60.06 \pm 0.14$ $m$as/yr and $\mu_{\delta} = 48.16 \pm 0.23$ $m$as/yr, respectively. The total on-sky proper motion is found to be 77 $m$as/yr. This can be used to identify any unrelated background source. We compared the sky region towards NY\,Hya as observed by the Palomar Observatory Sky Survey (POSS I and II)\footnote{\url{http://skyserver.sdss.org}} providing archive imaging data from the 1950s. Since then around 70 yr have elapsed implying NY\,Hya to have moved around 5 arcsec on the sky. This is too small a displacement in order to be able to ’look behind’ NY\,Hya, in an attempt to detect any unrelated background sources.

Combining the {\it Gaia} proper motion and distance results in a tangential velocity of $38.7 \pm 0.27$ km/s. From modelling spectroscopic observations (see Sect.~\ref{spectralanalysis}) the systemic RV to the binary barycentre was measured to be $40.79 \pm 0.16$ km/s. By combining the total proper motion, distance and RV of NY\,Hya we find the galactic (J2000.0) 3D space position $(X,Y,Z) = (-48.19 \pm 0.32, -79.24 \pm 0.52, 51.44 \pm 0.34)$ pc and space velocity $(U,V,W)_{\rm LSR}=(-41.90\pm 0.25,-3.01\pm 0.15,18.21\pm 0.11)$ km/s relative to the local standard of rest (LSR), assuming a solar motion of $(11.10,12.24,7.25)$ km/s from \citet{schoenrich2010}.

The position and kinematics can be used to make inferences of the galactic population of NY\,Hya \citep{bensby+2003,soubiran+2003} including an estimate of the system lifetime if identified with a stellar association. We applied {\sc galpy}\footnote{\url{https://docs.galpy.org/en/v1.6.0/}} and qualitatively searched the neighbourhood of NY\,Hya for any obvious cluster membership. A search out to a distance of 80 pc from NY\,Hya did not result in any obvious cluster membership. In the $U-V$ space we found a small over-density of stars, but it is located more than $\sim$10 pc from NY\,Hya. We also applied the {\sc banyan}-$\Sigma$ \citep{gagne2018} code\footnote{\url{http://www.exoplanetes.umontreal.ca/banyan/}} and found no cluster membership to any of the 27 known young associations within 150 pc. A probability of 99.9\% of NY\,Hya being a field star was returned. Therefore, no obvious relationship to any star association was found, significantly diminishing the possibility of age determination via main-sequence fitting. We also calculated the membership probability of NY\,Hya belonging to the thin, thick or halo population \citep{reddy+2006, bensby+2014}. From \cite{bensby+2014}, we find percentage probabilities of $(P_{\rm thin}, P_{\rm thick}, P_{\rm halo}) = (97.7, 2.3, <0.1)$\% indicating that NY\,Hya belongs to the galactic thin disc population. Stellar ages for this population are in the range from 0 to 10 Gyr and thus the lifetime of NY\,Hya is unconstrained from 6D kinematical considerations.

\section{Comparison with stellar evolution models}
\label{Comparisonwithstellarevolutionmodels}
NY\,Hya is among a small number of bright detached eclipsing binary systems for which most relevant physical properties for the component stars are constrained by observations \citep[masses, radii, temperatures, luminosities, metallicity;][]{torres2021}. This allows for stringent tests of stellar structure and evolution model predictions, including models that incorporate non-standard physical ingredients (e.g. star spots, magnetic fields). 

Comparisons with stellar structure and evolution models were performed using models from the Dartmouth series \citep{dotter2008,feiden2016}. Models used in this analysis deviate from the original Dartmouth Stellar Evolution Program release \citep[DSEP]{dotter2008} with changes that make them more suitable for studying M-dwarfs and pre-main-sequence (PMS) stars (see, e.g. \citealt{feiden2016}). There is also a difference in the adopted solar composition, with \citet{feiden2016} adopting a more recent solar $Z/X$ value \citep{grevesse2007} compared to \citet{dotter2008}. Differences in fundamental stellar properties in the solar-mass regime are small owing to both model sets calibrating to the Sun. This calibration mitigates differences in model predictions owing to variations in the adopted physics and solar composition. The primary reason for choosing this model set was the ability to perform controlled numerical experiments by calculating standard and non-standard (e.g. magnetic, reduced $\alpha_{\rm MLT}$) evolution models with otherwise consistent physics. 

We used a grid of stellar model mass tracks and isochrones that are described in \citet{feiden2021}. For this work, the grid was extended to include masses between 0.8~$M_\odot$ and 2.0~$M_\odot$ with a grid spacing of 0.05 $M_\odot$ above 1.0 $M_\odot$. Metallicity values in the grid ranged between $-0.7 \le$~[Fe/H]~$\le +0.5$ with a spacing of 0.1 dex. Simple isochrones were computed for ages between 0.1 Myr and 10 Gyr with an age spacing approximately equal to 10\% of the isochronal age (e.g. 0.1 Gyr for ages $>$1 Gyr).

\subsection{Model inference procedure}
Testing the validity of stellar model predictions was done using three procedures: (1) an initial by-eye comparison, (2) estimating the most probable model properties for each component in NY\,Hya using a Markov Chain Monte Carlo (MCMC) method, and (3) individual modelling of the two components based on measured properties (cf. Table~\ref{tab:physprop}). The latter approach was only used for conducting controlled experiments using non-standard model physics (cf. Sect.~\ref{sec:nonstd}).

By-eye relied on comparing a number of stellar model isochrones to the properties of NY\,Hya to establish an approximate model age and metallicity. The intention was to provide a check on the MCMC analysis and provide a baseline for performing controlled experiments with non-standard models, should they be needed. Results of the by-eye comparison were also used to set initial conditions in the MCMC algorithm (see below), though testing revealed the results were insensitive to this choice.

Posterior probability distributions (PPDs) for the inferred stellar evolution model properties were integrated using an MCMC method. We used \texttt{emcee} \citep{foreman-mackey2013} to implement an affine-invariant ensemble sampling algorithm. Our likelihood function is the same as that used by \citet{mannfeiden2015}. This included a uniform prior distribution for the component ages, a choice that could be improved in the future to disfavour phases of rapid stellar evolution. The set of unknown parameters was taken to include the system age, system metallicity, and the two component masses. Each of these parameters has an associated prior probability distribution that helps constrain the permitted model results. These four parameters are required to specify a unique set of model properties (radius, luminosity, and temperature) for each component that are constrained by observations.

Our analysis used 100 walkers taking 1000 steps yielding a total of 100\,000 random samples of the joint posterior probability distribution. We experimented with larger simulations containing 1000 walkers with 100\,000 steps, but found the results were nearly identical to the smaller simulation. Each walker was seeded using a value for each unknown parameter drawn from a Gaussian distribution centred on the results of our initial by-eye comparison. Simulations seeding the walker with random values for each unknown parameter did not affect our results, but did require more steps for convergence. For each random realisation of the unknown system properties, a corresponding set of individual component properties were found by linearly interpolating within our grid of stellar model isochrones \citep[see e.g.][]{mannfeiden2015,feiden2021}. The age sampling in our isochrone grid was fine enough that linear interpolation provided reliable results as long as the model stars were not ascending the red giant branch. Given the estimated $\log g$ for each star was greater than 4.0, our lack of resolution on the red giant branch was not a problem.

\subsection{Standard models}
\label{sec:stdmodel}
\begin{figure}
    \centering
    \includegraphics[width=1.0\linewidth]{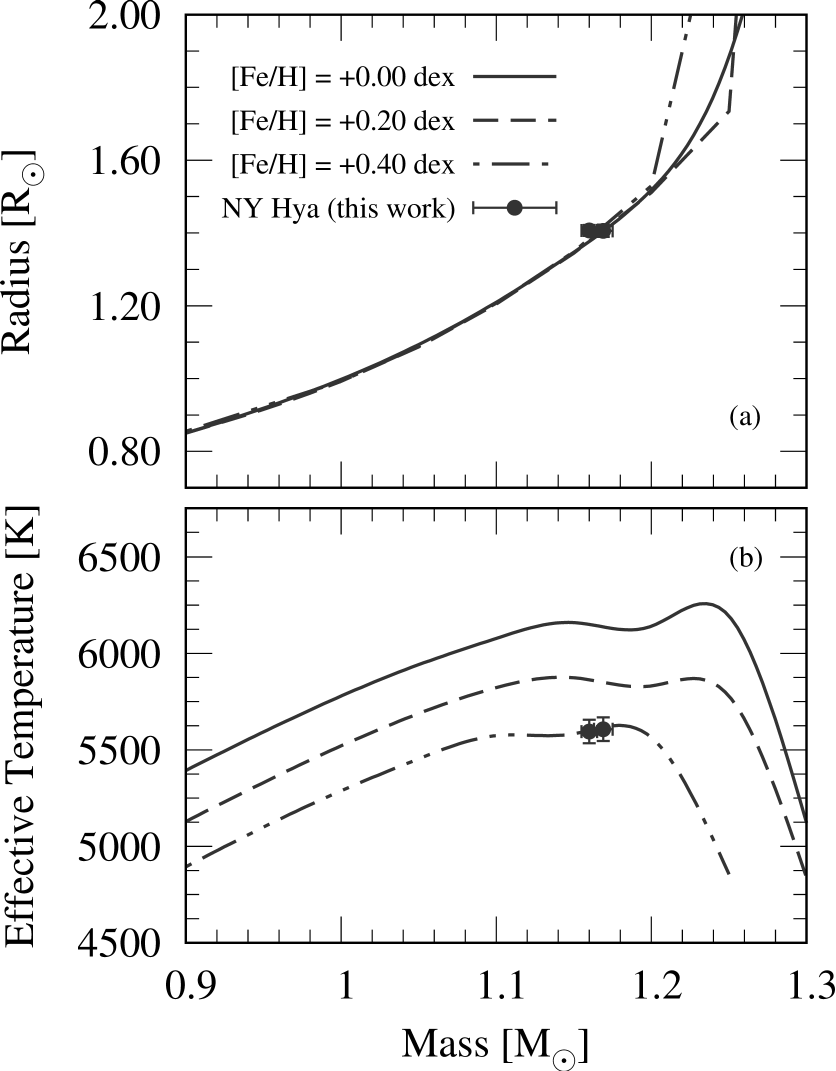}
    \caption{Mass-radius (\emph{top panel}) and mass-$\Teff$ (\emph{bottom panel}) diagrams showing the components of NY\,Hya (solid points) against three standard Dartmouth stellar evolution models. The age-metallicity combinations of the isochrones are ([Fe/H], $\tau_{\rm age}$) = (+0.0, 4.4 Gyr; solid line), (+0.2, 5.7 Gyr; dashed line), and (+0.4, 7.3 Gyr; dash-dotted line).}
    \label{fig:isochrones}
\end{figure}

\begin{figure}
    \centering
    \includegraphics[width=0.9\linewidth]{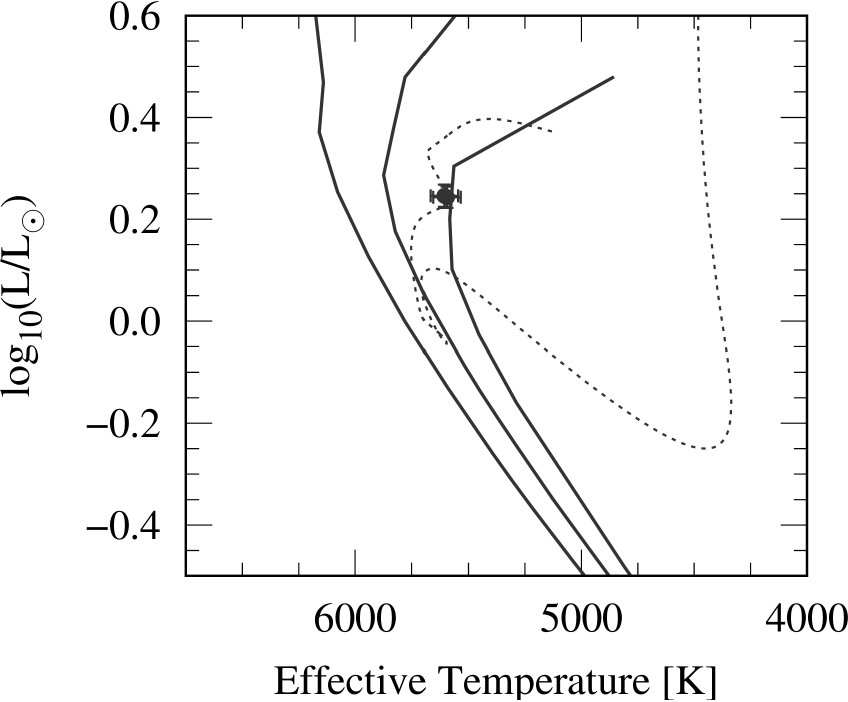}
    \caption{Hertzsprung-Russell diagram showing the two components of NY\,Hya (solid points) against three standard Dartmouth stellar evolution model isochrones (solid lines) and a single stellar evolution model mass track (dotted line). Isochrones are shown with the same line style to provide contrast with the individual mass track. From left to right (warmer to cooler), the isochrones are ([Fe/H], $\tau_{\rm age}$) = (+0.0, 4.4 Gyr), (+0.2, 5.7 Gyr), and (+0.4, 7.3 Gyr). The stellar evolution mass track is showing an $M = 1.160\ M_\odot$ model with [Fe/H] = +0.4.}
    \label{fig:model_hrd}
\end{figure}

Results from our initial by-eye comparison of NY\,Hya's observed component properties to standard stellar models are shown in Figs.~\ref{fig:isochrones} and \ref{fig:model_hrd}. Each figure plots three stellar model isochrones with different metallicities and ages that were found to best reproduce the observed data (filled points). Fig.\,\ref{fig:isochrones}(a) demonstrates that finding agreement in the mass-radius plane can be accomplished using models of almost any metallicity. Predictions for the system age differ, with lower metallicity models producing younger ages. We find a predicted age of about 4.4 Gyr at solar metallicity, 5.7 Gyr when [Fe/H] = +0.20, and 7.3 Gyr with [Fe/H] = +0.40.

To discriminate between models with different metallicities, we need information from the mass-$T_{\rm eff}$ plane. Fig.\,\ref{fig:isochrones}(b) shows that only super-solar metallicity models with [Fe/H]~$\sim +0.40$ are able to reproduce the observed \Teff s from spectral synthesis ($\Teff \approx 5600$~K). This result disagrees significantly with estimates that the system has a near-solar metallicity. However, results from SED fitting did suggest a higher metallicity and high \Teff\ ($T_{\rm eff,\,SED} \approx 5800$~K) was plausible. In this case, super-solar metallicity models are still required to find agreement with the observed \Teff s, with [Fe/H]~$\approx +0.20$.

To estimate uncertainties associated with our by-eye analysis and more rigorously explore the age-metallicity parameter space, we ran a series of MCMC simulations to estimate the joint posterior probability distribution for the model-predicted NY\,Hya properties. To constrain the system's age, we first performed a comparison with standard stellar evolution models at a fixed metallicity, [Fe/H] = +0.00. Resulting PPDs for the system parameters were bimodal, suggesting NY\,Hya could either be a young, PMS system ($\tau_{\rm age} \sim 10$~Myr) or an older main-sequence system ($\tau_{\rm age} \sim 5$~Gyr). PPDs for the component masses and  component radii were consistent within the 1$\sigma$ uncertainty of the observations. However, model \Teff s were inconsistent with the measured values in each case, similar to what was found in the by-eye analysis. Assuming a young 10~Myr age implied that model \Teff s were about 700\,K cooler than our measured \Teff s. Conversely, assuming a main-sequence age implied model \Teff s were about 700\,K warmer than the measured values. 

A second series of simulations relaxed the fixed-metallicity assumption to allow for any metallicity values available in the model grid. We applied a Gaussian prior probability distribution centred on the estimated value ([Fe/H] = $0.07\pm0.1$), but that only weakly constrained the simulation results. We did enforce that both stars should have the same age and metallicity, assuming that a single system age and metallicity should apply. The resulting PPDs for the system parameters were again bimodal, but with strong correlations between metallicity and age. Younger solutions ($\sim$10~Myr) were predicted to be sub-solar in metallicity, while older ages ($\sim$6~Gyr) were decidedly super-solar. 

This result is not surprising given the results of our fixed-metallicity simulations. Solutions suggesting the system has a young age were found to be too cool compared to the observations, and the opposite was true for older age solutions. Lowering the metallicity increases stellar \Teff s, while increasing the metallicity has a cooling effect. The joint PPDs from the variable-metallicity runs highlight this effect as the simulations attempt to balance the \Teff\ offsets by offsetting the system metallicity. 

The young sub-solar solution suggests the system has an age $\tau_{\rm age} = 12$~Myr with [Fe/H] $= -0.35 \pm 0.09$~dex. The measured properties of the individual components are largely reproduced within the 1$\sigma$ observational uncertainties (see Table~\ref{tab:physprop}). However, the radius of star~A is not well reproduced. The model solution predicts that NY\,Hya\,A is smaller than observed by about 2.7\% ($2.5\sigma$). Despite this, the solution could be viable. 

There are two main difficulties with a young, metal-poor solution. The first issue is that the PMS phase is rapid compared to the overall lifetime of the star. NY\,Hya is not definitively associated with a young group. Thus, the prior probability of NY\,Hya (a random field star system) being this young is relatively small. The second problem is that models predict a significant abundance of lithium for a star of this mass at 12~Myr, with A(Li)~$\approx$~3.3. This is a testable prediction. However, there are no clear Li absorption lines present in the spectra of NY\,Hya (cf. Sect.~\ref{stellaratmosphericparameters}). With no observational evidence of youth, a PMS solution is unlikely.

Alternatively, the older, super-solar metallicity solution suggests the system has an age $\tau_{\rm age} = 6.5$~Gyr with  [Fe/H]$_{\rm init}$ $= +0.30 \pm 0.07$~dex. The measured masses and radii of the individual components are reproduced within 1$\sigma$ of the measured values. \Teff s predicted by models are too warm by about 100~K, placing them within about $1.5\sigma$ of the measured \Teff s. While the initial metallicity is significantly higher than the solar metallicity estimate, gravitational settling and diffusion acting over the lifetime of the model star suggest the metallicity at the main-sequence turn-off is about 0.1 dex lower than the initial value. This appears to be a viable solution. While encouraging, the difficulty with an older super-solar metallicity solution is that both stars must be at or just older than their respective core hydrogen exhaustion ages. This solution is \emph{a priori} unlikely. 

The problem with the old age, super-solar metallicity solutions predicted from by-eye fitting and the MCMC integration is that the resulting models are found to be just at the start of a rapid phase of stellar evolution. This is illustrated in Fig.~\ref{fig:model_hrd}, where an individual model mass track for an $M = 1.160\ M_\odot$ star with [Fe/H]~$= + 0.40$ is shown as a dotted line. The stars in NY\,Hya are right at the nexus of the core hydrogen burning phase and a proceeding period of core contraction prior to the establishment of hydrogen shell burning, where the mass track increases in luminosity and \Teff. This phase of evolution is brief ($\tau \sim 0.26$~Gyr) compared to the overall lifetime of the star ($\tau = 8.5$~Gyr). While it is not unfathomable that both stars would be near the end of the core hydrogen burning phase, the a priori probability of finding a star at this critical phase is small. Further evidence is needed to definitively establish the evolutionary stage of NY\,Hya. 

\begin{figure*}
    \centering
    \includegraphics[width=0.47\textwidth]{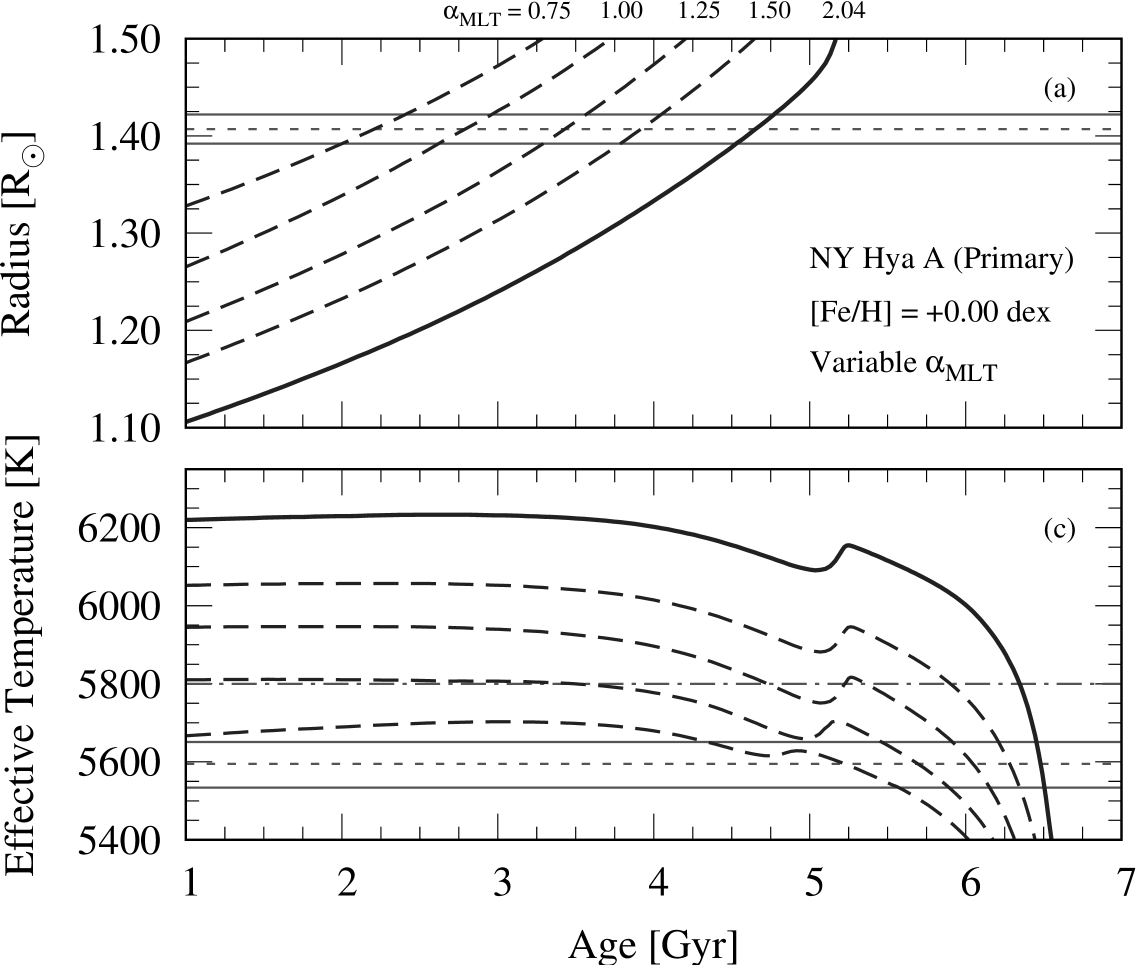} \qquad
    \includegraphics[width=0.47\textwidth]{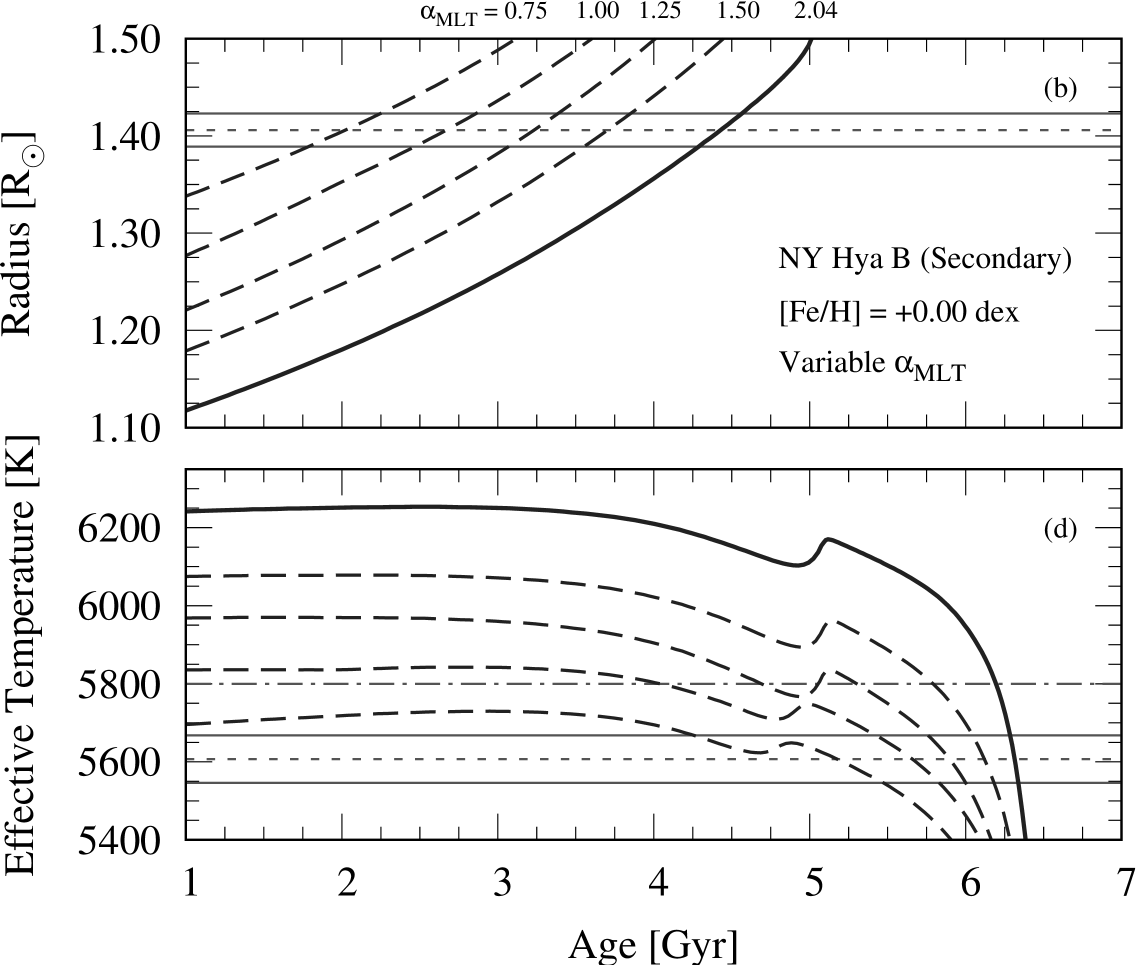} \\
    \vspace{\baselineskip}
    \includegraphics[width=0.47\textwidth]{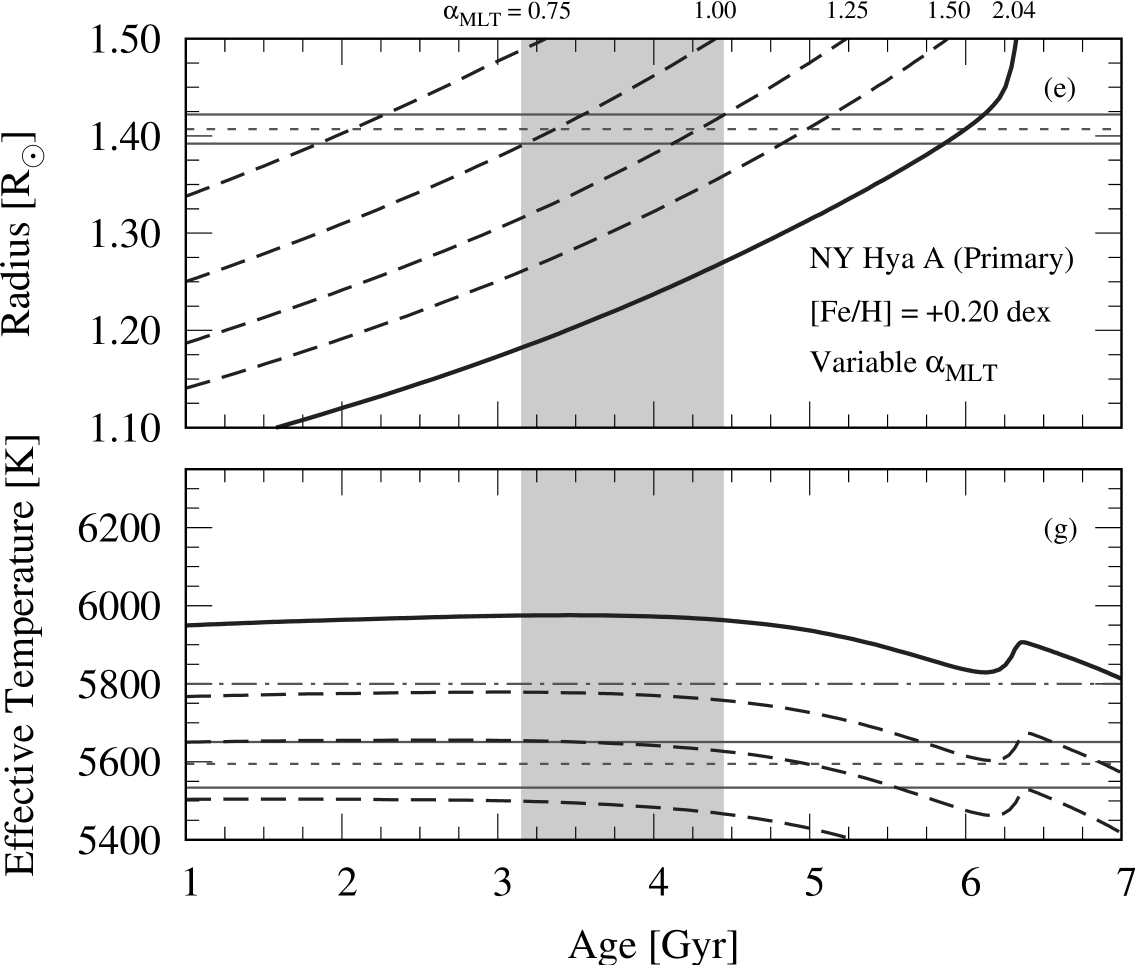} \qquad
    \includegraphics[width=0.47\textwidth]{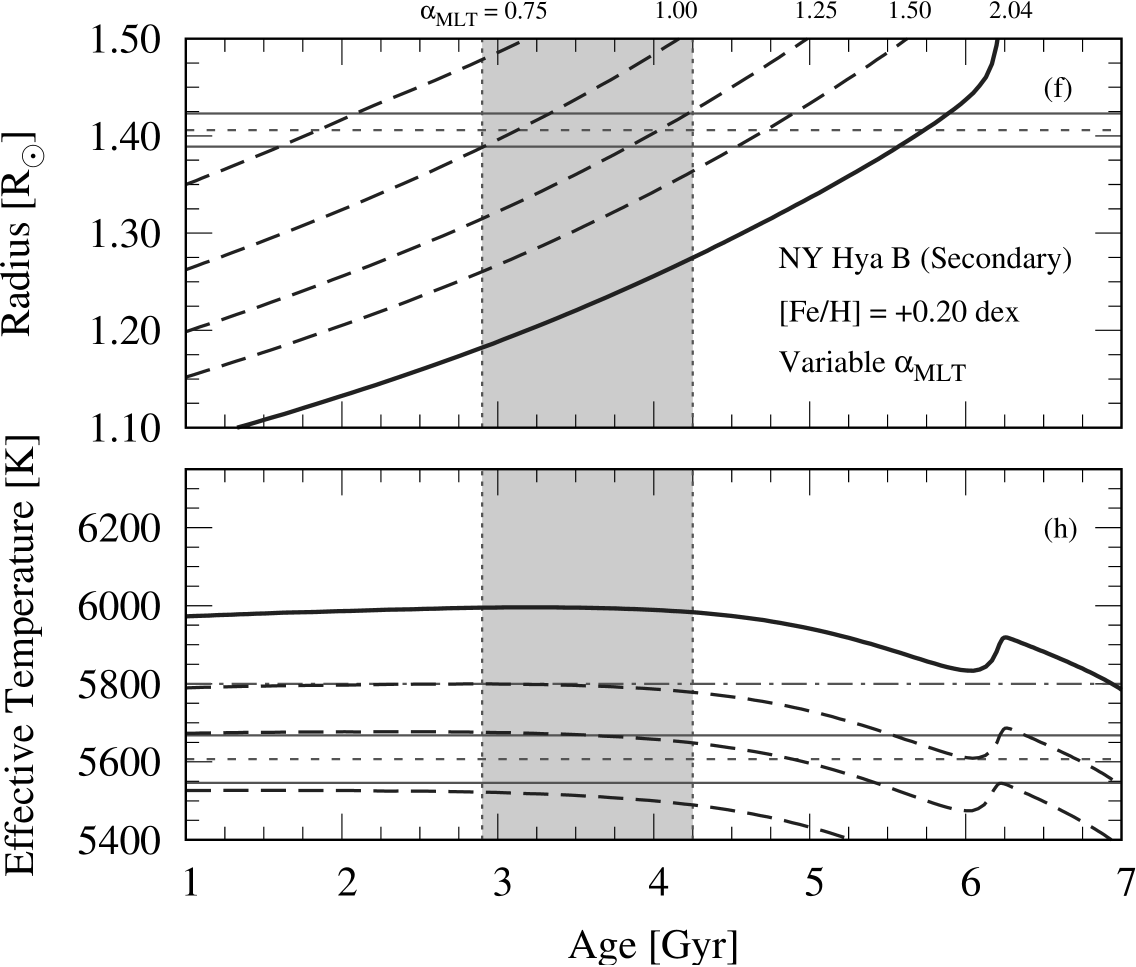}
    \caption{Stellar evolution mass tracks for NY\,Hya A and B computed with different values of the convective mixing length parameter, $\alpha_{\rm MLT} = 0.75,\, 1.00,\, 1.25,\, 1.50,\, 2.04$. Panels (a, c, e, g) show model radii and \Teff s as a function of age, respectively, for mass tracks with $M = 1.160 M_\odot$ (NY\,Hya A). Decreasing $\alpha_{\rm MLT}$ leads to larger radii and cooler temperatures at a given age. Panels (b, d, f, h) are the same, but computed with a model mass $M = 1.168 M_\odot$ (NY\,Hya B). Solar metallicity models are shown in panels (a) -- (d) and super-solar metallicity models are shown in panels (e) -- (h). The horizontal short-dashed lines indicate the measured values from Table~\ref{tab:physprop};  the horizontal solid lines indicate the corresponding $1\sigma$ uncertainty. The horizontal dot-dashed lines in the \Teff\ panels represent the \Teff\ measured from SED fitting. The grey shaded regions indicate a model-inferred age range.}
    \label{fig:mlt_model_results}
\end{figure*}

\subsection{Non-standard models}
\label{sec:nonstd}

A standard model solution that finds both stars on the main-sequence does not appear to exist without modifying the physical ingredients of the models. PMS and main-sequence turn-off solutions exist, but should be viewed with scepticism in the absence of further observational evidence. Instead, we investigate the possibility that non-standard physics are required to reproduce the observed properties of NY\,Hya. We make the explicit assumption that the stars should be on the main-sequence to avoid them existing at rapid phases of stellar evolution. Non-standard physics for this study include a modified convective mixing length, inhibition of convection due to magnetic fields, and the influence of star spots on stellar properties.

\subsubsection{Convective mixing length}

Convection in one-dimensional stellar evolution models is prescribed using a phenomenological approach known as mixing length theory (MLT). In MLT, a convective flow is described by means of a buoyant adiabatic fluid parcel traversing a distance $\ell = \alpha_{\rm MLT}H_p$ before mixing with its surroundings. Here, $H_p$ is a local pressure scale height and $\alpha_{\rm MLT}$ is a free parameter often chosen so that a $1\ M_\odot$ stellar model reproduces the solar properties at the solar age (4.56 Gyr). However, there is no strong \emph{a priori} justification for this parameter remaining constant with model mass, or through a single stellar model's evolution. As a first approach to exploring whether main-sequence solutions exist for NY\,Hya, we investigate the impact of the convective mixing length parameter $\alpha_{\rm MLT}$.

Results from a series of models with a reduced $\alpha_{\rm MLT}$ are shown in Fig.~\ref{fig:mlt_model_results}. Stellar evolution models were computed with the measured masses $M_A = 1.160\ M_\odot$ and $M_B = 1.168\ M_\odot$ at solar metallicity and [Fe/H]~$= +0.2$. We attempted to keep the metallicity as close to solar as possible while still finding a solution. For each mass and metallicity combination, a series of models were calculated with $\alpha_{\rm MLT} = 0.75,\, 1.00,\, 1.25,\, 1.50$ for comparison with our solar-calibrated $\alpha_{\rm MLT,\,\odot} = 2.04$. 

Fig.\,\ref{fig:mlt_model_results}(a-d) illustrate how the model radii and \Teff\ evolve with time for the standard case (solid line) and adjusted mixing length models (dashed lines) at solar metallicity. As is expected for inefficient convection, the stellar radius increases and the \Teff\ decreases at a given age as the mixing length decreases. The inferred age of NY\,Hya is reduced to 2 -- 4~Gyr based on radius measurements for the two components. However, an $\alpha_{\rm MLT} < 0.75$ is required to produce agreement with the \Teff\ (as determined from spectral synthesis) at an age consistent with stellar radii assuming a solar metallicity. To match the \Teff\ determined from SED fitting (dot dashed line), the mixing length parameter needs to be $\alpha_{\rm MLT} = 1.00$.

Models computed with a super-solar metallicity [Fe/H]~$= +0.2$ require  $1.00 < \alpha_{\rm MLT} < 1.25$, as is shown in Fig.~\ref{fig:mlt_model_results}(e -- h). One would infer an age for NY\,Hya of 3.1 -- 4.5~Gyr. Increasing the metallicity further allows for main-sequence model solutions with higher values of $\alpha_{\rm MLT}$. This is the result of higher metallicity models having naturally lower \Teff s prior to $\alpha_{\rm MLT}$ adjustments. It appears that artificially lowering the convective mixing length through modifications to $\alpha_{\rm MLT}$ can produce the intended results: agreement between models and observations at a main-sequence age. Unfortunately, changing $\alpha_{\rm MLT}$ carries no immediate physical meaning. It is not possible to immediately diagnose why a reduced $\alpha_{\rm MLT}$ is required. 

\begin{figure*}
    \centering
    \includegraphics[width=0.47\textwidth]{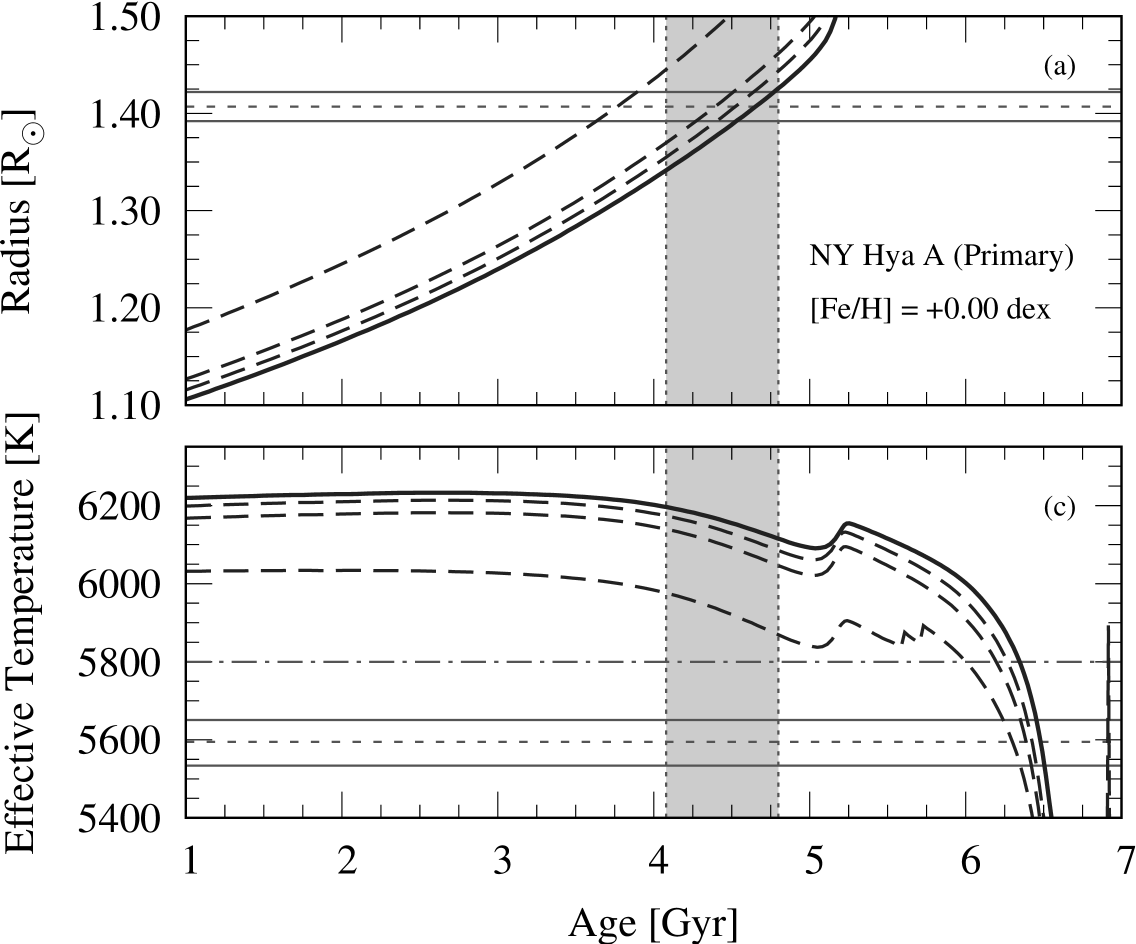} \qquad
    \includegraphics[width=0.47\textwidth]{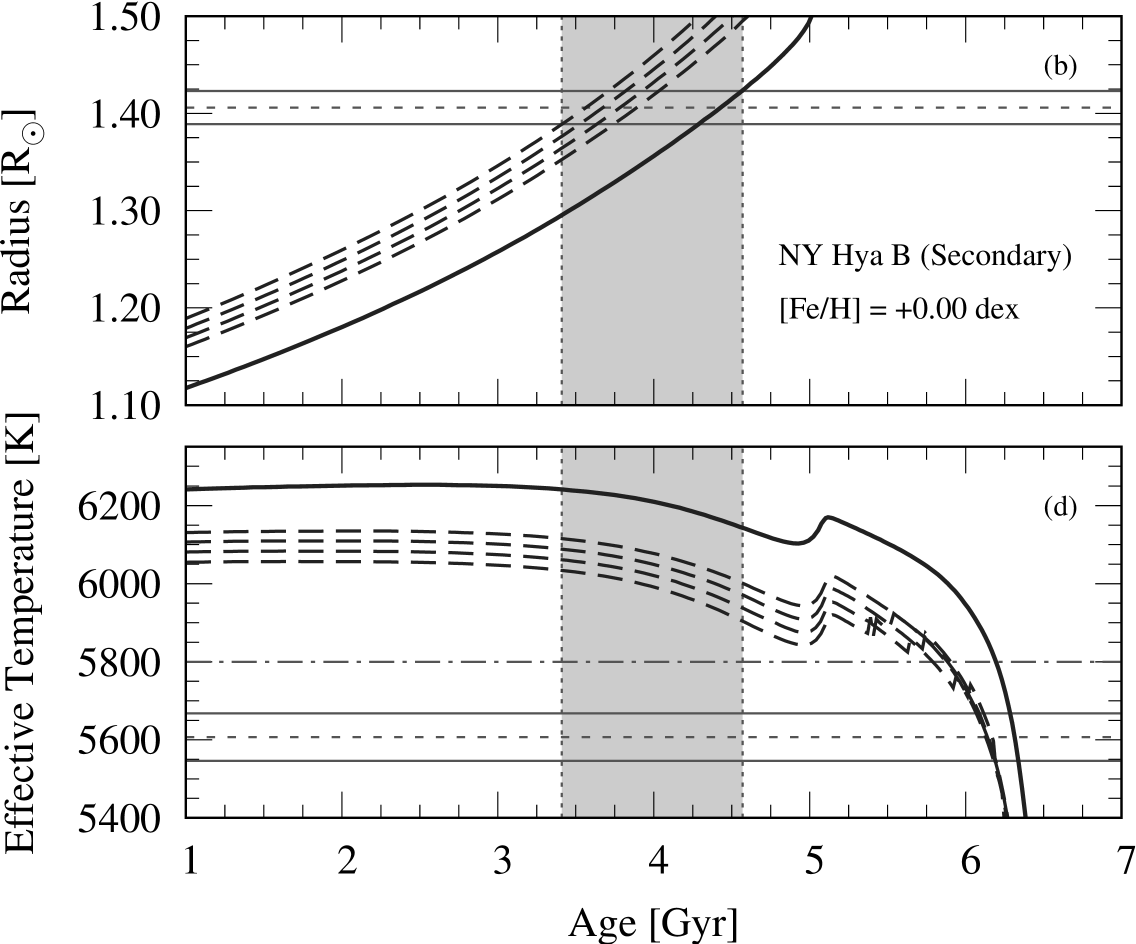} \\ \vspace{\baselineskip}
    \includegraphics[width=0.47\textwidth]{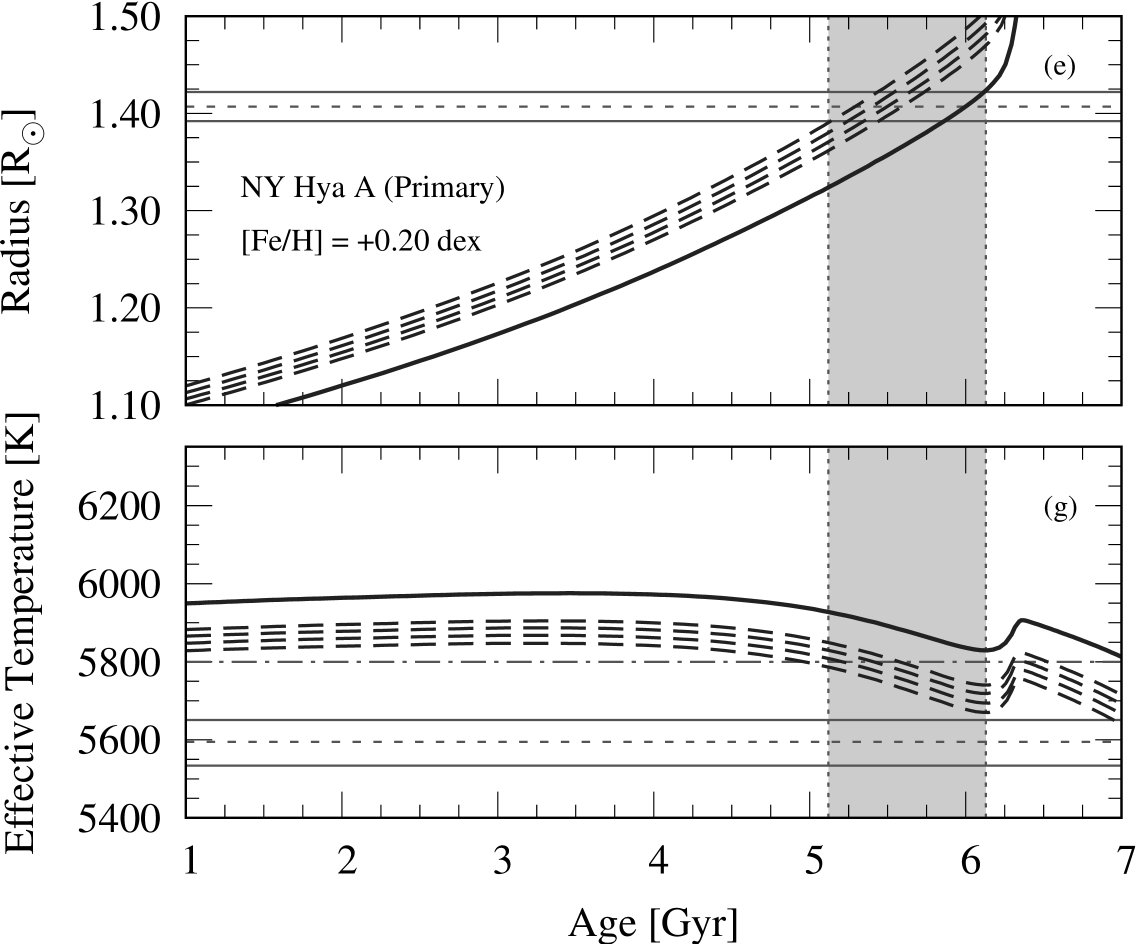} \qquad
    \includegraphics[width=0.47\textwidth]{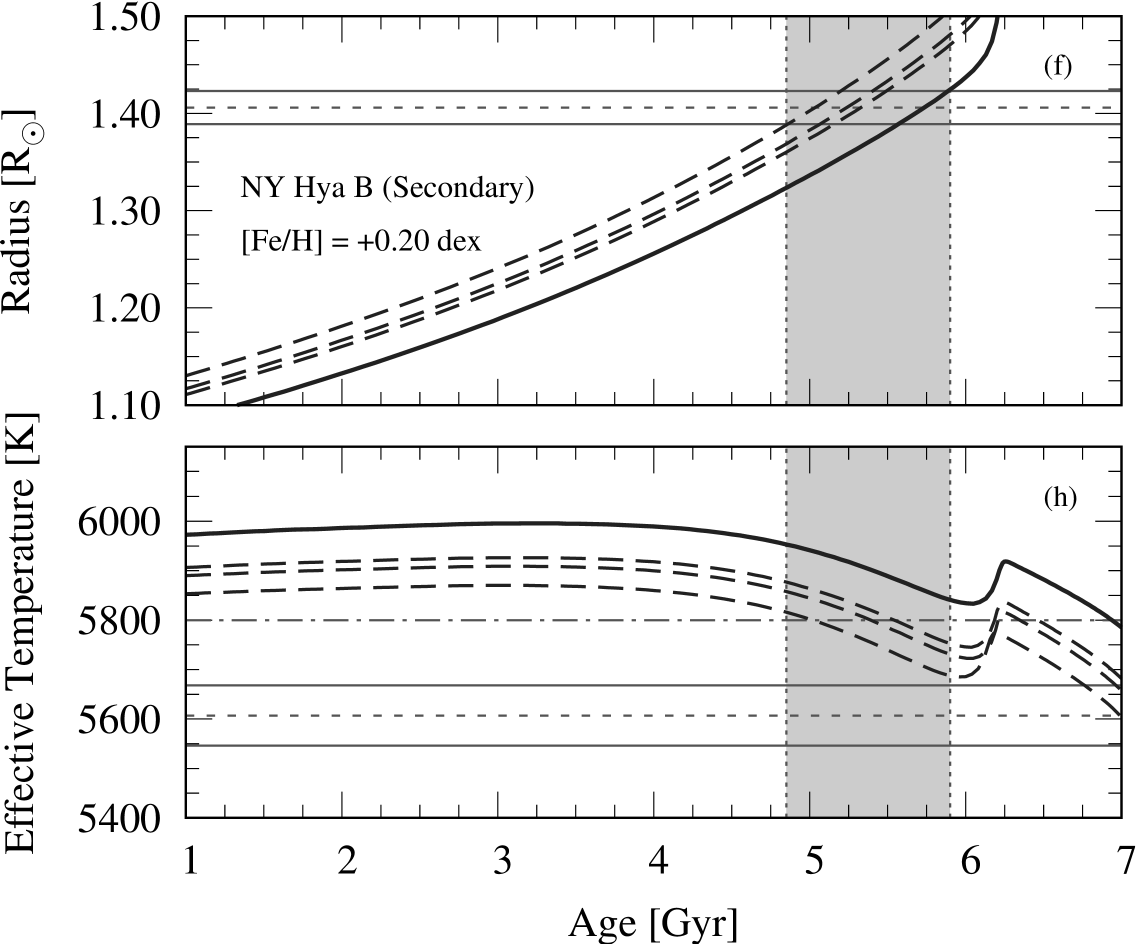}
    \caption{Stellar evolution mass tracks for NY\,Hya A and B showing the temporal evolution of model radii and \Teff s between 1 and 7 Gyr that include magnetic inhibition of convection. Mass tracks are computed with different values of the surface magnetic field strength, $B_{\rm surf} = 900,\, 1000,\, 1100,\, 1200$~G. Increasing surface magnetic field strengths leads to larger model radii and cooler \Teff \ values at a given age. Panels (a), (c), (e), and (g) are computed with $M = 1.160 M_\odot$ for NY\,Hya A, while panels (b), (d), (f), and (h) are computed with $M = 1.168 M_\odot$ for 
    NY\,Hya B. Solar metallicity models are shown in panels (a) -- (d) and super-solar metallicity models are shown in panels (e) -- (h).}
    \label{fig:mag_model_results}
\end{figure*}

\subsubsection{Magnetic inhibition of convection}
\label{magnetic_inhibition_of_convection}
Reducing the efficiency of convection by tuning the convective mixing length parameter ($\alpha_{\rm MLT}$) can bring model predictions into agreement with the measured properties of NY\,Hya A and B. This is particularly true if the two stars have a slightly super-solar metallicity, with [Fe/H]~$=+0.20$. However, this approach is phenomenological in nature and does not lend physical insight as to why convection may be inefficient in these stars. One testable explanation for inefficient convection is that strong magnetic fields are inhibiting convective flows thereby suppressing convective energy transport. Magnetic inhibition of convection has been previously shown to explain differences between model predictions and observations of the properties (radii and \Teff s) for the solar-type detached eclipsing binary EF Aqr \citep{vos2012,feidenchaboyer2012}.

The hypothesis that magnetic inhibition of convection is responsible for the noted disagreements between main-sequence stellar models and the measured properties of NY\,Hya A and B is tested using magnetic Dartmouth stellar evolution models \citep{feidenchaboyer2012}. Individual mass tracks were computed for each NY\,Hya component assuming multiple initial metallicity values ([Fe/H] = [0.0, 0.2] dex) and a range of surface-averaged magnetic field strengths ($\langle Bf \rangle$ = [0.3, 0.6, 0.9, 1.0, 1.1, 1.2] kG). The surface-averaged magnetic field is assumed to be constant throughout the model's evolution. While not necessarily realistic, the impact of magnetic inhibition of convection on the predictions of main-sequence stellar properties at a given age were found to be insensitive to the magnetic field strength evolution \citep{feidenchaboyer2012}. Masses for the individual components were taken to be the most probable values from Table \ref{tab:physprop}. Deviations from the most probable value of the order of the observed uncertainty are not large enough to affect the overall conclusions of our magnetic model analysis. 

Results from our magnetic stellar model analysis for a solar and super-solar metallicity are shown in Fig.~\ref{fig:mag_model_results}. We find that with magnetic models, model radii are consistent with the measured radii during the model's main-sequence lifetime. Estimated ages are between 4 and 5 Gyr for solar metallicity models (Fig.~\ref{fig:mag_model_results}a -- d), and between 5 and 6 Gyr for super-solar metallictiy models (Fig.~\ref{fig:mag_model_results}e -- h). However, assuming an initial solar metallicity produces models with \Teff s higher than the measured values by 300~K -- 400~K at the age where models reproduce the measured stellar radii. Adopting a super-solar initial metallicity, [Fe/H] = 0.2~dex, yields model \Teff s that are about 200~K higher than the measured values. In the latter case, magnetic models agree with \Teff s estimated from SED fitting ($\sim$5\,800~K), but they are nevertheless discrepant with estimates from the detailed spectroscopic analysis. 

Assuming the components NY\,Hya A and B have \Teff s more consistent with the SED fitting analysis ($T_{\rm eff} \sim 5800$~K),\footnote{This assumes a more modern solar composition for the SED fitting consistent with the adopted stellar model solar composition.} models suggest a super-solar metallicity for the system. An initial metallicity of [Fe/H] = 0.2~dex provides a viable main-sequence solution, which after diffusion has a final metallicity of [Fe/H] = 0.1~dex. Magnetic stellar models suggest that the best-fit surfaced-averaged magnetic field is $\langle Bf \rangle = 1.0 \pm 0.2$~kG to provide a main-sequence solution for both stars at the same age. Given the similarities in stellar mass, the two components have approximately the same surface-averaged magnetic field strength. This restricts the age of the NY\,Hya system to $\tau_{\rm age} = 5.3 \pm 0.3$~Gyr. 

\subsubsection{Star spots}

Star spots can potentially affect stars in multiple ways depending on the timescale over which the star spots exist on the stellar surface. If star spots are transient features that appear and disappear rapidly relative to the star's thermal timescale, spots should have a limited effect on the star's thermal structure. Transient star spots will cause the star's average surface temperature to appear cooler than the ambient photosphere and will cause a temporary decrease in the star's observed luminosity. However, if star spots are not transient features, they may cause the star to restructure to maintain thermal equilibrium \citep[e.g.][]{spruit1982,spruit1986,somers2020}.

We evaluate whether transient spots can produce the observed temperature disagreements between observational measurements (Table~\ref{tab:physprop}) and standard models (see Sect.~\ref{sec:stdmodel}) using a simple phenomenological model. Transient spots would leave the stellar radius unaffected and cause a decrease in bolometric luminosity due to a cooler average surface temperature. Mathematically,
\begin{equation}
    \left(\frac{T_{\rm eff}}{T_{\rm eff,\,0}}\right)^4 = 1 - \varrho\left(1 - \varpi^4\right),
\end{equation}
where $T_{\rm eff}$ is the \Teff\ of a spotted star, $T_{\rm eff,\,0}$ is the \Teff\ of a similar unspotted star, $\varrho$ is the areal surface coverage of star spots, and $\varpi$ is the temperature ratio between the spots and ambient photosphere. Assuming that star spots on NY\,Hya A and B have similar characteristics to sunspots, $\varpi \approx 0.65$ \citep{solanki2003}, each star requires an areal surface coverage of star spots $\varrho \approx 0.40$ to produce the observed 5\,600~K \Teff\ compared to solar metallicity model predictions of 6200~K. This surface coverage translates to an approximate mean surface magnetic field strength by assuming star spots are formed by equipartition strength magnetic fields \citep{torres2021}. This yields an estimate of $\langle Bf \rangle = 760$~G, which is roughly consistent with magnetic model predictions (see Sec.~\ref{magnetic_inhibition_of_convection}) and further supported from a semi-empirical estimate based on the observed X-ray luminosity (see Sec. \ref{semi-empiricalestimateofmagneticfieldstrengthandtotalXrayluminosity}).

If star spots affect a stars' internal structure, they produce effects similar to magnetic inhibition of convection \citep{somers2020}. We test whether long-lived star spots can explain the observed properties of the NY\,Hya stars using {\sc spots} models \citep{somers2020}. These models assume a solar metallicity and are calculated with a fixed temperature ratio of $\varpi = 0.8$ between spots and the stellar photosphere. {\sc spots} models are computed on a fixed-mass grid, so we adopt a mass $M = 1.15 M_\odot$ for our comparisons. Spotted models reproduce the radii of NY\,Hya A and B at ages between 1.75 Gyr and 4.75 Gyr, depending on the surface coverage of spots. High coverage fractions produce the observed radii at younger ages. However, only {\sc spots} models with a surface coverage $f > 0.80$ are able to reproduce the measured $T_{\rm eff} \sim 5600$~K. In this case, the inferred age for NY\,Hya is $\tau_{\rm age} = 2.0\pm0.3$~Gyr. By contrast, a surface coverage of $f\sim0.55$ is required to produce models with $T_{\rm eff} \sim 5800$~K, leading to $\tau_{\rm age} = 3.0\pm0.5$~Gyr.

There are a couple reasons to be sceptical of these results. First, the assumed temperature ratio $\varpi = 0.8$ is not necessarily representative of spot temperatures on the stars of NY\,Hya, which could be as low as $\varpi = 0.65$. Cooler spots would lead to a lower surface coverage required to reproduce the observed properties of NY\,Hya. Assuming luminosity is conserved during spot-driven inflation, the required surface coverages would decrease to $f \sim 0.55$ and $0.40$ for  $T_{\rm eff} \sim 5600$~K and $5800$~K, respectively. A second reason is rooted in the assumptions built into the {\sc spots} models, which prescribe surface boundary conditions for stellar models based on the temperature structure for the ambient photosphere. This works well for lower surface coverages, but is a questionable assumption when spots cover over half of the stellar surface, at which point the temperature structure for the spotted surface may be more appropriate (similar to a cool star with hot spots). However, this scepticism cannot be used to rule out the possibility that star spots are responsible for difference between standard stellar models and the measured properties of NY\,Hya A and B.

In summary, a surface coverage of approximately 40\% for each star is required for short-lived star spots to reproduce the properties of NY\,Hya. However, a surface coverage of 55\% -- 80\% is required if star spots are long-lived and affect stellar structure.

\subsection{Semi-empirical estimate of magnetic field strength and total X-ray luminosity}
\label{semi-empiricalestimateofmagneticfieldstrengthandtotalXrayluminosity}


We performed a consistency check on the best-fit magnetic field strength required by the non-standard stellar evolution models. An independent estimate of the surface averaged mean magnetic field strength $\langle Bf\rangle$ and total X-ray luminosity $\log L_X$ can be obtained by following the methodology described in \citet{torres2014}.

The method is semi-empirical partially relying on theory. A power-law relationship between 
$\langle Bf\rangle$ and the Rossby number $R_o = P_{\rm rot}/\tau_c$ was given in \citet{saar2001}
where $P_{\rm rot}$ is the stellar rotation period and $\tau_c$ is the convection turnover time.
For NY\,Hya, the measured rotation velocities and stellar radii for the two components are consistent with a
derived stellar rotation period equal to the orbital period within uncertainties. This implies that the
stars are tidally locked and rotate synchronously as noted earlier. We therefore set 
$P_{\rm rot} = P_{\rm orb}$. The convection turnover time is estimated from theory using
the relationship\footnote{\citet{gilliland1986} fig. 7 was digitised and a cubic spline was determined for
increased accuracy and proper error propagation.} in \citet{gilliland1986} for a given measured effective
temperature.

For star~A we found $\tau_c = 20.9 \pm 1.2$ days, and for star~B $\tau_c = 20.6 \pm 1.3$ days, resulting in Rossby numbers of $R_o = 0.229 \pm 0.012$ and $R_o = 0.231 \pm 0.012$ for stars A and B, respectively. The \citet{saar2001} calibration relation then yields a mean magnetic field strength
for star~A of $\langle Bf\rangle_{A} = 327 \pm 209$ G and star~B $\langle Bf\rangle_{B} = 323 \pm
206$ G confirming a near-identical mean field strength of the two components. The total uncertainty was
found from MC error propagation added in quadrature to the scatter of the relationship as provided by \citet{saar2001}. 

Another estimate of the magnetic field strength may be obtained from applying a scaling relation
using the core emission feature in the Ca II K line \citep{schrijver1989,feidenchaboyer2012}. The relation was obtained from solar observations correlating magnetic field measurements with Ca II K core emissions of local photospheric active regions. The scaling relation is independent of any Doppler-shift that might occur as a result of the binary motion. We applied the relation given by \citet{feidenchaboyer2012} assuming its validity to the components of NY\,Hya. Rough estimates of the intensities of the core emission feature ($I_c$) and at the Ca II K wing ($I_{w};7.4$ \si{\angstrom} to the red) were determined. We found these intensities from a normalised FEROS spectrum at phase 0.5 that allowed us to estimate   the mean magnetic field strength of $\langle Bf \rangle = 959 \pm 431$ G of star~A. This estimate is about a factor of 3 larger than the estimate obtained from \citet{saar2001}, while in closer agreement with the 1.0 kG field strength derived from the non-standard stellar evolution model.

As a final consistency check, the mean magnetic field strength of $\langle Bf \rangle = 959$ G can be transformed to an X-ray luminosity using the empirical scaling relations provided in \citet{pevtsov2003} and \citet{feidenchaboyer2013}. This provides a consistency check allowing us to compare the X-ray luminosities obtained from the mean magnetic field of each component with the total X-ray emission detected by the {\it ROSAT} satellite. \citet{pevtsov2003} found a tightly
correlated power-law relationship between the X-ray luminosity $L_X$ and the magnetic flux at the stellar
surface $\Phi = 4 \pi R^2 \langle Bf\rangle$ with $R$ denoting the stellar radius. \citet{feidenchaboyer2013} expanded 
the \citet{pevtsov2003} data set and provide an updated scaling relation valid over 12 orders of
magnitude in both $\Phi$ and $L_X$. For NY\,Hya, we found $\log L_X = 29.7 \pm 2.1$ dex (with $L_X$ in 
erg/s) for star~A and $\log L_X = 29.9 \pm 2.0$ dex for star~B. These estimates are consistent
with the near-identical properties of the two components. The total X-ray luminosity was found to be $\log
L_{X,A+B} = 30.1 \pm 1.7$ dex. This estimate can now be compared with the {\it ROSAT} measurement.

NY\,Hya has a confirmed X-ray counterpart and is included in the second {\it ROSAT} all-sky survey 
(2RXS ID: J092121.7-064018) source catalogue by \cite{boller+2016} listing a count rate of 
$0.055 \pm      0.016\,{\rm counts}\,s^{-1}$ (in the 0.1 - 2.0 keV range) and a hardness ratio of
${\rm HR1} = 0.39 \pm 0.38$. These measurements were obtained from a 294\,s exposure (about 20\,s longer than the first all-sky catalogue). The total X-ray luminosity for NY\,Hya is calculated from
\citet{szczygiel2008} using the {\it Gaia} (DR2) parallax and we find 
$\log L_{X}({\rm {\it ROSAT}}) = 29.89 \pm 0.18$ dex. 

Comparing the {\it ROSAT} measurement with the 
semi-empirical derived total X-ray luminosity, we find an agreement at the $0.62\sigma$ level. We conclude that the observed X-ray flux is consistent with the expected flux for a magnetically active star with $\langle Bf \rangle \approx 1$ kG. As for the case studied in \citet{torres2014} this good agreement may 
be interpreted as an indication of the accuracy of the $\langle Bf \rangle_{A,B}$ values obtained above although their formal $1\sigma$ uncertainties are large. We conclude this section by stating that a mean magnetic field of 1 kG is plausible for each stellar component lending further credibility to non-standard magnetic stellar evolution models.

\section{Discussion and conclusions}
\label{Summaryandconclusions}

We have presented a complete analysis of astrometric, photometric, and spectroscopic data for the NY\,Hya eclipsing binary system. The analysis is based on the assumption that the data are correctly interpreted as being caused by the binary nature of two orbiting stars and that measurement errors follow normal statistics.

The eclipsing nature of NY\,Hya was discovered from {\it Hipparcos} photometry. The two stars were found to be identical to within the measurement errors. The classic differentiation of a primary and secondary component is therefore limited by surface activity via star spots. The combination of joint properties of two solar-twin stars and a long orbital period renders this system ideal for studying stellar evolution of solar-type stars based on the assumption that complicated tidal effects are small.

We measured the physical properties of NY\,Hya to high precision based on high-quality spectroscopic and light curve data. Recent astrometric measurements from {\it Gaia} provide a direct empirical, and accurate distance measurement to a high precision, enabling a reliable estimate of distance-dependent system properties,  thus replacing previous semi-empirical distance estimates from photometry. We utilised the latest version of the {\sc jktebop} code and obtained model-independent maximum fractional uncertainties of 0.8\% and 1.2\% in the stellar masses and radii, respectively. 

We confronted the physical properties with standard and recently developed non-standard stellar evolution models. A standard model solution that describes both stars on the main-sequence consistently for all observed properties does not appear to exist. NY\,Hya belongs to the group of dEB with a short orbital period. Those binaries usually fail to reproduce predictions from classic stellar evolution theories \citep{feidenchaboyer2012,vos2012}. The resulting parameter distributions were found to be bimodal: both PMS (young age) and main-sequence turn-off (old age) solutions exist. Their validity should be viewed with scepticism due to lack of further observational evidence, in order to avoid the possibility of declaring the two components to be in a rapid phase of stellar evolution. For example, while standard models with an old age and super-solar metallicity are encouraging, they seem unlikely because $i)$ the likelihood that both stars currently are observed just at the main-sequence turn-off (sub-giant) phase is low;  $ii)$ because the  model \Teff s were about 700 K cooler than the measured temperatures;  $iii)$ because observational data suggest a solar metallicity for the two components, which is in discrepancy with the super-solar metallicity as suggested by the model; and, finally, $iv)$ standard models pointing towards a pre-main-sequence solution seem unlikely without further evidence of stellar youth. We attempted to constrain stellar age from kinematical consideration, but were not able to show NY\,Hya to be associated with any known nearby stellar cluster of known age. Age determination from spectral indicators requires a careful in-depth analysis. The preliminary results presented here point  towards a young stellar age, but are deemed not trustworthy at the moment. Lithium absorption features were absent from the FEROS spectra.

Under the assumption that the two components are observed in their main-sequence evolutionary phase and invoking magnetic non-standard models (assuming a constant surface-averaged magnetic field throughout the simulated stellar evolution) seems to reconcile the observed tensions in \Teff s. The best-fit  theoretical model, however, reproduces observed properties at a super-solar metallicity and an age of approximately 5.3 Gyr. This age is consistent with the derived synchronisation and circularisation  timescale. However, the required super-solar metallicity is still in conflict with the metallicity derived from our spectral synthesis analysis pointing towards solar metallicity for both components. We recall that a solar metallicity for both components is supported by Str{\"o}mgren photometry, although with large uncertainties, via calibration relations.

From visual inspection of the Ca II K and H lines, we found chromospheric activity in both components. This is further substantiated with a measured X-ray excess emission by {\it ROSAT} and indicates enhanced stellar magnetic activity. The best-fit theoretical model was found at a mean magnetic field of approximately 1 kG and is found to be consistent with a semi-empirical estimate of a mean magnetic field from the detected X-ray flux.

Some tension and shortcomings are still present. First, the detailed spectroscopic analysis resulted in \Teff\ values that are too cool compared to the mean temperature obtained from the SED analysis. The difference is significant and cannot be explained by the inferred uncertainties. Temperatures obtained from Str{\"o}mgren photometry are not precise enough to provide a reliable verdict in pointing in one (5600 K) or the other direction (5800 K). The SED models also pointed towards a higher metallicity for each component, as were inferred from a direct spectral analysis. Second, the present analysis did not carry out a detailed chemical abundance study for a direct empirical estimate of metallicity for each component. Such an analysis would be advantageous to place further constraints on stellar evolution models. Although the FEROS spectra are of relatively high quality, we did experience complications when working with them in detail in an attempt to measure individual abundances. Finally, an estimate of age would be useful for the establishment of the current evolutionary stage of NY\,Hya. On that account age accuracy (reliability) is more important than precision to differentiate between the young and old dichotomy.

We encourage future work in an attempt to further constrain the nature (i.e. age and abundances) of this eclipsing binary.  We also encourage a proper treatment of surface activity via modelling star spots directly or by removing star spot activity from the original observations.

During the final stage of analysis, we learned about the existence of additional high-quality photometric {\it TESS} and spectroscopic (ESO/HARPS) data,  and will leave their analysis to a future follow-up study. The HARPS data, of likely higher quality, might be able to decide on the true \Teff\ for each component. Furthermore, the {\it Gaia} DR2 and EDR3 point to a close (6 arcsec) companion star ({\it Gaia} DR2/EDR3 5746104876937814912), 8.5 magnitudes fainter than NY\,Hya, which is not visible in the PanSTARRS images (situated within the PanSTARRS NY\,Hya PSF). The contaminating flux contribution to the TESS target aperture is negligible, and is slightly fainter than the brightest of the three contaminating stars found in the PanSTARRS images. In addition, due to the companion star being 8.5 magnitudes fainter, it was not detected in the contrast curve obtained from the TCI images, which have a sky-background contrast of 8 magnitudes fainter than NY\,Hya. An interesting note is that the object 5746104876937814912 seems to be a bound companion to NY\,Hya. The measured parallaxes agree at a $0.41\sigma$ level, consistent with the stars being at the same distance. The proper motions are different by $\sim$4.7 mas/yr, which is statistically significant. This gives a projected separation of approximately 500 AU and a projected velocity difference of $\sim$0.5 AU/yr, while using the masses and separation would give an expected orbital velocity of $\sim$0.4 AU/yr. Thus, NY\,Hya seems to be part of a triple system.

\begin{acknowledgements}

A. Kaufer, O. Stahl, S. Tubbesing, and B.\ Wolf kindly observed nearly all the spectra of NY\,Hya 
during the Heidelberg/Copenhagen guaranteed time at FEROS in 1998 and 1999.

This research made use of {\sc lightkurve}, a {\sc python} package for {\it Kepler} and {\it TESS} data analysis \citep{Lightkurve}. J. T.-R. acknowledges financial support from CONICYT/FONDECYT  in terms of a 2018 Postdoctoral research grant, project number: 3180071. 
T.C.H. would like to acknowledge fruitful conversations with the following individuals: Arne Henden, Patrick Lenz, Eric Mamajek, Guillermo (Willie) Torres and Ignasi Ribas who also kindly provided details of CORAVEL data. This research has received funding from the Europlanet 2024 Research Infrastructure (RI) programme. The Europlanet 2024 RI provides free access to the world’s largest collection of planetary simulation and analysis facilities, data services and tools, a ground-based observational network and programme of community support activities. Europlanet 2024 RI has received funding from the European Union’s Horizon 2020 research and innovation programme under grant agreement No. 871149.
This research has made use of the SVO Filter Profile Service (http://svo2.cab.inta-csic.es/theory/fps/) supported from the Spanish MINECO through grant AYA2017-84089.
J. V. Clausen participated fully in the data collection and analysis up to the time of his death, but bears no responsibility for the final text of this paper.
B. E. Helt participated fully in the data collection and analysis up to the time of her death in 2023, but bears no responsibility for the final text of this paper.
E. H. Olsen participated fully in the data collection and analysis (especially data described in Section \ref{stromgrenphotometry}) up to the time of his death on March 1, 2024, but bears no responsibility for the final text of this paper.
N.P.’s work was supported by Funda\c{c}\~ao para a Ci\^{e}ncia e a Tecnologia (FCT) through the research grants UIDB/04434/2020 and UIDP/04434/2020.
U.G.J. acknowledges funding from the Novo Nordisk Foundation Interdisciplinary Synergy Programme 
grant no. NNF19OC0057374 and from the European Union H2020-MSCA-ITN-2019 under Grant no. 860470 (CHAMELEON).
J.V. acknowledges support from the Grant Agency of the Czech Republic (GA\v{C}R 22-34467S). The Astronomical Institute Ond\v{r}ejov is supported by the project RVO:67985815.

J.C.M. acknowledges financial support by Spanish grants PID2021-125627OB-C31 funded by MCIU/AEI/10.13039/501100011033 and by “ERDF A way of making Europe”, PGC2018-098153-B-C33 funded by MCIU/AEI, by the programme Unidad de Excelencia María de Maeztu CEX2020-001058-M, and by the Generalitat de Catalunya/CERCA programme.

P.L.P. was partly funded by Programa de Iniciaci{\'o}n en Investigaci{\'o}n-Universidad de Antofagasta. INI-17-03

\end{acknowledgements}

\bibliographystyle{aa}
\bibliography{references}

\onecolumn

\begin{appendix}

\section{Spectroscopic observations of NY Hya}
\label{appendix1}

\begin{table}[hbt!]
\caption{Log notes on spectroscopic (FEROS) observations of NY\,Hya. See tables notes for more details.}
\centering
\begin{tabular}{ccccccc}
\hline
BJD(TDB)           & ID      & Obs     & Orb.  & $t_{\rm exp}$ & $S/N$ \\
+2,450,000.0      & [fero*] &         & phase & [s]                    &       \\
\hline
\hline
1141.83831 & 1334    &  1  & 0.03  & 900			 & 195\\
1142.81181 & 1372    &  1  & 0.24  & 900			 & 221\\
1143.78362 & 1423    &  1  & 0.44  & 600			 & 210\\
1144.76583 & 1467    &  1  & 0.64  & 600			 & 213\\
1145.78391 & 1515    &  1  & 0.86  & 600			 & 214\\
1146.80692 & 1559    &  1  & 0.07  & 600			 & 225\\
1147.78716 & 1603    &  1  & 0.28  & 600			 & 253\\
1148.73485 & 1638    &  2  & 0.48  & 600			 & 220\\
1149.75746 & 1678    &  2  & 0.69  & 600			 & 187\\
1150.72278 & 1719    &  2  & 0.89  & 600			 & 206\\
1151.75561 & 1764    &  2  & 0.11  & 600			 & 261\\
1155.73697 & 1954    &  2  & 0.94  & 600			 & 245\\
1171.83471 & 2266    &  2  & 0.31  & 600			 & disc. \\
1173.79548 & 2346    &  2  & 0.73  & 600			 & 219\\
1174.75602 & 2385    &  2  & 0.93  & 600			 & 203\\
1175.87387 & 2436    &  2  & 0.16  & 600			 & 216\\
1176.79909 & 2452    &  2  & 0.35  & 600			 & 230\\
1177.79159 & 2478    &  2  & 0.56  & 600			 & 194\\
1178.76368 & 2501    &  2  & 0.77  & 600			 & 220\\
1180.81055 & 2558    &  2  & 0.20  & 420			 & 192\\
1181.81465 & 2591    &  2  & 0.41  & 600			 & 228\\
1182.80487 & 2621    &  2  & 0.61  & 600			 & 221\\
1183.79834 & 2663    &  2  & 0.82  & 600			 & 218\\
1185.82246 & 2738    &  2  & 0.24  & 600			 & 232\\
1186.68894 & 2765    &  2  & 0.43  & 600			 & 206\\
1187.80567 & 2810    &  2  & 0.66  & 600			 & 213\\
1188.70072 & 2838    &  2  & 0.85  & 600			 & 185\\
1190.69662 & 2912    &  2  & 0.27  & 600			 & disc. \\
1191.81861 & 2962    &  2  & 0.50  & 600			 & 196\\
1192.73725 & 2998    &  2  & 0.69  & 600			 & 182\\
1193.72282 & 3038    &  2  & 0.90  & 600			 & 203\\
1194.88422 & 3077    &  2  & 0.14  & 600			 & 174\\
1197.77888 & 3154    &  2  & 0.75  & 600			 & 207\\
1199.81885 & 3225    &  2  & 0.18  & 600			 & 215\\
1200.80023 & 3267    &  2  & 0.38  & 600			 & 190\\
\hline
\end{tabular}
\label{specdata} 
\tablefoot{The BJD(TDB) time stamps are the exposure mid-time and were converted from original HJD(UTC) time stamps. ID refers to the filename as returned by the FEROS Heidelberg database. Observers: Obs=1 (Kaufer), Obs=2 (Kaufer/Tubbesing/Wolf/Szeifert/Rivinius). $t_{\rm exp}$ is the exposure time. The $S/N$ was measured from narrow bands at continuum between 5360~\si{\angstrom} and 5600~\si{\angstrom} and averaged. Two spectra were discarded. See main text for details.}
\end{table}

\onecolumn

\section{RV measurements from {\sc todcor}}
\label{appendix2}

\begin{table}[hbt!]
\caption{RV measurements from {\sc todcor}.}
\centering
\label{tab:todcor}
\setlength\tabcolsep{4pt}
\begin{tabular}{lcccc}  
\hline
BJD(TDB) - & ID & Orb.      & RV${_{\rm A}}$             & RV${_{\rm B}}$           \\
2,450,000.0  & [fero*] & phase &  [km/s]          & [km/s]       \\
\hline
\hline
1141.8383 & 1334 & 0.03 & $23.20\pm0.76$ & $58.05\pm0.75$ \\ 
1142.8118 & 1372 & 0.24 & $-42.44\pm1.92$ & $123.54\pm1.80$ \\ 
1143.7836 & 1423 & 0.44 & $10.05\pm1.64$ & $70.69\pm1.54$ \\ 
1144.7658 & 1467 & 0.64 & $108.32\pm3.38$ & $-25.31\pm3.41$ \\ 
1145.7839 & 1515 & 0.86 & $105.58\pm3.02$ & $-24.37\pm2.85$ \\ 
1146.8069 & 1559 & 0.07 & $3.95\pm1.90$ & $78.11\pm1.81$ \\ 
1147.7872 & 1603 & 0.28 & $-40.78\pm1.96$ & $122.52\pm5.31$ \\ 
1148.7349 & 1638 & 0.48 & $29.10\pm0.63$ & $52.32\pm0.63$ \\ 
1149.7575 & 1678 & 0.69 & $119.96\pm4.70$ & $-36.81\pm1.99$ \\ 
1150.7228 & 1719 & 0.89 & $90.94\pm2.15$ & $-11.57\pm2.09$ \\ 
1151.7556 & 1764 & 0.11 & $-12.28\pm2.42$ & $93.61\pm2.48$ \\ 
1155.7370 & 1954 & 0.94 & $69.88\pm1.83$ & $12.51\pm1.78$ \\ 
1173.7955 & 2346 & 0.73 & $123.41\pm5.62$ & $-41.93\pm1.67$ \\ 
1174.7560 & 2385 & 0.93 & $77.00\pm1.80$ & $3.95\pm1.80$ \\ 
1175.8739 & 2436 & 0.16 & $-30.06\pm3.55$ & $112.51\pm3.61$ \\ 
1176.7991 & 2452 & 0.35 & $-25.34\pm3.00$ & $106.56\pm3.05$ \\ 
1177.7916 & 2478 & 0.56 & $73.78\pm1.63$ & $7.89\pm1.66$ \\ 
1178.7637 & 2501 & 0.77 & $123.33\pm1.88$ & $-42.44\pm1.67$ \\ 
1180.8105 & 2558 & 0.20 & $-38.09\pm4.36$ & $119.71\pm4.25$ \\ 
1181.8147 & 2591 & 0.41 & $-5.89\pm2.00$ & $87.23\pm2.04$ \\ 
1182.8049 & 2621 & 0.61 & $96.48\pm2.59$ & $-13.61\pm2.50$ \\ 
1183.7983 & 2663 & 0.82 & $116.41\pm4.46$ & $-33.96\pm2.14$ \\ 
1185.8225 & 2738 & 0.24 & $-43.24\pm1.62$ & $123.92\pm1.51$ \\ 
1186.6889 & 2765 & 0.43 & $3.93\pm1.77$ & $76.88\pm1.71$ \\ 
1187.8057 & 2810 & 0.66 & $112.71\pm3.68$ & $-30.07\pm3.55$ \\ 
1188.7007 & 2838 & 0.85 & $109.53\pm3.23$ & $-26.39\pm3.13$ \\ 
1191.8186 & 2962 & 0.50 & $41.11\pm0.79$ & $41.11\pm0.98$ \\ 
1192.7372 & 2998 & 0.69 & $120.95\pm1.83$ & $-36.41\pm1.71$ \\ 
1193.7228 & 3038 & 0.90 & $89.29\pm2.43$ & $-7.69\pm2.36$ \\ 
1194.8842 & 3077 & 0.14 & $-24.98\pm3.37$ & $107.02\pm3.45$ \\ 
1197.7789 & 3154 & 0.75 & $125.02\pm1.67$ & $-42.55\pm1.56$ \\ 
1199.8189 & 3225 & 0.18 & $-34.55\pm1.80$ & $116.22\pm1.74$ \\ 
1200.8002 & 3267 & 0.38 & $-14.20\pm2.85$ & $95.70\pm2.94$ \\ 
\hline
\end{tabular}
\tablefoot{We excluded the spectrum at orbital phase 0.50 (\# 2962) since it is deemed unreliable due to line blending. The inclusion of this spectrum would have the effect to add scatter and thus make the RVs look slightly worse than they actually are.}
\end{table}

\onecolumn

\section{Corner plots of parameter uncertainty covariances and parent distribution}
\label{appendix3}

\begin{figure}[hbt!]
\centering
\includegraphics[width=1.0\textwidth]{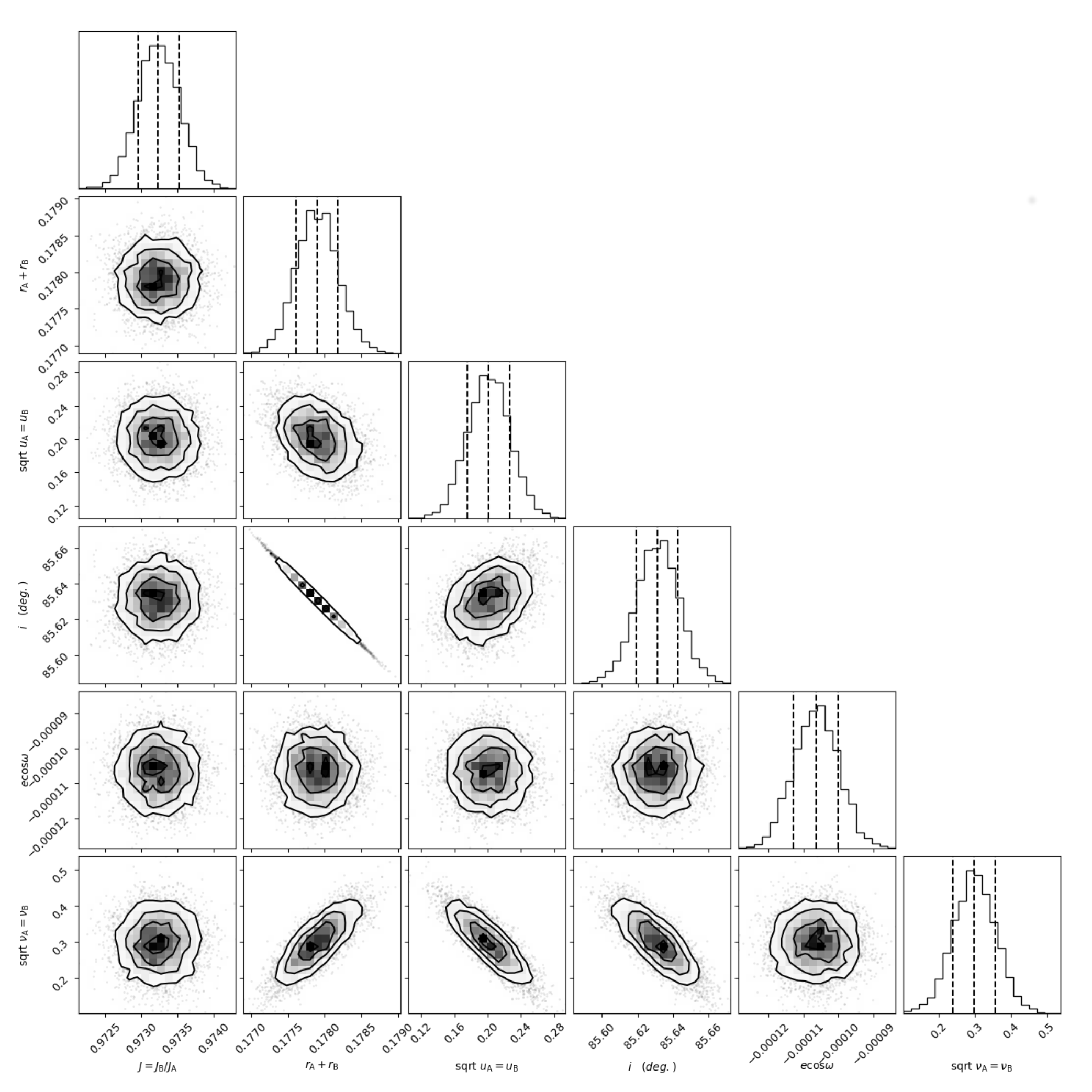}
\caption{Parameter uncertainty covariances and parameter parent distributions of light curve parameter based on {\it TESS} data analysis. We show the results from {\sc jktebop} task8 (MC, 5000 samples). We plot 16\%, 50\% and 84\% quantiles for each histogram and $1\sigma, 2\sigma$ and $3\sigma$ confidence regions as contours. A strong negative correlation is found between the inclination $(i)$ and sum of fractional radii $(r_A + r_B)$. We omit the explicit calculation of the correlation coefficients. The orbital period $(P)$ and reference epoch $(T_0)$ have been omitted since their covariance has been minimized by construction already. We do not show the results from PB-RP due to a low number of samples resulting in technical complications in rendering correct confidence limits. However, parameter covariances between MC and PB-RP are similar. The plots were generated with {\sc corner} \citep{corner}.}
\label{fig:corner_lc_task8}
\end{figure}

\begin{figure}[hbt!]
\centering
\includegraphics[width=1.0\textwidth]{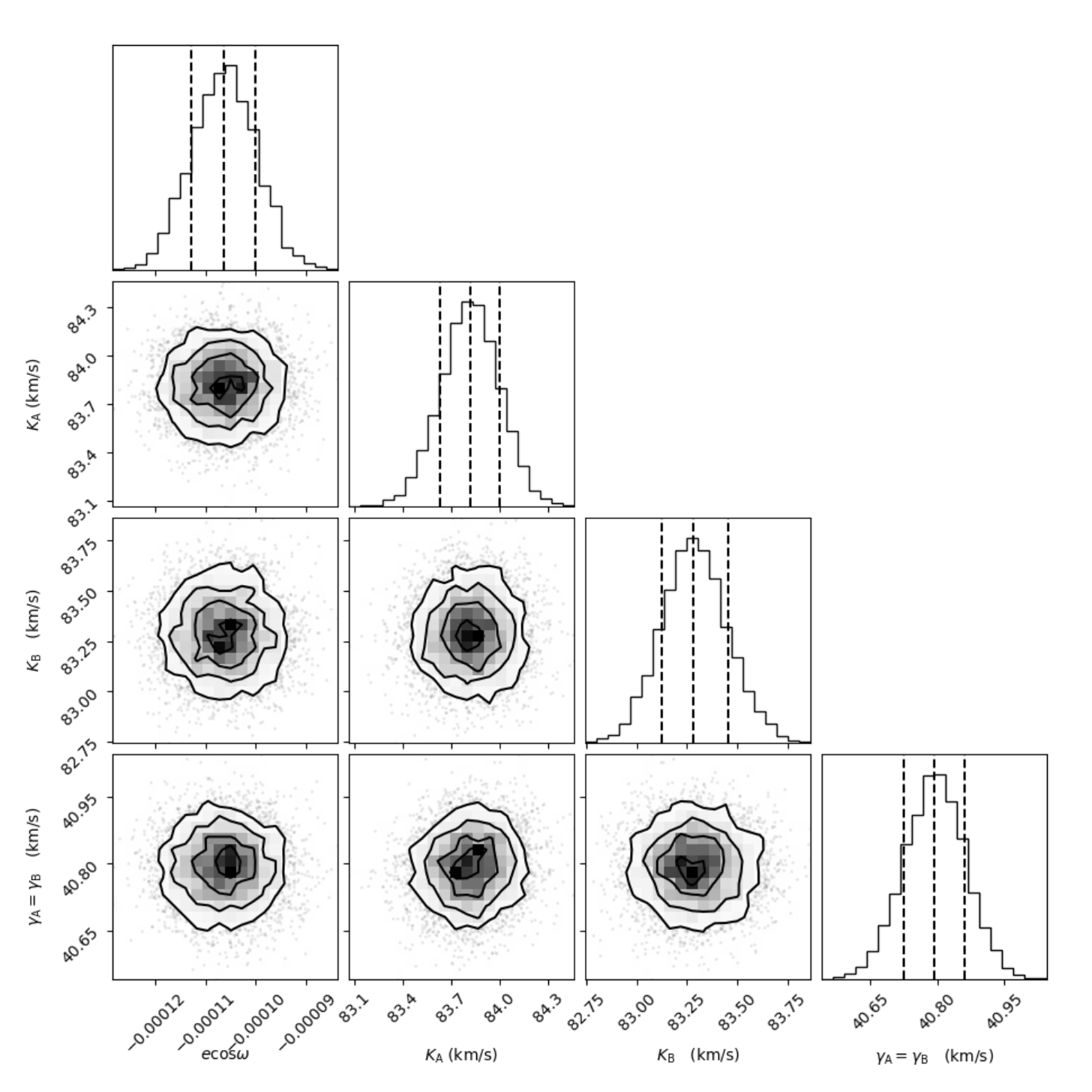}
\caption{Same as Fig.~\ref{fig:corner_lc_task8} but this time we consider parameters extracted from FEROS RV data.}
\label{fig:corner_rv_task8}
\end{figure}

\end{appendix}

\end{document}